\documentclass[12pt]{iopart}
\usepackage{iopams}
\expandafter\let\csname equation*\endcsname\relax
\expandafter\let\csname endequation*\endcsname\relax
\usepackage{amsmath}
\usepackage{amsfonts}
\usepackage{amssymb,bm}
\usepackage{mathtools}
\usepackage{mathrsfs}
\usepackage{pzccal}
\usepackage[dvipsnames]{xcolor}
\usepackage{graphicx}
\usepackage{tensor}
\usepackage{upgreek}
\usepackage[hidelinks]{hyperref}
\usepackage{tikz}
\usetikzlibrary{positioning}
\usetikzlibrary{tikzmark,fit}
\usetikzlibrary{er,positioning}
\usetikzlibrary{decorations.pathreplacing}
\usetikzlibrary{decorations.pathmorphing,patterns}
\usetikzlibrary{calc,patterns,decorations.markings}
\usetikzlibrary{shapes.geometric}
\usepackage[makeroom]{cancel}
\usepackage{appendix}
\usetikzlibrary{decorations.pathmorphing}
\tikzset{snake it/.style={decorate, decoration=snake}}
\usepackage{subcaption}
\usepackage[perpage,symbol*]{footmisc} % renew the footnote symbol on each page - error fixing
\usepackage[sort&compress,numbers]{natbib} 
\DeclareSymbolFont{matha}{OML}{txmi}{m}{it}% txfonts
\newcommand{\one}{\scalebox{0.5}{{\textcircled{1}}}}
\newcommand{\two}{\scalebox{0.5}{{\textcircled{2}}}}
\def\inner#1{\langle #1 \rangle}

\definecolor{MyGreen}{rgb}{0.04, 0.475, 0.435}
\definecolor{MyLightGreen}{rgb}{0.67, 0.94, 0.82}
\definecolor{MyGray}{rgb}{0.70, 0.70, 0.70}
\definecolor{MyRed}{rgb}{0.77, 0.12, 0.23}
\def\centerarc[#1](#2)(#3:#4:#5)% Syntax: [draw options] (center) (initial angle:final angle:radius)
    { \draw[#1] ($(#2)+({#5*cos(#3)},{#5*sin(#3)})$) arc (#3:#4:#5); }
\begin{document}

\title[Gaussian beams and caustic avoidance in gravitational optics]{Gaussian beams and caustic avoidance in gravitational optics}

\author{Nezihe Uzun}

\address{Center for Theoretical Physics, Polish Academy of Sciences, Al. Lotników 32/46, 02-668 Warsaw, Poland}
\ead{nuzun@cft.edu.pl}

\begin{abstract}
In this study, we consider a beam summation method adapted from the semiclassical regime of quantum mechanics to study the classical properties of thin light bundles in gravity. In Newtonian paraxial optics, this method has been shown to encapsulate the wave properties of the light beams. In our case, the wave function assigned to the light bundle can be viewed as a coarse-grained description that captures information about the dynamics of superposed bundles within the geometric optics regime. We investigate two solutions of the null bundle wave function that differ by their origin: (i) a point source and (ii) a finite source. It is shown that while the wave function in the point source case contains the same information as the standard thin null bundle framework, the finite source case corresponds to a Gaussian beam. The novel aspect of this work arises from our geometric construction of covariant Gaussian beams, which can be applied in any spacetime. Additionally, the effects of a finite source on cosmological distances are discussed. With this framework, one can model light propagation from coherent sources while avoiding the mathematical singularities of the standard thin null bundle formalism. We explicitly demonstrate the caustic-avoidance property of Gaussian beams in the analytically tractable example of a Barriola–Vilenkin monopole spacetime.
\end{abstract}

\vspace{2pc}
\noindent{\it Keywords}: gravitational optics, cosmology, luminosity distance, Gaussian beams, general relativity.

%\submitto{\CQG}
%\maketitle
%

\maketitle
%%%%%%%%%%%%%%%%%%%%%%%%%%%%%%%%%%%%%%%%%%%%%%%%%%%%%%%%%%%%%%%%%%%%

\section{Introduction}
Determining the exact solutions of Maxwell's equations defined on a generic curved spacetime is known to be a formidable task. Light propagation is often studied in terms of its geometric optics limit in astrophysical and cosmological applications. Essentially, this is a sufficiently good approximation, as the characteristic length scales in cosmos are much larger than the typical wavelength of light. 

On the other hand, geometric optics limit fails to operate at caustics. In that case, one   needs to go beyond the ray picture in order to propagate light, estimate brightnesses and calculate distances. Accordingly, many studies have been introduced in the literature for the last few decades that go beyond the geometric optics limit and which focus on the wave properties of fields \footnote{The gravitational field can also be considered in this respect as in the linear regime the propagation vectors of electromagnetic and gravitational waves follow null geodesics in the high frequency limit \cite{Misner:1973}.} on curved spacetimes \cite{Anile:1976gq, Gosselin:2007, Frolov:2011,Harte:2018wni,Harte:2019tid,Frolov:2020uhn, Oancea:2020, Andersson:2020gsj,Dolan:2018ydp,Koksbang:2022}.

In the current work, we provide a method to study the wave-like effects of light bundles to avoid caustics by using phase space techniques adapted from quantum mechanics. This is different in construction compared with the aforementioned studies. We mainly focus on the geometry of null congruences rather than on solutions defined on individual null geodesics. This choice is based on the fact that physically meaningful quantities, such as distances, brightnesses, or image distortions are defined in a covariant manner only via null bundles in general relativity. In addition, we focus on Gaussian beams which are studied only by a few in the context of field propagation on curved spacetimes (cf. \cite{Sbierski:2013mva,Torres:2022}). 

Caustics in general relativity can be studied in two main categories regarding the global or local properties of wavefronts \cite{Perlick:2010}: (i) null cone caustics and (ii) thin null bundle caustics. The former are relevant when one studies the set of \textit{all} points where a lightlike submanifold of a spacetime ceases to be an immersed submanifold \cite{Hasse:1996}. The latter are relevant for our current work, in which the main focus is on \textit{instantaneous wavefronts} and the geodesic deviation equation of the thin null bundles. What we would like to highlight here is that the existence of caustics is usually neglected in cosmological light propagation calculations with only a few exceptions. For example, previously, it was argued that sources at high redshifts are expected to be demagnified due to the effects of caustics \cite{Dyer:1988,Holz:1997ic,Hadrovic:1997fi,Ellis:1998qga}. As a result of this, the angular diameter distances are expected to be different than the one of the standard homogeneous and isotropic cosmological model. In \cite{Ellis:1998ha}, the authors estimated light caustics to be in the order of around $10^{22}$ on the past null cone until the surface of last scattering within a realistic, inhomogeneous universe. The main outcome of that study was that with the inclusion of caustics, the all-sky-average of the angular diameter distance is estimated to be higher by a significant amount when compared to the distance calculations in the standard cosmology.

In the current work, we suggest a method to calculate intensities and distances while avoiding the caustic singularities. Specifically, we focus on the \textit{Gaussian beam} solutions which avoid the mathematical singularities of the standard light bundle propagation. Those solutions are sometimes referred to as \textit{Gaussian wave packets} in the literature. However, in the current work, we reserve the two types of naming for different constructions. The reason for such a distinction will be more clear once we compare the works in the literature with ours.

In order to make the introduction coherent, we first remind the geometric optics limit in Section \ref{sec:Geometric optics limit}. We summarize the Gaussian wave packet approximation that has already been used in the literature in Section \ref{sec:Work in the literature}. We believe the motivation and the methodology of our work, outlined in Section \ref{sec:Current work}, can be better understood in relation to those works in the literature. 

The synopsis of the rest of the paper is as follows. The text book preliminaries of the null bundle geometry and its evolution equations are presented in Section \ref{sec:Preliminaries} in relation to the geodesic deviation equation. The current work is based on our previous work, \cite{Uzun:2018}, in which the Hamiltonian phase space formulation of the observable null bundles was presented. We provide a brief summary of the aforementioned work in Section \ref{sec:Summary of [30]: Reduced Hamiltonian optics}. The reader who is familiar with the standard null bundle formulation and our previous work, \cite{Uzun:2018}, can skip those sections. 

The main body of the paper starts at Section \ref{sec:Wavization} in which the quantization techniques of the semi-classical physics are adapted to study the classical evolution of the null bundle wave function. This wave function obeys an equation similar to the Schr{\"o}dinger equation and it is referred to as the \textit{paraxial wave equation} in the Newtonian optics. In our case, this equation dictates how superposed bundles, each defined in the geometric optics limit, evolve. The solution of the so-called paraxial wave equation for a wave function initiated from a point source  is presented in Section \ref{sec:Point sources}. It is observed that the evolution equation of the wave function contains the same set of information as the standard thin null bundle evolution. In Section \ref{sec:Gaussian beams}, we introduce Gaussian beams as a solution of the paraxial wave equation which initiates from a small yet finite extend. We should note that in dimensions higher than one, Gaussian beams are usually given in their general form in equation (\ref{eq:Gaussian_form}). What is different in the current paper is that we provide a specific covariant geometric construction starting from a specific set of initial conditions relevant for observations in cosmology. The welldefinedness of this geometric construction is proven in \ref{sec:Welldefinedness of the Gaussian beam for finite sources}. 

Comparison of the Gaussian beams with the point source bundles shows that the equations of motion of the former are different than the ones of the latter by the gradient of an extra term. This potential-like term is statistical in nature and it is responsible for the coupling of the amplitude of the wave function with its phase. It vanishes once the point source limit is taken. We provide qualitative and quantitative analyses in Section \ref{sec:Cosmological distances and caustic avoidance} where we investigate the effect of caustics on cosmological distances. In addition, an analytically tractable example of caustic avoidance for a Barriola-Vilenkin monopole is provided. The key insights of our framework are discussed in Section \ref{sec:Key Insights and Limitations}. Those include the expansion of the current formalism to study generic wavefronts in relation to gravitational lensing studies. We also include \ref{sec:Evolution of Gamma and Q of a Gaussian beam} and \ref{sec:Determination of the principal curvatures and the widths} in order to make our paper self-contained. Finally, we conclude with Section \ref{sec:Conclusions} which is essentially a summary of the current work. 
%%%%%%%%%%%%%%%%%%
\subsection{Geometric optics limit}\label{sec:Geometric optics limit}
Let us consider a $4$-dimensional spacetime manifold $\left(M,g\right)$ where $g$ is the Lorentzian metric with signature $\left( -, +, +, +\right)$. Let an associated covariant derivative operator be denoted by $\mathcal{D} _{\mu}$ with respect to the spacetime coordinates $x^\mu$. We denote the inner product of two objects, with respect to the spacetime metric, via the angled brackets, $\inner{.,.}$. In order to denote 4-dimensional vectors in their abstract form, we use the over-arrow, $\Vec{}$ .

The Maxwell field is often studied by a method similar to the ones of Jeffreys \cite{Jeffreys:1925}, Wentzel \cite{Wentzel:1926}, Kramers \cite{Kramers:1926} and Brillouin \cite{Brillouin:1926blg} (The JWKB method from thereon.). The main idea behind this method is to expand the electromagnetic field tensor (or the associated vector potential) with respect to a smallness parameter in order to find the approximate solutions of Maxwell's equations. When the smallness parameter is taken to be the wavelength of light, the leading order part of the solution is known as the geometric optics limit.

In order to show this explicitly, let us start with introducing the skew-symmetric electromagnetic field tensor, $\tensor{\mathcal{F}}{_\mu_\nu}$. The
Maxwell equations in curved spacetime are represented by the following set of equations,
\begin{equation}\label{eq:Maxwell}
\mathcal{D} _{\beta}\tensor{\mathcal{F}}{^\alpha ^\beta}=0,\qquad {\rm{with}}\qquad \mathcal{D} _{\left[{\mu}\right.}\tensor{\mathcal{F}}{_\alpha _{\left. \beta\right]}}=0,
\end{equation}
in the absence of electromagnetic sources. Here, the square brackets in the second equality represent the anti-symmetrization operator. As the field tensor satisfies the Bianchi identity, it can be expressed by a locally exact form. This allows it to be  represented in terms of a complex vector potential, $\Vec{\mathcal{A}}$, i.e.,
\begin{equation}\label{eq:Fraday_vector_potential}
\tensor{\mathcal{F}}{_\mu_\nu}=\mathcal{D} _{\mu}\mathcal{A}_{\nu}-\mathcal{D} _{\nu}\mathcal{A}_{\mu}.
\end{equation}
Substitution of (\ref{eq:Fraday_vector_potential}) in to (\ref{eq:Maxwell}) under the Lorenz gauge\footnote{Indeed, one also requires the harmonic condition, $\mathcal{D} ^{\mu}\mathcal{D} _{\mu}f=0$, in the gauge transformation $\mathcal{A}_{\nu}\rightarrow \mathcal{A}_{\nu} + \mathcal{D} _{\nu}f$, for an arbitrary function $f$, in order to fix a potential.}, $\mathcal{D} _{\mu}\mathcal{A}{^\mu}=0$, results in the following wave equation
\begin{equation}\label{eq:Maxwel_w_Vector_potential}
\mathcal{D} ^{\nu}\mathcal{D} _{\nu}\mathcal{A}^{\mu}-\tensor{R}{^\mu_\nu} \mathcal{A}^{\nu}=0,
\end{equation}
where $\tensor{R}{_\mu_\nu}$ is the Ricci tensor. In order to reach the geometric optics limit via a JWKB-like method, one can start by assigning an \textit{ansatz} for the Faraday tensor, for example, 
\begin{equation}\label{eq:Faraday_tensor_ansatz}
    \tensor{\mathcal{F}}{^\mu^\nu}={\rm{Re}} \left\{\tensor{\mathcal{f}}{^\mu^\nu}\exp{i\mathcal{S}/\epsilon}\right\} \,\,\,{\rm{with}}\,\,\, \tensor{\mathcal{f}}{^\mu^\nu}=\sum_{n=0} \epsilon ^n\tensor{\mathcal{f}}{_n^\mu^\nu}.
\end{equation} 
Here, ${\rm{Re}} \left\{\cdot\right\}$ signals that one should take the real part of a given object and $\epsilon$ is the so-called \textit{smallness} or \textit{book-keeping} parameter. When the ansatz (\ref{eq:Faraday_tensor_ansatz}) is substituted into Maxwell's equation, (\ref{eq:Maxwell}), the smallness parameter allows one to obtain a series solution order-by-order in an iterative manner for each value of $n$. Note that the tensor components $\tensor{\mathcal{f}}{_\mu_\nu}$ can take complex values in general. However, the phase function $\mathcal{S}$ that appears in (\ref{eq:Faraday_tensor_ansatz}) is restricted to be \textit{real}. Such a restriction results in the approximate solutions being represented by \textit{locally plane waves}. The wave vector, $\vec{k}$, of such a locally plane wave is a gradient field of the phase function, i.e., $k_\mu=\mathcal{D} _\mu \mathcal{S}$.
Accordingly, the wave vector in the above is defined in the \textit{real} domain. We will return to this fact in further discussions of the current section. For now, let us emphasize that assigning an ansatz on the Faraday tensor allows one to directly obtain the gauge invariant solutions.

Alternatively, one can also start with assigning a similar ansatz for the electromagnetic vector potential, $\Vec{\mathcal{A}}$. In that case, one needs to additionally impose the Lorenz gauge. Nevertheless, the solutions obtained through this alternative method is gauge invariant as well. For instance, let us consider the following ansatz for the vector potential,
\begin{equation}\label{eq:Vector_potential_ansatz}
\Vec{\mathcal{A}}= {\rm{Re}} \left\{\Vec{\alpha}\, e^{i\mathcal{S}/\epsilon}\right\}, \,\,\,{\rm{with}}\,\,\, \Vec{\alpha}=\sum_{n=0} \epsilon ^n\Vec{\alpha}_n.
\end{equation}
Here, $\mathcal{S}$ is again the real phase function and we neglect the tail terms\footnote{Tail terms in $\vec{A}$ represent the contribution of the electromagnetic field which propagate off the null cone, i.e., they represent timelike propagation. Those terms do not contribute to the electromagnetic field propagation only in (conformally) flat or (conformally) plane-wave spacetimes. Nevertheless, they are usually neglected in the literature in arbitrary spacetimes as well. See \cite{Friedlander:1975, Ellis:2012, Harte:2013dba}  for more details.}.
 Once one substitutes the lowest order of the ansatz in (\ref{eq:Vector_potential_ansatz}) into Maxwell's equation, (\ref{eq:Maxwel_w_Vector_potential}), and reads off the coefficients of the first three lowest orders in the resultant equation, one gets
 \begin{equation}\label{eq:Lowest order_Maxwell}
\inner{\vec{k},\vec{k}}=0,\qquad \mathcal{D}_{\vec{k}}\,\vec{\alpha}{_0}=-\frac{1}{2}\left(\mathcal{D}_\mu k^\mu\right)\vec{\alpha}{_0}, \qquad \mathcal{D}_\mu\mathcal{D} ^\mu \tensor{\alpha}{_0^\nu}-\tensor{R}{^\nu_\mu} \tensor{\alpha}{_0^\mu}=0,
\end{equation}
with the corresponding Lorenz gauge conditions,
\begin{equation}\label{eq:Lorenz_lowest_order}
\inner{\vec{k}, \vec{\alpha}_{0}}=0, \qquad \mathcal{D} _\mu \tensor{\alpha}{_0^\mu}=0.
\end{equation}
Then, in the geometric optic limit \cite{Misner:1973,Schneider:1992, Ellis:2012,Dolan:2018}, 
\begin{itemize}
\item [(i)] The wave vector is null due to the first equality of (\ref{eq:Lowest order_Maxwell}). In addition, as it is a gradient field, it satisfies
\begin{equation}
     \mathcal{D} _{\vec{k}} \,\vec{k}=0, \qquad {\rm{with}} \qquad k^\mu=\frac{dx^\mu}{dv}.
\end{equation}
That is, $\vec{k}$ is tangent to a real, null geodesic
where $v$ is the affine parameter that parametrizes the null curve.
\item [(ii)] Polarization vector, 
\begin{equation}\label{eq:Polarization_vector}
\Vec{V} = \frac{\Vec{\alpha}{_0}}{\alpha }, 
\end{equation}    
is \textit{transverse} to the propagation vector $\vec{k}$ due to the Lorenz gauge condition, (\ref{eq:Lorenz_lowest_order}). Here, we denote $\alpha=\inner{\Vec{\alpha}{_0},\Vec{\bar{\alpha}}{_0}}^{1/2}$.
\item [(iii)] According to the third equality of (\ref{eq:Lowest order_Maxwell}), the leading order part of the solution, $\Vec{\alpha}_0$, satisfies the same wave equation $\Vec{\mathcal{A}}$ obeys. Note that the information about the curvature of the underlying generic spacetime manifests itself not only through the covariant derivative operator but also through the Ricci tensor in (\ref{eq:Lowest order_Maxwell}).
\item [(iv)] The squared amplitude of the wave obeys
\begin{equation}\label{eq:conserv_photon_flux_density}
\mathcal{D} _\mu \left(\alpha^2 k^\mu\right)=0,
\end{equation}
due to the \textit{locally} conserved photon flux density, $\Vec{j} \propto \alpha^2 \Vec{k}$. 
\item [(v)] Polarization vector $\Vec{V}$ in (\ref{eq:Polarization_vector}), is parallel propagated throughout the propagation, i.e., $\mathcal{D}_{\Vec{k}}\, \Vec{V}=0$, due to the second equality in (\ref{eq:Lowest order_Maxwell}) and (\ref{eq:conserv_photon_flux_density}). 
\item [(vi)] The electromagnetic stress-energy tensor, $T^{\rm{EM}}_{\mu \nu}$, is approximated as the one of a null dust, i.e., $T^{\rm{EM}}_{\mu \nu}\propto \alpha ^2 k_\mu k_\nu.$
\end{itemize}
%%%%%%%%%%%%%%%%%%
\subsection{Work in the literature: Gaussian wave packets near null geodesics}\label{sec:Work in the literature}
Investigation of wave equations on Lorentzian manifolds has been a long lasting problem and Maslov's complex JWKB-like method is known to be the pillar of the most of the existing methods \cite{Maslov:1994}. This method is mostly known in semiclassical physics \cite{Buldyrev:1981, Belov:1989, Shapovalov:1999, Belov:2002, Babich:2009}.  When it comes to the Gaussian solutions obtained through such asymptotic techniques, Ralston's construction \cite{Ralston:1982} is the most well-known one in the mathematics community in which the aim is to solve a generic massless scalar field wave equation, $\mathcal{D}_\mu \mathcal{D} ^\mu {\mathcal{A}}=0$.
The idea behind this method is to consider a Gaussian wave ansatz rather than the locally plane wave ansatz outlined in the previous subsection. Namely, one first considers 
\begin{equation}\label{eq:Gaussian_ansatz}
{\mathcal{A}}= \left\{{\alpha}\, e^{i\mathcal{\tilde{S}}/\epsilon}\right\}, \,\,\,{\rm{with}}\,\,\, {\alpha}=\sum_{n=0} \epsilon ^n{\alpha}_n.
\end{equation}
Here, $\epsilon$ represents the smallness parameter as before. In addition, both ${\alpha}_n$ and $\mathcal{\tilde{S}}$ are \textit{complex} as opposed to the case of the locally plane wave ansatz in (\ref{eq:Vector_potential_ansatz}). Then, one defines a bicharacteristic, whose projection on the spacetime is a null geodesic, $\gamma \left(x^\mu(v)\right)$. The complex eikonal, $\mathcal{\tilde{S}}$, takes real values on the null geodesic and its complex part grows off the bicharacteristic. This allows one to have a decaying amplitude off the bicharacteristic if $\mathcal{\tilde{S}}$ is restricted to satisfy ${\rm{Im}}\left\{\mathcal{D}_\mu \mathcal{D} _\nu \mathcal{\tilde{S}}\right\}> 0$, at the intersection of the null geodesic, $\gamma$, with the spatial sections of the spacetime. 

The next curiosity is the long time behavior of such solutions. In \cite{Sbierski:2013mva}, Sbierski adapts the energy method of Morawetz \cite{Morawetz:1968} in order to study the Gaussian solutions of the scalar wave equation. In our view, this is a restricted starting point to estimate the Gaussian wave solutions of an electromagnetic field in general spacetimes due to the absence of the Ricci term and the scalar nature of the equation as opposed to Maxwell's equation, (\ref{eq:Maxwel_w_Vector_potential}). Nevertheless, an energy function is assigned to the Gaussian wave packets via,
\begin{equation}\label{eq:Energy_function_Sbierski}
E=\int _{\Sigma}\tensor{T}{_\mu_\nu}u^\mu n^\nu d\Sigma,
\end{equation}
where $\tensor{T}{_\mu_\nu}$ is the stress-energy tensor of the field in question, $\vec{u}$ is the observer 4-velocity, $\Sigma$ corresponds to the spacelike hypersurfaces and $\vec{n}$ is its normal vector. Clearly, this method requires the global hyperbolicity of the underlying spacetime. This is typical for many studies of partial differential equations defined on Lorentzian manifolds due to its relation to the Cauchy problem. Then, \textit{approximate} Gaussian solutions of the wave equation is defined through the so-called energy function in (\ref{eq:Energy_function_Sbierski}). For Gaussian wave packets, $E \approx \inner{\vec{u},\vec{k}}$ holds in some neighborhood of the null geodesics where ${\rm{Im}\gamma}\cap \Sigma$ \cite{Sbierski:2013mva}. Here, we use $\vec{k}$ to indicate the tangent vector of the null curve in question. In that case, the energy function, $E$, of the field is expected to remain in some relatively compact region of space similar to, for example, trapped surfaces or photon spheres of black holes in some finite time. Note that local conservation of the photon flux density in (\ref{eq:conserv_photon_flux_density}) is \textit{approximately} satisfied in that case. 

One of the open questions in relation to Ralston's method seems to be the ontology of the bicharacteristics in question. Even though one can interpret its projection on the spacetime as a real null geodesic, the representation of its complex section is an open question. The study of generic, complex worldlines on Lorentzian manifolds is a vast subject. Their physical interpretation seems to be relatively well understood only in the Minkowski spacetime \cite{Newman:2005ph, Adamo:2009, Adamo:2010}.

We should also mention another method of construction of Gaussian wave packets on curved spacetimes, even though their physical representation is different than the one of the Maxwell field. In \cite{Torres:2022}, Torres \textit{et al.} study the Gaussian beams of ocean waves in analogy with the gravitational waves. The reason we use the word ``beam'' here is that the authors use the locally plane ansatz rather than the Gaussian ansatz when defining the propagation path of the waves in question. Namely, they adapt the method of Popov \cite{Popov:1982} to define \textit{complex rays}. Those rays are obtained from the Hamiltonian formulation of the geometric optics limit which starts from assuming a locally plane wave ansatz. In the end, a 1-dimensional Gaussian profile is assumed to represent a scalar field. This is obtained through assigning \textit{complex initial conditions} on the Hamilton equations of the rays. The method of Torres \textit{et al.} is similar to the one in our work due to the underlying Hamiltonian formulation. What is different between the two approaches is the level of coarse-graining while studying the field propagation. Torres \textit{et al.} focus on single ray path and its Hamiltonian formulation whereas in the current work, we focus on the ray bundles and their phase space formulation.

In this respect, we can summarize this section as the following. The work of Sbierski \cite{Sbierski:2013mva} seems to represent a more fine-grained picture and thus, the author's method capture genuine wave effects. As the work of Torres \textit{et al.} incorporates complex rays defined through the locally plane wave ansatz, an additional level of coarse-graining is introduced. Our method here is the least granular among them as it focuses on the behaviour of a collection of null geodesics rather than just one ray path.
%%%%%%%%%%%%%%
\subsection{Scope of the current work: Clarifications and distinctions}\label{sec:Current work}
In general relativity, physically meaningful quantities are obtained through the relative motion of test particles and the measurements are done at the observer's local frame. When we make an observation on the sky, we receive information through a collection of light-like particles. This information resides on a null congruence rather than an individual geodesic. Therefore, in real life applications, light propagation is studied in a coarse-grained picture and it is the geodesic deviation equation of the bundle 
that represents the collective dynamics of a light beam.

When one speaks about the geodesic deviation equation in relativity one usually refers to its linear order approximation. To be more specific, connecting two curves requires the introduction of a bi-local object, i.e., a bi-vector that depends on two spacetime points. Only if the neighboring geodesics are very close to each other, then one can take the local limit and assign a connecting vector to this bundle. In the end, the connecting vector is represented by a Jacobi field, i.e., it satisfies the leading order part of the geodesic deviation equation (See \cite{Vines:2014} for a detailed analysis.). 

In a previous work \cite{Uzun:2018}, we studied observable thin null bundles via a Hamiltonian formalism. Starting form an action principle for the geodesic deviation vector, we showed that its dynamics can be studied on a reduced phase space. The phase space vector in question is constructed through the physically meaningful components of the geodesic deviation vector, i.e., its 2-dimensional projections on a local screen and the corresponding derivatives along the null propagation vector of the bundle. With this method, we were able to bring the problem of \textit{thin null bundle} propagation to a form which is in complete analogy with the \textit{paraxial limit} of the Newtonian optics. 

In the standard literature of the paraxial Newtonian optics, there exists a symplectic matrix which transforms the initial transverse positions and the angles of a ray to their final values. One studies the properties of an optical device (or an inhomogeneous medium) through the properties of this transformation matrix. In our formalism \cite{Uzun:2018}, we also obtained a symplectic matrix which transforms the initial screen-projected deviation vector and its derivatives to their final values. One can study the optical properties of a spacetime in analogy with optical devices via this symplectic transformation matrix. 

 In the current work, we extend this analogy to study the wave-like effects of a null bundle. For this, we assign a classical wave function to the thin null bundle. Note that in the geometric optics limit, a congruence of light rays is represented by a null dust \cite{Schneider:1992,Ellis:2012}, thus we can assume that the wave function represents non-interacting light-like particles. In our framework, we use the correspondence between the symplectic matrices and the metaplectic operators \cite{Moshinsky:1971, Wolf:1979, Littlejohn:1985}. This means that for every symplectic null bundle transformation matrix, there exists an integral kernel that transforms the initial wave function to a final one. The resultant wave function obeys a Schr{\"o}dinger-like equation which is known as the \textit{paraxial wave equation} in the Newtonian optics. We obtain the Gaussian solutions of this equation. Similar to the paraxial Newtonian optics, we have two smallness ``parameters'' in our construction. First one is the wavelength of light, $\lambda$, as in the geometric optics regime. The second one is the width of the Gaussian beam.  However, in our case, the latter can not simply be referred to as a parameter since it is obtained through the deviation vector, $\vec{\xi}$, of the null bundle as we will see in Section \ref{sec:Gaussian beams}. It is a dynamical variable. In other words, with $L$  being the characteristic length scale corresponding to the curvature of a spacetime, we have the following scale hierarchy, 
 \begin{equation}
\lambda \ll  |\vec{\xi}| \ll L \qquad\forall \, v.
 \end{equation}
Here, $v$ is the affine parameter that parametrizes the geodesics of the null bundle. We then develop a geometric set up in which Gaussian beams can be used to study the finite source effect in a relativistic context. Next, we show that those solutions avoid caustics, making the distance estimations possible beyond the singular points of the standard null bundle propagation approach.

Now, we would like to clarify what this work is and is not about with the following list.
\begin{itemize}
\item [(i)] The method we propose here is not completely new. Rather, it has been known since the 1960s that the phase space quantization techniques can be adapted to paraxial optics to recover some of the wave behavior of light in the classical regime \cite{Dragoman:2002}. The application of somewhat similar techniques to study wave properties of light-like or time-like particles in Riemannian space and spacetime was considered before \cite{Buldyrev:1981, Dahl:2006,Dahl:2007}. Those works follow coordinate approaches. Often, there exists a predefined time parameter and accordingly an inherent slicing of the underlying spacetime to study the evolution. Neither the covariance, nor the relevance of the results for the standard cosmological observables are discussed. What is new in the current work is the application of similar techniques to gravitational optics in a covariant manner while keeping track of the cosmological observables measured in a local frame.

\item [(ii)] We assign a wavefunction to the entire light beam. This wavefunction involves information about both the Maxwell field and the dynamics of the null bundle deviation. 

\item [(iii)] We try to be careful in distinguishing what is a genuine wave effect and what is a wave-like effect throughout the paper. The former seems to be more relevant for, for example, approaches that go beyond the geometric optics limit \cite{Anile:1976gq, Gosselin:2007, Frolov:2011,Harte:2018wni,Harte:2019tid,Frolov:2020uhn, Oancea:2020, Andersson:2020gsj,Dolan:2018ydp,Koksbang:2022} or Ralston and Sbierski's method \cite{Sbierski:2013mva}, which starts from the Gaussian ansatz for a null field and satisfies the local photon number conservation approximately (See  Section \ref{sec:Work in the literature}.). In all of those approaches, one considers the approximate solutions of the Maxwell field at a point or along the integral curves of the propagation vector. In our case, we are after the behavior of a collection of light rays which are defined in the geometric optics regime. We will see in Section \ref{sec:Gaussian beams} that the method we propose here essentially provides a technique to superpose null bundles without allowing them to interact. Meaning, (i) the null dust approximation of the electromagnetic stress-energy tensor is preserved, (ii) the locally plane wave ansatz, that is associated with the individual rays of the bundle, remains applicable. Moreover, the power contained in the bundle, i.e., intensity times the cross-sectional area, is conserved throughout the propagation similar to the case in the standard approach. In this respect, one can use our method to study the statistical nature of a null congruence of rays rather than to capture genuine wave effects. One can refer to Figure \ref{fig:Coarse_graining} to compare the different coarse-graining levels discussed here. 
\begin{figure}[ht]
    \centering
    \includegraphics[width=1\textwidth]{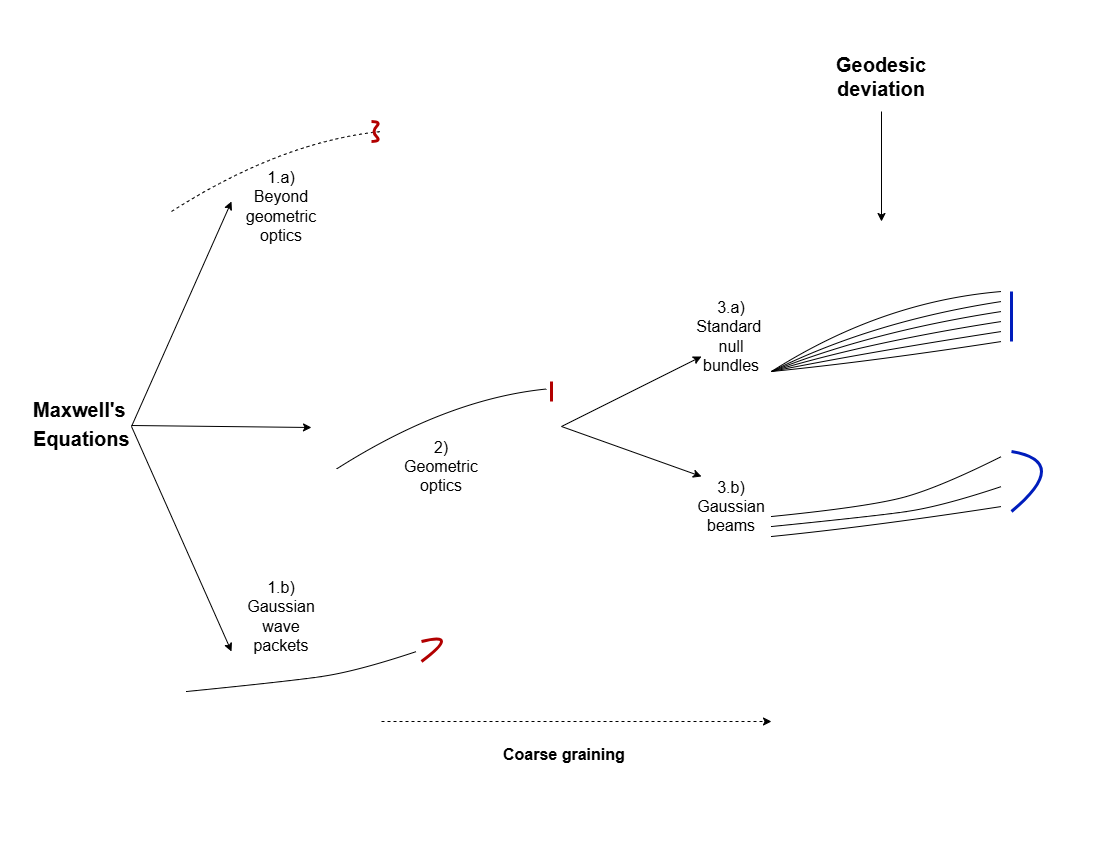}
    \caption{Coarse-graining levels of light propagation. 1a) and 1b) involves the (complex) JWKB method to study the approximate solutions of the Maxwell equation as in \cite{Anile:1976gq, Gosselin:2007, Frolov:2011,Harte:2018wni,Harte:2019tid,Frolov:2020uhn, Oancea:2020, Andersson:2020gsj,Dolan:2018ydp,Koksbang:2022, Sbierski:2013mva, Torres:2022}. Level 2) is the geometric optics limit studied with a locally plane  wave ansatz assigned on the points of a single geodesic. Level 3a) involves standard light bundles where the wave function of the central ray is assumed to represent the entire bundle. The intensity profile is homogeneous. Our work involves 3a) and 3b) in which the latter is defined through the superposition of light rays. The intensity profile is Gaussian.}
    \label{fig:Coarse_graining}
\end{figure}
\item [(iv)] When the Hamiltonian formulation of light ray propagation is addressed, it is customary to consider the Hamiltonian $H(\vec{x},\vec{p})=g^{\mu \nu}p_\mu p_\nu/2$ in vacuum \cite{Perlick:2000}. Here, $\vec{p}$ is the 4-momenta of the photon and $\vec{x}$ represents the spacetime coordinates. In cosmology, for example, after a relevant $3+1$ decomposition of the spacetime coordinates, a 6-dimensional phase space is constructed via the induced spatial coordinates and the 3-momenta of the photon \cite{Ellis:2012}. In the current work, the Hamiltonian in question is a reduced Hamiltonian assigned to the transverse space of the null bundle (See Section \ref{sec:Summary of [30]: Reduced Hamiltonian optics}). The canonical positions and momenta are chosen through the screen-projection of the deviation vector and its derivative along the central null curve of the bundle. Evolution parameter is not a timelike parameter, rather it is an affine parameter that parametrizes the congruence. Thus, our approach is different than seemingly similar approaches in the literature both mathematically and physically. 
\item  [(v)] The mathematical novelty of this work follows from the geometric formulation of Gaussian beams constructed from certain initial conditions for any spacetime. This is outlined in Section \ref{sec:Finite sources}. We prove the welldefinedness of those Gaussian beams in \ref{sec:Welldefinedness of the Gaussian beam for finite sources}.  As we will explicitly show in Section \ref{sec:Cosmological distances and caustic avoidance}, the Gaussian beams avoid caustics. We refer to this property as a wave-like effect as the standard bundles are known to have singularities at the caustic points/lines. 
\item  [(vi)] The Gaussian intensity profiles discussed here are not expected to be \textit{directly} observable. We will discuss in Section \ref{sec:Fourier transform vs. Gaussian beam decomposition} that Gaussian beams can be treated as effective elementary waveforms to decompose arbitrary wavefronts in generic spacetimes. This serves as an alternative to locally plane waves which are used as the elementary waveforms of Fourier transforms to decompose a field in (conformally) flat spacetimes in the realm of gravitational lensing.
\end{itemize}
%%%%%%%%%%%%%%%%%%%%%%%%%%%%%%%%%%%%%%%%%%%%%%
\section{\label{sec:Preliminaries} Preliminaries}
In this section, we give a brief summary of the preliminaries required to follow the rest of the paper:
\begin{itemize}
\item[(i)] In Section \ref{sec:The orthonormal tetrad and the screen basis}, we set up the orthonormal frame of an observer adjusted for an observation on the sky.
\item[(ii)] In Section \ref{sec:First order deviation of a null bundle}, we introduce the connecting vector of a null bundle and present its evolution equation. We discuss the necessary condition to define an instantaneous wavefront consistently.
\item[(iii)] In Section \ref{sec:Wavefront curvature matrix and its Riccati evolution equation}, we introduce the wavefront curvature matrix. Its relation to the deformation parameters of a null bundle and to the spacetime curvature is also presented.
\item[(iv)] In Section \ref{sec:cross-sectional area and the principal curvatures of the wavefront}, determination of the cross-sectional area of a null bundle in addition to the principal curvatures of the wavefront is outlined. 
\item[(v)] Finally, in Section \ref{sec:Power conservation and distances}, the relationship between the geodesic deviation variables of the null bundle, cosmological distances and the power conservation is summarized.
\end{itemize}
%%%%%%%%%%%%%%%%%
\subsection{\label{sec:The orthonormal tetrad and the screen basis}
The orthonormal tetrad and the screen basis}
Let us consider an observer with a future directed $4-$velocity, $\vec{u}=d\vec{x}/d\tau$, and a spacelike, outward directed line of sight vector, $\vec{r}=d\vec{x}/d\ell$, with 
\begin{equation}
\inner{\,\vec{u}, \vec{u}\,}=-1, \qquad \inner{\,\vec{r}, \vec{r}\,}=1, \qquad \inner{\,\vec{u}, \vec{r}\,}=0.
\end{equation}
Here, the proper time and the proper length are denoted by $\tau$ and $\ell$, respectively.
As discussed before, most observations on the sky are obtained through bundles of null geodesics in the geometric optics limit. We denote a null vector that is tangent to the central geodesic of the bundle as $\vec{k}$. In the geometric optics limit, it satisfies $k_\mu=\mathcal{D} _\mu \mathcal{S}$, $\inner{\,\vec{k}, \vec{k}\,}=0$ and $\mathcal{D} _{\vec{k}}\vec{k}=0$ as discussed earlier. We assume that the null vector is past directed, $\inner{\,\vec{k}, \vec{u}\,}>0$, and outgoing, $\inner{\,\vec{k}, \vec{r}\,}>0$, such that it can be decomposed in the form
\begin{equation}\label{eq:k_decompose_u_r}
k^\alpha = -\omega \left(u^\alpha  - r^\alpha \right).
\end{equation}
The observed frequency of light, $\inner{\,\vec{k}, \vec{u}\,}$, is denoted by $\omega$ here.
The integral curves of $\vec{k}$ are parametrized by an affine parameter, $v$, whose relation to the proper time and the proper length is given respectively by,
\begin{equation}\label{eq:affine_clocks_rulers}
\omega dv=|d\tau|=|d\ell|.
\end{equation}
In order to define an observer's local screen in a consistent manner, one considers a dyad basis $s^\alpha_{a}$ that spans a 2-dimensional, spacelike screen 
with $\{a, b\} = \{1, 2\}$ and $\inner{\vec{s}_a, \vec{s}_b}=\delta _{a b}$. We assume that this basis is $C^{\infty}$ along the null geodesics and it satisfies,
\begin{equation}\label{eq:Sachs_basis}
\inner{\,\vec{u},\vec{s}_{a}\,}=0,
\qquad
\inner{\,\vec{r}, \vec{s}_{a}\,}=0,
\qquad
\mathcal{D} _{\vec{k}}{\vec{s}}_{a}=0.
\end{equation}
In the literature, a screen basis defined as in (\ref{eq:Sachs_basis}) is known as the Sachs basis \cite{Sachs:1961,Perlick:2010}. It is defined uniquely, up to 2-dimensional rotations around the observation direction vector, $\vec{r}$. Here, we set the local orthonormal frame of an observer. Note that it is straightforward to reverse the situation and set up another frame adopted by a source/emitter. In either of the cases, the geometric setting outlined in this section allows one to write the evolution equations of a null bundle with respect to observable quantities.
%%%%%%%%%%%%%%%%%%%
\subsection{\label{sec:First order deviation of a null bundle}
Null bundle deviation}
The propagation properties of a light bundle are often studied via its connecting vector, $\vec{\xi}$. This vector is known to satisfy a geodesic deviation equation which follows as
\begin{equation}\label{eq:first_ord_dev}
\mathcal{D} _{\vec{k}}\mathcal{D} _{\vec{k}}{\xi}^\alpha=\tensor{R}{^\alpha_{\vec{k}}_{\vec{k}}_{\vec{\xi}}}\,.
\end{equation}
Here, $\tensor{R}{_\alpha_{\beta}_{\mu}_{\nu}}$ is the Riemann curvature tensor. We assume that torsion is zero in this work.
Within the setting of linear geodesic deviation, (\ref{eq:first_ord_dev}), the integrability conditions imply that $\vec{k}$ and $\vec{\xi}$ are Lie dragged along each other, i.e.,
\begin{equation}\label{eq:integ_cond}
\mathcal{D} _{\vec{k}}\,\vec{\xi}=\mathcal{D} _{\vec{\xi}}\,\vec{k}.
\end{equation}
Condition (\ref{eq:integ_cond}) and equation (\ref{eq:first_ord_dev}) guarantee that $\inner{\,{\vec{k}},\vec{\xi}\,}$ remains constant along $\vec{k}$.
If this constant is chosen to be zero, then the deviation vector of the bundle and the tangent vector of the central null geodesic remain orthogonal to each other, i.e.,
\begin{equation}\label{eq:k_xi_orthog}
\inner{\,{\vec{k}}, \vec{\xi}\,}=0.   
\end{equation}
This choice is equivalent to demanding that the instantaneous wavefront (intersection of a constant-phase surface of the field with the observer's 3-dimensional space) remains orthogonal to $\vec{k}$. In that case, one ensures that $\vec{\xi}$ connects simultaneous events for the observer. In other words, the observed section of the infinitesimal wave-front is composed of rays which reach the observer simultaneously. This is a necessary condition to define a \textit{beam} unambiguously \cite{Schneider:1992}. 

Due to the orthogonal correspondence, the deviation vector can be decomposed into components that are parallel and transverse to the propagation vector \cite{Perlick:2010}, i.e.,
\begin{equation}\label{eq:4dev_vec_decomp}
\vec{\xi}=\xi^{k}\vec{k}+\vec{\xi}_{\perp},
\end{equation}
where $\vec{\xi}_{\perp}$ spans the 2-dimensional spacelike screen space. In Sachs basis, it follows as
\begin{equation}\label{eq:dev_vec_screen_decomp}
\vec{\xi}_{\perp}=\xi^{{1}}\vec{s}_{{1}}+\xi^{{2}}\vec{s}_{{2}},
\end{equation}
where $\xi^{{1}}$ and $\xi^{{2}}$ are the dyad basis components of $\vec{\xi}_{\perp}$. From now on we denote the screen basis components of objects with bold letters. For example, we denote
\begin{equation}
\boldsymbol{\xi}:=\left[
\begin{array}{c}
\xi^{{1}}  \\
\xi^{{2}}
\end{array}
\right].
\end{equation}
In the next subsection, we observe the relationship between the null bundle geometry and the wavefront curvature given through the deformation matrix. We also outline the matrix evolution equation for the wavefront curvature and its relation to the spacetime curvature.
%%%%%%%%%%%%%%%%
\subsection{\label{sec:Wavefront curvature matrix and its Riccati evolution equation} Wavefront curvature matrix and its Riccati evolution equation}
The shape of a thin light bundle wavefront  can be determined via the screen basis components of the deviation vectors introduced in the previous section. 
In order to show this, let us first consider the screen projection of the integrability condition given in (\ref{eq:integ_cond}), and write $\inner{\vec{s}_a,\mathcal{D} _{\vec{k}}\,\vec{\xi}\,}=\inner{\vec{s}_a,\mathcal{D} _{\vec{\xi}}\,\vec{k}}$,
in which $\Vec{s}_a$ is parallel propagated along the null bundle as in (\ref{eq:Sachs_basis}) and $\Vec{\xi}$ is in normal correspondence with $\Vec{k}$ as in (\ref{eq:k_xi_orthog}).
Then,
\begin{equation}\label{eq:deform_matrix_defn}
    \mathcal{D} _{\vec{k}}\,\xi_a=\frac{d\xi_a}{dv}:=\dot{\xi}_a=\inner{\vec{s}_a,\xi ^b \mathcal{D} _{\vec{s}_b}\vec{k}}=\Gamma_{ab}\xi ^b,
\end{equation}
where we use the overdot to denote the standard derivative with respect to the affine parameter, $v$. Note that $\mathcal{D} _{\vec{k}}\,\xi^a=\dot{\xi}^a$ holds as the dyad basis components act like scalars under the covariant derivative. Here, $\Gamma_{ab}$ denote the components of the \textit{deformation matrix},
\begin{equation}\label{eq:deform_matrix_components}
\boldsymbol{\Gamma}=\left[\begin{array}{c c}
\inner{\vec{s}_1,\mathcal{D}_{\vec{s}_1}\vec{k}} & \inner{\vec{s}_2,\mathcal{D}_{\vec{s}_1}\vec{k}} \\
\inner{\vec{s}_1,\mathcal{D}_{\vec{s}_2}\vec{k}} & \inner{\vec{s}_2,\mathcal{D}_{\vec{s}_2}\vec{k}}
\end{array}\right].
\end{equation}
 Clearly, this is the \textit{wavefront curvature matrix} of the bundle. $\boldsymbol{\Gamma}$ can be decomposed into its symmetric and anti-symmetric parts. The symmetric part can be further decomposed into pure-trace and trace-free portions. Namely,
\begin{equation}\label{eq:Gamma_decomp}
\boldsymbol{\Gamma}=
\left[
\begin{array}{c c}
\theta & \, \, 0 \\
0 & \theta
\end{array}
\right]
+
\left[
\begin{array}{c c}
-{\rm{Re}}\{\sigma\} & \, \, {\rm{Im}}\{\sigma\} \\
{\rm{Im}}\{\sigma\} & {\rm{Re}}\{\sigma\}
\end{array}
\right]
+
\left[
\begin{array}{c c}
0 & w \\
-w & 0
\end{array}
\right],\nonumber\\
\end{equation}
where $\theta$, $\sigma={\rm{Re}}\{\sigma\}+i{\rm{Im}}\{\sigma\}$ and $w$ represent the expansion, the shear and the twist of the bundle respectively. They are defined through,
\begin{equation}\label{eq:expansion}
\theta =\frac{1}{2}\left(\inner{\vec{s}_1,\mathcal{D}_{\vec{s}_1}\vec{k}}+\inner{\vec{s}_2,\mathcal{D}_{\vec{s}_2}\vec{k}}\right), 
\end{equation}
\begin{equation}\label{eq:re_shear}
{\rm{Re}}\{\sigma\}=\frac{1}{2}\left(\inner{\vec{s}_2,\mathcal{D}_{\vec{s}_2}\vec{k}}-\inner{\vec{s}_1,\mathcal{D}_{\vec{s}_1}\vec{k}}\right),
\end{equation}
\begin{equation}\label{eq:im_shear}
{\rm{Im}}\{\sigma\}=\frac{1}{2}\left(\inner{\vec{s}_1,\mathcal{D}_{\vec{s}_2}\vec{k}}+\inner{\vec{s}_2,\mathcal{D}_{\vec{s}_1}\vec{k}}\right),
\end{equation}
\begin{equation}\label{eq:twist}
w=\frac{1}{2}\left(\inner{\vec{s}_2,\mathcal{D}_{\vec{s}_1}\vec{k}}-\inner{\vec{s}_1,\mathcal{D}_{\vec{s}_2}\vec{k}}\right).
\end{equation}
Note that the twist term, $w$, in (\ref{eq:twist}) is a zero due to the geometric optics approximation. Namely, $k_\mu=\mathcal{D}_\mu \mathcal{S}$ is a gradient field and a null bundle with its central ray represented by such a gradient field is twist-free. Therefore, $\boldsymbol{\Gamma}$ is a \textit{symmetric matrix}.

In order to obtain the evolution equation of $\boldsymbol{\Gamma}$, one first takes the derivative of (\ref{eq:deform_matrix_defn}) along the null vector $\Vec{k}$. Then, one gets
\begin{equation}\label{eq:derv_xi_dot}
\ddot{\boldsymbol{\xi}}=\left(\boldsymbol{\dot{\Gamma}}+\boldsymbol{\Gamma}\boldsymbol{\Gamma}\right)\boldsymbol{\xi}.
\end{equation}
Also, the screen basis components of the deviation equation, (\ref{eq:first_ord_dev}),can be written as,
\begin{equation}\label{eq:proj_first_dev_eq}
\ddot{\boldsymbol{\xi}}=\boldsymbol{\mathcal{R}}\boldsymbol{\xi},\qquad{\rm{with}}\qquad \tensor{{\mathcal{R}}}{_{\,}_{a}_{b}}:=\tensor{R}{_{\vec{s}_a}_{\vec{k}}_{\vec{k}}_{\vec{s}_b}},
\end{equation}
where $\mathcal{R}$ is known as the \textit{optical tidal matrix} in the literature \cite{Seitz:1994, Perlick:2010}. 
Comparison of (\ref{eq:derv_xi_dot}) and (\ref{eq:proj_first_dev_eq}) gives
\begin{equation}\label{eq:Real_Riccati}
\boldsymbol{\dot{\Gamma}}+\boldsymbol{\Gamma}\boldsymbol{\Gamma}-\boldsymbol{\mathcal{R}}=\boldsymbol{0},
\end{equation}
which is a real, non-linear Riccati equation. Thus, (\ref{eq:Real_Riccati}) is an evolution equation for the wavefront curvature matrix. We must emphasize that solving this Riccati equation is equivalent to solving the evolution equation for the screen basis components of the deviation vector due to (\ref{eq:derv_xi_dot}). Moreover, (\ref{eq:Real_Riccati}) is also equivalent to the well-known Sachs optical equations
\begin{equation}\label{eq:Sachs_optical}
    \dot{\rho}=\rho ^2+|\sigma|^2+\Phi _{00}\qquad {\rm{and}}\qquad
\dot{\sigma}=\left(\rho+\bar{\rho}\right)\sigma+\Psi _0.
\end{equation}
This is because of the fact that the decomposition in (\ref{eq:Gamma_decomp}) can be rewritten as,
\begin{equation}
\boldsymbol{\Gamma}=-\frac{1}{2}
\left[
\begin{array}{c c}
\left(\rho+\bar{\rho}\right) +\left(\sigma+\bar{\sigma}\right) & i\left(\sigma-\bar{\sigma}\right)+i\left(\rho-\bar{\rho}\right) \\
i\left(\sigma-\bar{\sigma}\right)-i\left(\rho-\bar{\rho}\right) & 
\left(\rho+\bar{\rho}\right) -\left(\sigma+\bar{\sigma}\right)
\end{array}   
\right],\nonumber\\
\end{equation}
with the notation of the Newman-Penrose formalism\footnote{Note that in this work we choose the signature of the metric to be $(-,+,+,+)$. Therefore, when compared to Newman and Penrose’s original work \cite{Newman:1961}, our spin coefficients and curvature scalars have an extra negative
sign. Eventually, we are using the notation of \cite{ONeill:1995}.}\cite{Newman:1961}. Here, $\rho=-\theta+iw$ with $w=0$, due to the bundle being twist-free. The overbar denotes complex conjugation and
\begin{equation}\label{eq:optical tidal matrix}
\boldsymbol{\mathcal{R}}=-
\left[
\begin{array}{c c}
\Phi_{00}+{\rm{Re}}{\Psi_0} & -{\rm{Im}}{\Psi_0} \\
-{\rm{Im}}{\Psi_0} & \Phi_{00}-{\rm{Re}}{\Psi_0}
\end{array}
\right].
\end{equation}
The real function $\Phi_{00}=\tensor{R}{_\mu_\nu}k^\mu k^\nu/2$ is one of the Ricci scalars and the complex function $\Psi_0=\tensor{C}{_\mu_\nu_\alpha_\beta}k^\mu m^\nu k^\alpha m^\beta$ is one of the Weyl scalars of the Newman-Penrose formalism. We denote the Ricci tensor by $\tensor{R}{_\mu_\nu}$ and the Weyl tensor by $\tensor{C}{_\mu_\nu_\alpha_\beta}$. The relationship between the complex spatial vector, $\Vec{m}$, of Newman and Penrose and our Sachs basis, $\tensor{\Vec{s}}{_a}$, is given via $\vec{m}=\left(\Vec{s_1}-i\Vec{s_2}\right)/\sqrt{2}$.

We can summarize the current subsection as the following. The wavefront curvature matrix, $\boldsymbol{\Gamma}$, is obtained through the deformation matrix of the null bundle. It is a symmetric matrix due to the underlying geometric optics approximation, i.e., due to the propagation vector being a gradient field. The evolution equation of $\boldsymbol{\Gamma}$ is a real, non-linear Riccati equation which is equivalent to the screen projection of the geodesic deviation equation or the well-known Sachs optical equations. In the next subsection, we briefly outline the relationship between $\boldsymbol{\Gamma}$ and the principal curvatures of the wavefront. A similar summary can be found in \cite{Perlick:2010}.
%%%%%%%%%%%%%%%%%
\subsection{\label{sec:cross-sectional area and the principal curvatures of the wavefront}Cross-sectional area and the principal curvatures of the wavefront}
The cross-sectional area, $\delta \mathcal{X}$, of a thin bundle is obtained through the solutions of the geodesic deviation vector aligned with the  semi-major and the semi-minor axes of the cross-sectional ellipse.
We denote those two solutions as,
\begin{equation}\label{eq:Y_vectors}
\left[
\begin{array}{c}
\vec{Y}{_-}  \\
\vec{Y}{_+} 
\end{array}
\right]
=
\mathbb{D}
\left[
\begin{array}{c}
\Vec{e}{_-}  \\
\Vec{e}{_+} 
\end{array}
\right],\qquad
{\rm{with}}\qquad
\mathbb{D}
:=
\left[
\begin{array}{c c}
|D_-| & 0 \\
0 & |D_+|
\end{array}
\right],
\end{equation}
where $\{\Vec{e}{_-},\Vec{e}{_+}\}$ is the unit basis aligned with the semi-minor and semi-major axes. The corresponding magnitude of the deviation vector components are denoted by $|D_-|$ and $|D_+|$.

Note that the semi-minor and the semi-major axes of the cross-section of the bundle do not necessarily align with the Sachs basis. It is the latter one that we use in our calculations. We often have two arbitrary solutions of the screen-projected geodesic deviation equation, $\vec{\xi}_{\perp}$ and $\vec{\tilde{\xi}}_{\perp}$, whose components are calculated in the Sachs basis. Nevertheless, one can use those two arbitrary linearly independent solutions to study the geometry of the cross section; instead of the solutions, $\vec{Y}{_-}$ and $\vec{Y}{_
+}$, aligned with the semi-minor and the semi-major axes \cite{Perlick:2010}. To see this, let us denote the angle between the Sachs basis, $\{\Vec{s}{_1},\Vec{s}{_2}\}$, and the basis that is aligned with the semi-minor and semi-major axes of the cross-sectional ellipse, $\{\Vec{e}{_-},\Vec{e}{_+}\}$, by $\Theta$. We denote the angle between the two sets of solutions with $\Upsilon$ as in Figure \ref{fig:Solutions_null bundle}.
\begin{figure}[ht]
\centering
\scalebox{0.8}{
\begin{tikzpicture}
    \draw[rotate around={45:(0,0)}]  (0,0) ellipse (2cm and 4cm);

    % Sachs basis
    \draw  (0,0) -- (4,0);
    \draw  (0,0) -- (0,4);
    \node[draw=none] at (-0.25,3.8) {$\vec{s}_2$};
    \node[draw=none] at (3.8,-0.25) {$\vec{s}_1$};

    % Semi-major and semi-minor axes of the ellipse
    \draw[thick,dotted]  (0,0) -- (4,4);
    \draw[thick,dotted]  (0,0) -- (-3.5,3.5);
    \node[draw=none] at (4.2,3.8) {$\vec{e}_{-}$};
    \node[draw=none] at (-3.6,3.2) {$\vec{e}_{+}$};
    
    % Vectors aligned with the semi-major and semi-minor axes
    \draw[->,ultra thick,MyGreen]  (0,0) -- (1.4,1.4);
    \draw[->,ultra thick,MyGreen]  (0,0) -- (-2.8,2.8);
    \node[draw=none,MyGreen] at (1,1.5) {$\vec{Y}_{-}$};
    \node[draw=none,MyGreen] at (-2.8,2.3) {$\vec{Y}_{+}$};
    %\node[draw=none,MyGreen] at (-1.2,0.6) {$|{D}_{-}|$};
    %\node[draw=none,MyGreen] at (1.2,1.8) {$|{D}_{+}|$};
    
    % Two arbitrary solutions
    \draw[->,ultra thick,MyRed]  (0,0) -- (1.9,0.91);
    \draw[->,ultra thick,MyRed]  (0,0) -- (-1.5,3.16);
    \node[draw=none,MyRed] at (-1.6,3.5) {$\vec{\boldsymbol{\tilde{\xi}}}_{\perp}$};
    \node[draw=none,MyRed] at (2.3,1) {$\vec{\boldsymbol{\xi}}_{\perp}$};

    % Angles
    % Angle between the Sachs basis and the semi-axes
    \draw [thick,MyGreen] (1,0) arc (0:45:1);
    \node[draw=none,MyGreen] at (1.2,0.3) {$\Theta$};
    % Angle between two solutions
    \draw [thick,MyRed] (1.2,0.6) arc (17:40:1);
    \node[draw=none,MyRed] at (1.35,0.85) {$\Upsilon$};
\end{tikzpicture}}
\caption{Screen-projected solutions of the geodesic deviation equation of a null bundle.}
\label{fig:Solutions_null bundle}
\end{figure}
The relationship between the two sets of solutions follows as
\begin{equation}\label{eq:xi_to_Y}
\left[
\begin{array}{c}
\vec{Y}{_-}  \\
\vec{Y}{_+}
\end{array}
\right]
=
R^{-1}(\Upsilon)
\left[
\begin{array}{c}
\vec{\xi}_\perp  \\
\vec{\tilde{\xi}}_\perp 
\end{array}
\right],
\end{equation}
where $R(\Upsilon)$ is a $2\times 2$ rotation matrix. The components of $\vec{\xi}_\perp$ and $\vec{\tilde{\xi}}_\perp$ in the  $\{\vec{e}_{-}, \vec{e}_{+}\}$ basis is related to its components in the Sachs basis, $\{\vec{s}_{1}, \vec{s}_{2}\}$, as
\begin{equation}\label{eq:Q_defn}
\left[
\begin{array}{c c}
{\xi}_\perp{^-} & \, \, \tilde{\xi}_\perp{^-} \\
{\xi}_\perp{^+} & \, \, \tilde{\xi}_\perp{^+}
\end{array}
\right]
= R^{-1}(\Theta)\,\mathbf{Q},\qquad {\rm{with}}\qquad \mathbf{Q}:=
\left[
\begin{array}{c c}
\xi{^1} & \, \, \tilde{\xi}{^1} \\
\xi{^2} & \, \, \tilde{\xi}{^2}
\end{array}
\right].
\end{equation}
Then, following (\ref{eq:Y_vectors}), (\ref{eq:xi_to_Y}) and (\ref{eq:Q_defn}) we have
\begin{equation}\label{eq:D_to_Q}
\mathbf{Q}=R(\Theta)\,\mathbb{D}\, R^{-1} (\Upsilon).
\end{equation}
The value of the cross-sectional area can now be obtained through two arbitrary linearly independent projected solutions via,
\begin{equation}\label{eq:X-section-detQ}
    \delta \mathcal{X}:=|D_-D_+|={\rm{det}}\mathbb{D}={\rm{det}}\mathbf{Q},
\end{equation}
due to (\ref{eq:Y_vectors}) and (\ref{eq:D_to_Q}).

Similarly, the wavefront curvature matrix and the principal curvatures of the wavefront can be obtained through, any two linearly independent solutions. In order to show this, let us first define a $2\times 2$ matrix through the derivatives of the projected solutions of the geodesic deviation equation, i.e.,
\begin{equation}\label{eq:P_matrix_defn}
\mathbf{P}:=\mathbf{\dot{Q}}=
\left[
\begin{array}{c c}
\dot{\xi}{_1} & \, \, \dot{\tilde{\xi}}{_1} \\
\dot{\xi}{_2} & \, \, \dot{\tilde{\xi}}{_2}
\end{array}
\right].
\end{equation}
Here, the overdot represents the standard derivative with respect to the affine parameter, $v$, as before. Now, recall that the wavefront curvature matrix is defined via (\ref{eq:deform_matrix_defn}). Then,
\begin{equation}\label{eq:Gamma_defn}
\mathbf{\Gamma}:=\mathbf{PQ}^{-1}, 
\end{equation}
is another way of representing the wavefront curvature matrix. It can be written in terms of Sachs' optical scalars through (\ref{eq:deform_matrix_components}).

Previously, Kantowski gave an interpretation of the optical scalars by relating them to the wavefront curvature \cite{Kantowski:1968}. He proved that in the rest frame of the observer, the principal curvatures of a 2-dimensional surface that lies on the intersection of the null cone and the 3-space of the observer are given through\footnote{In fact, in his original work \cite{Kantowski:1968}, Kantowski uses a spatial normal to the 2-dimensional wavefront ($\vec{r}$ in (\ref{eq:k_decompose_u_r}) in this work). Therefore, the principal curvatures are defined with respect to the proper length, $d\ell$. In this work, we consider the null vector $\vec{k}$ as the normal vector to the screen-space. Thus, Kantowski's result has an extra $\omega=\inner{\vec{k},\vec{u}}$ term when compared to ours. This follows from the relationship between the affine parameter, $v$, and the proper length, $\ell$, as in (\ref{eq:affine_clocks_rulers}).} $\Lambda _-=\theta -|\sigma|$ and $\Lambda _+=\theta +|\sigma|$.
Those principal curvatures can be obtained through the eigenvalue equation of the wavefront curvature matrix. Namely, as $\mathbf{\Gamma}$ is symmetric it can be diagonalized through $\mathbf{\Gamma}=\mathbf{M}\mathbf{\Lambda} \mathbf{M}^{-1}$, where $\mathbf{M}$ is the modal matrix of $\mathbf{\Gamma}$ and $\mathbf{\Lambda}$ is the eigenvalue matrix with $\Lambda _-$ and $\Lambda _+$ being its diagonal elements. 

At this point, we should also emphasize the observer independence of the shape and the size of a null bundle. To be more specific, let us again consider any two solutions, $\vec{\xi}$ and $\vec{\tilde{\xi}}$, of the geodesic deviation equation which are in orthogonal correspondence with the propagation vector. Neither the length of the corresponding vectors, $|\,\vec{\xi}\,|$ and $|\,\vec{\tilde{\xi}}\,|$, nor the angle, $\phi$, between those projected solutions depend on the frame of the observer \cite{Kermack:1934, Sachs:1961}. In other words, $\inner{\,\vec{\xi},\vec{\tilde{\xi}}\,}=\inner{\,\vec{\xi}_{\perp},\vec{\tilde{\xi}}_{\perp}\,}=|\vec{\xi}_{\perp}||\vec{\tilde{\xi}}_{\perp}|\cos{\phi}$ is observer independent.

We conclude this subsection with the following summary. Given any two linearly independent solutions that span the 2-dimensional screen space of a thin null bundle, its cross-sectional area can be determined through (\ref{eq:X-section-detQ}). The wavefront curvature matrix, $\mathbf{\Gamma}$, can be obtained through (\ref{eq:Gamma_defn}) where $\mathbf{Q}$ and $\mathbf{P}$ are defined in (\ref{eq:Q_defn}) and (\ref{eq:P_matrix_defn}) respectively. The principal curvatures of the wavefront, $\Lambda _-$ and $\Lambda _+$ can then be obtained via the diagonalization procedure of $\mathbf{\Gamma}$. 

In the next subsection, we remind how the cross-sectional area of a bundle and the wavefront curvature matrix are related to the cosmological distances and the power conservation in the bundle.
%%%%%%%%%%%%%%%%%%%%%%%%%%%
\subsection{\label{sec:Power conservation and distances} Power conservation and the cosmological distances}
In this subsection, we first recall how the power, i.e., intensity times the cross-sectional area, contained in a null bundle evolves in the propagation direction, $\vec{k}$. We observe from Section \ref{sec:Geometric optics limit} that in the geometric optics limit, the vector potential, $\Vec{\alpha}{_0}$, evolves according to (\ref{eq:Lowest order_Maxwell}) at the leading order of part the JWKB-like approximation. Then, the derivative of the squared amplitude, $\alpha ^2=\inner{\Vec{\alpha}{_0},\Vec{\bar{\alpha}}{_0}}$, obeys
\begin{equation}\label{eq:evol_ampl_sq}
\frac{d \alpha ^2}{dv} =- \left(\mathcal{D} _\mu k^ \mu\right) \alpha ^2=-2\theta \alpha ^2.
\end{equation}
Here, the second equality follows from (i) the definition of the expansion scalar in (\ref{eq:expansion}) and (ii) Sachs basis satisfying $\inner{\vec{s}_a, \vec{s}_b}=\delta _{a b}$. The equation above shows how the intensity of the bundle evolves along the integral curves of $\Vec{k}$.

In the previous subsection, we observed that two linearly independent solutions of the screen-projected geodesic deviation equation can be incorporated in a matrix $\mathbf{Q}$ and the value of the cross-sectional area can be obtained through ${\rm{det}}\mathbf{Q}$. Then, the change of the cross-sectional area of the bundle along the propagation direction can be calculated via
\begin{equation}\label{eq:evol_cross_sec}
    \frac{d\,\delta \mathcal{X}}{dv}={\rm{det}}\mathbf{Q}\,{\rm{tr}}\left(\mathbf{Q}^{-1}\mathbf{\dot{Q}}\right)={\rm{det}}\mathbf{Q}\, {\rm{tr}}\mathbf{\Gamma}=2\theta \delta \mathcal{X},
\end{equation}
by making use of (\ref{eq:Gamma_decomp}) and definitions (\ref{eq:P_matrix_defn})-(\ref{eq:Gamma_defn}). Then, (\ref{eq:evol_ampl_sq}) and (\ref{eq:evol_cross_sec}) imply that
\begin{equation}\label{eq:cons_photon_flux}
\mathcal{D}_{\vec{k}}\,\mathcal{P}:=\mathcal{D}_{\vec{k}}\left(\alpha ^2\delta \mathcal{X}\right)=0.
\end{equation}
This means that the power, $\mathcal{P}$, is conserved within the bundle throughout the evolution. In other words, the total number of photons contained in the bundle or the number of rays that pierce the cross-sectional area is constant. Then, the thin bundle approximation and the geometric optics limit imply that there is no net emission or absorption throughout the propagation of an observable light bundle.

Let us now recall the definitions of the cosmological distances.
The angular diameter distance, $D_{A}$, is obtained by taking the ratio of the estimated proper cross-sectional area of an object at the source point, $s$, to the measured solid angle at the observation point, $o$. The luminosity distance\footnote{Here, we reserve the name \textit{luminosity distance} for the distance definition given in (\ref{eq:distances_X-sec_angle}) which is sometimes referred to as \textit{corrected luminosity distance} in the literature. We refer to the distance estimated through the comparison of absolute and apparent magnitudes of an object as \textit{uncorrected luminosity distance} as it does not involve the relativistic correction.}, $D_{L}$, is defined in a similar way when the situation is reversed, i.e.,
\begin{equation}\label{eq:distances_X-sec_angle}
    D_{A}=\left(\frac{\delta \mathcal{X}_s}{\delta \Omega_o}\right)^{1/2}\qquad {\rm{and}} \qquad
    D_{L}=\left(\frac{\delta \mathcal{X}_o}{\delta \Omega_s}\right)^{1/2}.
\end{equation}
The solid angle, $\delta \Omega$, is described at a vertex point and is given by
\begin{equation}\label{eq:solid_angle}
    \delta \Omega := \Big|\frac{dD_+}{d\ell}\frac{dD_-}{d\ell}\Big|=\frac{1}{\omega ^2} \Big|\frac{dD_+}{dv}\frac{dD_-}{dv}\Big|.
\end{equation}
As before, $D_+$ and $D_-$ correspond to the magnitudes of the solutions of the geodesic deviation equation that are aligned with the semi-major and semi-minor axes of the observational screen.
The second equality in the above follows from the relationship between the affine parameter and the proper length, (\ref{eq:affine_clocks_rulers}).

In summary, the cross-sectional areas and the subtended solid angles are estimated via the projected geodesic deviation vectors and their derivatives along the propagation. In the geometric optics limit, the evolution of intensity and cross-sectional area always balance each other out. As a result of this, the power contained in the bundle is conserved throughout the propagation.

At this point, we finish the Preliminaries section of the paper. In the next section, we outline the action principle of the null bundle evolution and its underlying Hamiltonian formalism. Note that the next section is essentially a summary of \cite{Uzun:2018} in which the phase space formulation of thin bundles were presented in analogy to the paraxial regime of the Newtonian optics.
%%%%%%%%%%%%%%%%%%%%%%%%%%%%%%%%%%%%%%%%%%%%%%%%%%%%%%%%%%%%%%%%%
\section{\label{sec:Summary of [30]: Reduced Hamiltonian optics}A Summary: Reduced Hamiltonian optics on a symplectic phase space}
In this section, we give a short summary of our previous work \cite{Uzun:2018} in which the light bundle propagation on a reduced phase space was presented via a Hamiltonian formalism. The resultant construct is similar to the one in Newtonian paraxial ray optics and it follows from the geodesic deviation action. As the current section presents no new results, the reader who is familiar with \cite{Uzun:2018} can skip this section. On the other hand, we use the Hamiltonian formalism of \cite{Uzun:2018} in the main body of the current paper. Specifically, the Hamilton-Jacobi equation of the null bundle deviation vectors plays the main role in the new framework we introduce. Therefore, the formalism summarized here is crucial for the understanding of the new results presented in Section \ref{sec:Wavization} and onward. 
%%%%%%%%%%%
\subsection{\label{sec:Action of a thin bundle} Action of a thin null bundle}

Let us represent a null bundle by: (i) an outer most null geodesic, $\zeta(v)$, with a tangent vector $\vec{k}'$, (ii)
the central null geodesic, $\Theta(v)$, with a tangent vector $\vec{k}$. Those two curves are parametrized with the same affine parameter $v$. Such type of parametrization is called the \textit{isosynchronous parametrization} in the literature. In order to find the equations of motion of $\vec{k}'$ and $\vec{k}$ we consider the action functionals
\begin{equation}\label{eq:Geod_actions}
S_{k'}=\int \frac{1}{2}{k'}^2 dv,\qquad {\rm{and}} \qquad S_{k}=\int \frac{1}{2}{k}^2 dv,
\end{equation}
on curves $\zeta(v)$ and $\Theta(v)$ respectively. The extremized actions $S_{k'}$ and $S_{k}$ provide the corresponding equations of motion which are the geodesic equations, $\mathcal{D}_{\vec{k'}}\vec{k'}=0$ and $\mathcal{D}_{\vec{k}}\vec{k}=0$ under the affine parameterization. 

Instead of considering the action functional of each null curve within the bundle, we would like to assign an effective action to the entire light bundle. One can follow various methods to accomplish this. In \cite{Uzun:2018}, we followed Vines' construction \cite{Vines:2014} in which the author obtains an action functional corresponding to the deviation variables of generic geodesic curves. The procedure starts with expanding $S_{k'}$ by $S_{k}$ so as to obtain an effective action for the bundle.
\begin{figure}
\hspace*{1cm}
\begin{center}
\begin{tikzpicture}[thick, scale=1]
%\draw [help lines,step=0.5] (0,0) grid (9,3);
\draw [MyGreen] (0.25,0.5) to[bend left=20] (7.75,3);
\draw [MyGreen,dashed] (0.25,0.5) to[bend left=10] (7.75,1.5);
\draw [MyRed,thick] (5,0.75) to[bend right] (4,3.25);

\node [draw=none] at (4.8,1.3) {m};
\node [draw=none] at (4.23,2.80) {n};

\node [draw=none] at (4.4,2.1) {$\Sigma(\lambda)$};
\node [draw=none] at (2,1.2) { $\Theta (v)$};
\node [draw=none] at (2,2) {$\zeta (v)$};

\draw [->] (5,1.465) to (5.75,1.6);
\draw [->] (5.03,1.465) to (5.02,2.2);
\draw [->] (4.53,2.74) to (5.05,2.95);

\node [draw=none] at (5.9,1.8) {$\vec{k}$};
\node [draw=none] at (5.2,3.2) {$\vec{k}'$};
\node [draw=none] at (5.2,2.25) {$\vec{t}$};
\end{tikzpicture}
\end{center}
\caption{A null bundle with central null geodesic $\Theta (v)$. The red curve represents $\Sigma(\lambda)$ given by Synge's spacelike world function, $\sigma (m,n)$. The outermost null geodesic $\zeta (v)$ can be uniquely obtained through $\Theta (v)$ and $\Sigma(\lambda)$.}\label{fig:Syngesworldfun}
\end{figure}
For an arbitrarily thick bundle, two geodesics can not be connected locally. Rather, its dynamics can be studied via a unique geodesic curve, $\Sigma(\lambda)$, that is represented by a function $\sigma (m,n)$ (See Figure \ref{fig:Syngesworldfun}.). This bi-local function depends on two spacetime points $m$ and $n$ and it corresponds to the Synge's world function \cite{Synge:1960}. In this framework, the null vector $\vec{k}'$ at point $n$ can be expanded through a perturbative method via the null vector $\vec{k}$ at point $m$ in addition to $\sigma (m,n)$ and its derivatives along the null bundle. The derivative of the world function can be interpreted as a bi-local deviation vector, $\vec{\xi} (m,n)$ \cite{Poisson:2003, Vines:2014}.

With this geometric construction, the geodesic action of the outer most curve, $S_{k'}$, was written in terms of the geodesic action of the central curve. Namely, consider the expansion of $\vec{k}'$ in terms of $\vec{k}$, $\vec{\xi} (m,n)$ and its derivatives. Substitution of $\vec{k}'$ written in the aforementioned form into the geodesic action, $S_{k'}$, in (\ref{eq:Geod_actions}) allows one to rewrite this action in the form $S_{k'}=S_{k}+S_{\xi}$, up to the desired order in $\vec{\xi}$ and its derivatives. Meaning, $S_{k'}$ can be written in terms of $S_{k}$ and another action functional $S_{\xi}$. Note that the coincidence limit, i.e., $m\rightarrow n$, defines a \textit{infinitesimally thin bundle} \cite{Uzun:2018,Vines:2014}. In that case, $\vec{\xi}$ is small and it represents a geodesic deviation \textit{vector} that is defined locally. Accordingly, one can consider terms up to quadratic order in $S_{\xi}$. This results in a geodesic deviation action for a thin bundle which was written as\footnote{We remind that one can refer to Section 3.2 of 
\cite{Uzun:2018} and Vines' original work \cite{Vines:2014} for the rigorous definitions and the detailed calculations for equations presented here.}
\begin{equation}\label{eq:Deviation_action}
S_{\xi}\approx \int \left(\frac{1}{2}\inner{\,\mathcal{D}_{\Vec{k}}\,\vec{\xi}, \mathcal{D}_{\Vec{k}}\,\vec{\xi}\,}+\frac{1}{2}\tensor{R}{_{\vec{\xi}}_{\vec{k}}_{\vec{k}}_{\vec{\xi}}}\right)dv.
\end{equation}
When the geodesic deviation action, (\ref{eq:Deviation_action}), is extremized with respect to locally defined $\vec{\xi}$ and $\mathcal{D}_{\vec{k}}\,{\vec{\xi}}$, the corresponding equations of motion are the \textit{linear} geodesic deviation equations, i.e., $\mathcal{D} _{\vec{k}}\mathcal{D} _{\vec{k}}\,{\xi}^\alpha=\tensor{R}{^\alpha_{\vec{k}}_{\vec{k}}_{\vec{\xi}}}$ and $\vec{\xi}$ can be represented by the Jacobi fields. 
%%%%%%%%%%%%%%
\subsection{The reduced Lagrangian and its invariance under the Lorentz transformations}\label{sec:The reduced Lagrangian and its invariance under Lorentz transformations}
In the previous subsection, we summarized the procedure to obtain the geodesic deviation action. It is important to notice that this functional represents the action of the entire bundle without distinguishing the individual curves in the bundle at the lowest order. The aforementioned coarse-grained nature of our method follows from this fact. 

Previously, we treated the integrand of the geodesic deviation action in (\ref{eq:Deviation_action}) as the Lagrangian of the deviation system, i.e.,
\begin{equation}\label{eq:Lagrangian}
L:=\frac{1}{2}\inner{\,\mathcal{D}_{\Vec{k}}\vec{\xi}, \mathcal{D}_{\Vec{k}}\vec{\xi}\,}+\frac{1}{2}\tensor{R}{_{\vec{\xi}}_{\vec{k}}_{\vec{k}}_{\vec{\xi}}}\,.    
\end{equation}
Let us recall from Section \ref{sec:First order deviation of a null bundle} that the orthogonality condition of the deviation vector and the propagation vector, i.e., $\inner{\,{\vec{k}}, \vec{\xi}\,}=0$ is the necessary condition to define a beam unambiguously. With this condition, $\vec{\xi}$ represents a congruence of rays which reach the observer simultaneously. In that case, the deviation vector can be decomposed into two portions: (i) a component along $\vec{k}$, (ii) components transverse to $\vec{k}$. The explicit form of the deviation vector under this decomposition is given in (\ref{eq:4dev_vec_decomp}) as $\vec{\xi}=\xi^{k}\vec{k}+\vec{\xi}_\perp$.
Substitution of this decomposition into the Lagrangian (\ref{eq:Lagrangian}) gave us a reduced Lagrangian function
\begin{equation}\label{eq:Red_Lagrangian}
L=L_{\rm{red}}=\frac{1}{2}\tensor{\delta}{_{a}_{b}} \dot{\xi}^{a}\dot{\xi}^{b}+\frac{1}{2}\tensor{\mathcal{R}}{_{\,}_{a}_{b}}\xi^{a}\xi^{b},
\end{equation}
due to the symmetries of the Riemann tensor 
and $\vec{k}$ satisfying the geodesic equation \cite{Uzun:2018}. Here, $\{a,b\}=\{1,2\}$ represent the Sachs basis components and $\tensor{\mathcal{R}}{_{\,}_{a}_{b}}(v)$ are the components of the symmetric optical tidal matrix introduced in (\ref{eq:proj_first_dev_eq}). They are composed of the Ricci curvature, $\Phi _{00}$, and the Weyl curvature, $\Psi _{0}$. 

Then, we understand that the Lagrangian function in (\ref{eq:Lagrangian}) is reduced by two degrees of freedom. In that case, the geodesic deviation action can be rewritten entirely through the screen projections of the deviation vector and their derivatives with respect to the affine parameter $v$, i.e., 
\begin{equation}\label{eq:action}
S_{\xi}=\int L_{\rm{red}} \,dv,
\end{equation}
with $L_{\rm{red}}$ given in (\ref{eq:Red_Lagrangian}).

Since we are interested in a process of null dust emission from a source to an observer, we are only interested in those Lorentz transformations that keep the propagation direction unchanged. Those types of transformations are known as the Type-I Lorentz transformations within the context of the Newman-Penrose formalism \cite{Newman:1961} and they are given through 
\begin{align}\label{eq:TypeI_Lorentz_trans}
    \vec{k}&\rightarrow \vec{k},\qquad \qquad \qquad \qquad \qquad \qquad \vec{m}\rightarrow f\vec{k}+\vec{m},\nonumber \\
    \vec{n} &\rightarrow f\bar{f}\vec{k}+\vec{n}+\bar{f}\vec{m}+f\vec{\bar{m}},\qquad \qquad \vec{\bar{m}}\rightarrow \bar{f}\vec{k}+\vec{\bar{m}}.
\end{align}
Here, $f$ is an arbitrary complex scalar and $\{\vec{k}, \vec{n}, \vec{m}, \vec{\bar{m}}\}$ complete the semi-null tetrad of Newman and Penrose. The only non-vanishing inner products associated with this tetrad are 
$\inner{\vec{k},\vec{n}}=-1$ and $\inner{\vec{m},\vec{\bar{m}}}=1$ where $\{\vec{m}, \vec{\bar{m}}\}$ span the screen-basis. They are related to the (parallel-propagated) Sachs basis via $\vec{m}=\left(\Vec{s_1}-i\Vec{s_2}\right)/\sqrt{2}$ and $\vec{\bar{m}}=\left(\Vec{s_1}+i\Vec{s_2}\right)/\sqrt{2}$ up to a $v$-independent 2-dimensional rotation. It is known that \cite{Stewart:1993}
\begin{equation}\label{eq:Curvature scalars under Type I Lorentz}
\Phi_{00}\rightarrow \Phi_{00}\qquad {\rm{and}}\qquad \Psi_0 \rightarrow  \Psi_0, 
\end{equation}
under the transformation given in (\ref{eq:TypeI_Lorentz_trans}). Then, it is easy to show that, under Type-I Lorentz transformations,
\begin{align}
\tensor{\delta}{_{a}_{b}} \dot{\xi}^{a}\dot{\xi}^{b}\rightarrow  \tensor{\delta}{_{a}_{b}} \dot{\xi}^{a}\dot{\xi}^{b},\qquad{\rm{and}}\qquad
\tensor{\mathcal{R}}{_{\,}_{a}_{b}}\xi^{a}\xi^{b}\rightarrow \tensor{\mathcal{R}}{_{\,}_{a}_{b}}\xi^{a}\xi^{b}
\end{align}
holds.
The second relation in the above follows from the definition of the optical tidal matrix, (\ref{eq:optical tidal matrix}), and the transformation properties of the Ricci and Weyl scalars given in (\ref{eq:Curvature scalars under Type I Lorentz}).

Thus, we conclude that the Lagrangian, (\ref{eq:Red_Lagrangian}), and the action functional, (\ref{eq:action}), of the system are invariant under the Lorentz transformations that keep the propagation direction unchanged.

In the next subsection, we give a summary of the Hamiltonian formulation of the null bundle dynamics defined on a reduced symplectic phase space in \cite{Uzun:2018}.
%%%%%%%%%%%
\subsection{\label{sec:Symplectic evolution}Symplectic evolution in phase space}
We reminded in the previous subsection that  the action functional of the null geodesic deviation can be reduced by two dimensions in the case where the deviation vector and the propagation vector are orthogonal to each other. Then, the Lagrangian is composed of only the screen projections of the deviation vector and their derivatives along the propagation. In this subsection we give a summary of the Hamiltonian dynamics to study the evolution of a bundle on a reduced phase space. 

In our previous work \cite{Uzun:2018}, we defined a $4$-dimensional phase space vector via the following Darboux coordinates, 
%$\mathbf{z}:=\left[q ^{a}, p _{b}\right]^\intercal=\left[\xi ^{{1}}\, \xi ^{{2}} \,\dot{\xi} _{{1}} \,\dot{\xi} _{{2}}\right]^\intercal$.
\begin{equation}\label{eq:Phase_coords}
\mathbf{z}=
\left[\begin{array}{c }
 q ^{a} \\
 p _{b} 
\end{array}\right]
=\left[\begin{array}{c }
 \xi ^{{1}} \\
 \xi ^{{2}} \\
 \dot{\xi} _{{1}} \\
 \dot{\xi} _{{2}}
\end{array}\right].
\end{equation}
Here, the screen basis components, $\boldsymbol{\xi}$, of the deviation vector act as canonical coordinates and their derivatives, $\boldsymbol{\dot{\xi}}$, act as canonical momenta.

The Hamiltonian of the system, corresponding to the reduced Lagrangian was found through
\begin{equation}\label{eq:Red_Hamiltonian}
H=p_{a}\dot{q}^{a}-L_{\rm{red}}=\frac{1}{2}\tensor{\delta}{^{a}^{b}} \dot{\xi}_{a}\dot{\xi}_{b}+\mathcal{V}\left(\boldsymbol{\xi},v\right),
\end{equation}
with
\begin{equation}\label{eq:raybundle_potential}
\mathcal{V}\left(\boldsymbol{\xi},v\right)= -\frac{1}{2}\tensor{\mathcal{R}}{_{\,}_{a}_{b}}\xi^{a}\xi^{b},  
\end{equation}
which is analogous to a time dependent Hamiltonian of a classical oscillator. In our case, the evolution parameter is the affine parameter, $v$, and we refer to the term $\mathcal{V}\left(\boldsymbol{\xi},v\right)$ as the \textit{ray bundle potential}. Due to the linearity of the problem, the Hamiltonian function was rewritten as, 
\begin{equation}\label{eq:Hamiltonian_matrix}
H\left(\mathbf{z},v\right)= \frac{1}{2}\mathbf{z}^{\intercal}\bm{\Omega}^\intercal\mathbf{H}\,\mathbf{z},\qquad{\rm{where}}\qquad \mathbf{H}=
\left[
\begin{array}{c|c}
\mathbf{0_2} & \, \, \delta ^{ab} \\
\hline
\tensor{\mathcal{R}}{_{\,}_{a}_{b}} & \, \, \mathbf{0_2}
\end{array}
\right],
\end{equation}
is the Hamiltonian matrix of the system. It satisfies $\mathbf{\Omega}{\mathbf{{H}}}=\left(\mathbf{\Omega}{\mathbf{{H}}}\right)^\intercal$ due to the optical tidal matrix, $\boldsymbol{\mathcal{R}}$, being symmetric. Matrix $\mathbf{\Omega}$ is known as the \textit{standard/fundamental symplectic matrix} in the literature \cite{deGosson:2011,Treves:2022}, whose components are given by,
\footnote{Matrices $\mathbf{0_2}$ and $\mathbf{I_2}$ refer to the $2$-dimensional zero and identity matrices.}
\begin{equation}
\Omega ^{ij}=
\left[
\begin{array}{c|c}
\mathbf{0_2} & \, \, \mathbf{I_2} \\
\hline
\mathbf{-I_2} & \, \, \mathbf{0_2}
\end{array}
\right],\qquad{\rm{with}}\qquad \boldsymbol{\Omega}^\intercal= \boldsymbol{\Omega}^{-1}=-\boldsymbol{\Omega},\qquad \boldsymbol{\Omega}^{2}=-\mathbf{I_4},\qquad {\rm{det}}\boldsymbol{\Omega}=1,
\end{equation}
where ``$\rm{det}$'' refers to the determinant of a matrix. The Hamiltonian equations were written as a matrix equation,
\begin{equation}\label{eq:Ham_eqs_matrix}
\dot{\mathbf{z}}=\mathbf{H}\,\mathbf{z},
\end{equation}
whose solution is given by a linear transformation
\begin{equation}\label{eq:TransferGr}
\mathbf{z}=\mathbf{T}\left(v,v_0\right)\mathbf{z_0}.
\end{equation}
The $4\times 4$ transformation matrix, $\mathbf{T}$, was obtained by an ordered exponential (OE) map of the Hamiltonian matrix, ${\mathbf{{H}}}$, i.e.,
\begin{equation}\label{eq:Transfer_OE}
\mathbf{T}\left(v,v_0\right)={\rm{OE}}\left[{\int _{v_0}^{v}{\mathbf{{H}}}dv}\right],\qquad {\rm{with}} \qquad \mathbf{T}\left(v_0,v_0\right)=\mathbf{I_4}.\nonumber \\
\end{equation} 

In \cite{Uzun:2018}, we put $\mathbf{T}$ in a block form by making use of $2\times 2$ submatrices $\{\mathbf{A},\mathbf{B},\mathbf{C},\mathbf{D}\}$, i.e,
\begin{equation}\label{eq:Transfer_block}
\mathbf{T}=
\left[
\begin{array}{c|c}
\mathbf{A} & \mathbf{B} \\
\hline
\mathbf{C} & \mathbf{D}
\end{array}
\right],
\end{equation}
in order to emphasize the analogy between the symplectic $ABCD$ transformation matrices of the Newtonian paraxial optics \cite{Torre:2005} and our ray bundle transformation matrix, $\mathbf{T}$. We should mention that this matrix was previously discovered in \cite{Fleury:2013sna} via a Wro\'{n}skian method in order to solve the geodesic deviation equation for arbitrary initial conditions without the realization of the underlying Hamiltonian structure. It was also used in other applications in order to propagate light in realistic scenarios \cite{Fleury:2014gha}.

In our previous work, we realized that the light bundle transformation matrix, $\mathbf{T}$, satisfies \cite{Uzun:2018}
\begin{equation}\label{eq:Symplectic_T}
\mathbf{T}^{\intercal}\,\mathbf{\Omega}\,\mathbf{T}=\mathbf{\Omega}, \qquad{\rm{with}}\qquad \rm{det}\,\mathbf{T}=1,
\end{equation}
as exponential map of Hamiltonian matrices are symplectic matrices. 
The symplecticity condition (\ref{eq:Symplectic_T}) imply that
\begin{align}\label{eq:symp_conds}
\mathbf{A}\mathbf{B}^{\intercal},\,\mathbf{A}^{\intercal}\mathbf{\mathbf{C}},\,\mathbf{B}^{\intercal}\mathbf{\mathbf{D}}\,\,\rm{and}\,\, \mathbf{\mathbf{C}}\mathbf{\mathbf{D}}^{\intercal}\,\, \rm{are\,\,symmetric,\,and}\,\,
\mathbf{A}\mathbf{\mathbf{D}}^{\intercal}-\mathbf{B}\mathbf{\mathbf{C}}^{\intercal}=\mathbf{I_2}.
\end{align}
Moreover, the Hamiltonian equations, (\ref{eq:Ham_eqs_matrix}), are equivalent to,
\begin{align}\label{eq:four_sets}
&\mathbf{\dot{A}}=\mathbf{C},\qquad \mathbf{\dot{C}}=\bm{\mathcal{R}}\mathbf{A}, \qquad \mathbf{A}\left(v_0,v_0\right)=\mathbf{I_2},\qquad\mathbf{C}\left(v_0,v_0\right)=\mathbf{0_2}, \nonumber \\
&\mathbf{\dot{B}}=\mathbf{D},\qquad \mathbf{\dot{D}}=\bm{\mathcal{R}}\mathbf{B},\qquad \mathbf{B}\left(v_0,v_0\right)=\mathbf{0_2},\qquad  \mathbf{D}\left(v_0,v_0\right)=\mathbf{I_2}.
\end{align}
due to (\ref{eq:Hamiltonian_matrix}) and (\ref{eq:TransferGr}).

In this subsection, we outlined the phase space dynamics and the Hamiltonian formulation of the light bundle propagation. We reminded that the screen-projected geodesic deviation variables can be linearly transformed from an initial point to a final point via a symplectic matrix, $\mathbf{T}$.

In the next subsection, we summarize the procedure to reach the Hamilton-Jacobi equation. Note that this is pivotal in the construction of the main body of the paper where we introduce the so-called paraxial wave equation of the bundle.
%%%%%%%%%%%
\subsection{\label{sec:Hamilton-Jacobi equations of the ray bundle} Hamilton-Jacobi equation of the ray bundle}
In this subsection, we summarize the relationship between the action functional of the null geodesic deviation and the Hamilton-Jacobi equation. In order to achieve this, for our generic canonical transformations,
\begin{align}\label{eq:gen_transf}
\boldsymbol{\xi '} \rightarrow \boldsymbol{\xi}=\boldsymbol{\xi}(\boldsymbol{\xi '},\boldsymbol{\dot{\xi '}};v),\qquad{\rm{and}}\qquad
\boldsymbol{\dot{\xi '}} \rightarrow \boldsymbol{\dot{\xi}}=\boldsymbol{\dot{\xi}}(\boldsymbol{\xi '},\boldsymbol{\dot{\xi}'};v),
\end{align}
we assume that 
\begin{equation}\label{eq:free_can_trans}
{\rm{det}}\left[\frac{\partial\left(\boldsymbol{\xi},\boldsymbol{\xi '}\right)}{\partial \left(\boldsymbol{\dot{\xi '}},\boldsymbol{\xi '}\right)}\right]={\rm{det}}\left[\frac{\partial \boldsymbol{\xi}}{\partial \boldsymbol{\dot{\xi}'}}\right]={\rm{det}}\mathbf{B}\neq 0,
\end{equation}
holds. Matrix $\mathbf{B}$ in the above is the upper-right block of the symplectic ray bundle transfer matrix, $\mathbf{T}$. The physical meaning of this assumption will become clear in a while. For the time being, we can take it as a mathematical assumption which guarantees that our ray bundle transformation is a \textit{free canonical transformation} \cite{Arnold:1978}. The associated $1$-form, $d{S}(\boldsymbol{\dot{\xi}'},\boldsymbol{\xi '};v)=\boldsymbol{\dot{\xi}'}d\boldsymbol{\xi '}-\boldsymbol{\dot{\xi}}d\boldsymbol{\xi}$,
of the transformation given in (\ref{eq:gen_transf}) is exact. Then, it can be locally expressed as ${S}(\boldsymbol{\dot{\xi}},\boldsymbol{\xi};v)\rightarrow S(\boldsymbol{\xi},\boldsymbol{\xi '};v)$
via a Legendre transformation where $S(\boldsymbol{\xi},\boldsymbol{\xi '};v)$ is given by
\begin{equation}\label{eq:gen_fnc_action}
S(\boldsymbol{\xi},\boldsymbol{\xi '};v)=\int _{\boldsymbol{\xi '}, v_0}^{\boldsymbol{\xi }, v}\boldsymbol{\dot{\xi}}d\boldsymbol{\xi}-Hdv.
\end{equation}
Note that the action given in (\ref{eq:gen_fnc_action}) is exactly equal to the reduced action, $S_{{\xi}}$, of the ray bundle given in (\ref{eq:action}). Therefore, $S=S_{\xi}$ is indeed the generating function of our free canonical transformation and from now on, we will drop the suffix $_{\xi}$ from the action functional of the geodesic deviation.

It can also be shown that for the corresponding free canonical transformation, the quadratic generating function $S(\boldsymbol{\xi},\boldsymbol{\xi'};v)$ can be written by matrix inner products as \cite{Moshinsky:1971,deGosson:2006}
\begin{align}\label{eq:gen_fun}
S(\boldsymbol{\xi},\boldsymbol{\xi'};v)=\frac{1}{2}\boldsymbol{\xi}^\intercal \mathbf{D}\mathbf{B}^{-1}\boldsymbol{\xi}-\boldsymbol{\xi'}^\intercal\mathbf{B}^{-1}\boldsymbol{\xi}
+\frac{1}{2}\boldsymbol{\xi'}^\intercal\mathbf{B}^{-1}\mathbf{A}\boldsymbol{\xi'}.\nonumber \\
\end{align}
In the Appendix of \cite{Uzun:2018}, we provided the explicit calculations to show that $S(\boldsymbol{\xi},\boldsymbol{\xi '};v)$ satisfies 
\begin{equation}
\boldsymbol{\dot{\xi}}=\frac{\partial S}{\partial \boldsymbol{\xi}}, \qquad \boldsymbol{\dot{\xi}'}=-\frac{\partial S}{\partial \boldsymbol{\xi}'},
\end{equation}
and the Hamilton-Jacobi equation
\begin{equation}\label{eq:HamiltonJacobi_cl}
\frac{\partial S}{\partial v}+H=0.    
\end{equation}
This is expected from any such generating function of a free linear canonical transformation \cite{deGosson:2006}. Similar proofs can also be found in other works in which one focuses on the linear canonical transforms of systems guided by quadratic Hamiltonians.

Let us now return to assumption (\ref{eq:free_can_trans}) and its physical relevance.
Recall that the angular diameter distance, ${D}_A$, and the luminosity distance, ${D}_L$, between an emitter and an observer are obtained via taking the ratios of cross-sectional areas and solid angles at the source and the observation points. For example, one can estimate the cross-sectional area at point $s$ via the measured solid angle at a vertex point $o$. Then, the linear canonical transform (symplectomorphysm) one needs to consider is
\begin{equation}\label{eq:Transfer_o_to_s}
\left[\begin{array}{c}
 \boldsymbol{\xi} \\
 \boldsymbol{\dot{\xi}}
\end{array}\right]_{s}= 
\left[
\begin{array}{c|c}
\mathbf{A} & \, \, \mathbf{B} \\
\hline
\mathbf{C} & \, \, \mathbf{D}
\end{array}
\right]_{(v_s,v_o)}
\left[\begin{array}{c}
 \boldsymbol{0} \\
 \boldsymbol{\dot{\xi}}
\end{array}\right]_{o},\,
\end{equation}
in order the calculate the angular diameter distance. Note that the initial deviation vector, $\vec{\xi'}=\vec{\xi}|_o$, is zero in that case as the observer is located at the vertex of the bundle. Then, the angular diameter distance is obtained via \cite{Perlick:2010,Uzun:2018}
\begin{equation}\label{eq:D_A_B}
{D}_A=\omega _o{\rm{det}}\left|\mathbf{B}\left(v_s,v_o\right)\right|^{1/2}.
\end{equation}
Likewise, one can reverse the situation and consider the case where the bundle is initiated at the source location. Then, the luminosity distance is given by
\begin{equation}\label{eq:D_L_B}
{D}_L=\omega _s{\rm{det}}\left|\mathbf{B}\left(v_o,v_s\right)\right|^{1/2}.
\end{equation}
The frequency factors in (\ref{eq:D_A_B}) and (\ref{eq:D_L_B}) appear due to the solid angles being defined with respect to proper length as in (\ref{eq:solid_angle}). Namely, one needs to use the relationship $\omega dv= dl$ in order to use the correct parameterization in the definition of the solid angle \footnote{We should also note that, matrix $\mathbf{B}$ is often denoted by the letter $\boldsymbol{\mathcal{D}}$ in the literature and it is referred to as the Jacobi matrix. This is mainly  because of the fact that only bundles starting from a vertex point are considered in the standard cosmological distance calculations. The physical importance of the total ray bundle transfer matrix and its extensions were appreciated only in a few studies \cite{Fleury:2013sna, Fleury:2014gha, Korzynski:2019oal, Grasso:2019, Serbenta:2021tzv, Korzynski:2024jqt}.}.

Now, it is clear that once the determinant of $\mathbf{B}$ becomes zero, the cross-sectional area of the bundle collapses to a point or a line and the distance estimations become impossible. Therefore, by assuming that ${\rm{det}}\mathbf{B} \neq 0$, one guarantees that there are no caustic points throughout propagation of the null bundle  \cite{Perlick:2010}. This is equivalent to the free canonical transform criterion given in (\ref{eq:free_can_trans}) which leads one to the Hamilton-Jacobi equation (\ref{eq:HamiltonJacobi_cl}). 

Here, we emphasize two things: (i) the caustics in question are the ones of an instantaneous wavefront rather than the global wavefront. The caustics of the latter is defined through the set of all points where the null cone ceases to be an immersion, i.e., a submanifold of the spacetime manifold. In that case, one needs to study the entire celestial sphere of the observer and its conjugate points. In our case, we are interested in only a small section of the wavefront and accordingly the criterion ${\rm{det}}\mathbf{B} = 0$ is sufficient to study the local effects. (ii) In general relativity, light bundles are non-twisting at the geometric optics limit. For generic null bundles that do not represent light, twist prevents the cross-sectional area to become zero \cite{Perlick:2010}. In that case ${\rm{det}}\mathbf{B} = 0$ would not provide the sufficient condition to define the caustics of the bundle in question.

In this section, we gave a brief summary of \cite{Uzun:2018} in which the dynamics of a thin null bundle was investigated on a reduced phase space. We outlined that the screen projections of the deviation vector act as canonical coordinates and their derivatives act as canonical momenta. We summarized the underlying Hamiltonian formalism of this linear system and showed how to obtain the Hamilton-Jacobi equation by standard techniques. 

In the next section, we present the main body of the paper which is essentially constructed on the correspondence between the linear canonical transformations of the null bundle and their associated metaplectic operators. We later show that this methodology allows one to define Gaussian beams on generic curved spacetimes which avoid caustics of the standard null bundles.
%%%%%%%%%%%%%%%%%%%%%%%%%%%%%%%%%%%%%%%%%%%%%%%%%%%%%%%%%%%%%%%%%
\section{\label{sec:Wavization} Wavization of an observed light beam}
In this section, we aim to answer the following question. Can one recover some of the wave-like properties of a light beam starting from its geometric optics limit?
In fact, a similar question was asked within the Newtonian optics community decades ago for light propagation in inhomogeneous refractive media \cite{Collins:1970, Mondragon:1986, Siegman:1986}. 
The main idea behind the answer to such question was to use the analogies between classical optics and quantum mechanics \cite{Dragoman:2002,Dragoman:2013}.

The procedure of obtaining a classical wave function by making use of quantum mechanical techniques is sometimes referred to as \textit{wavization} in the Newtonian optics community \cite{Castanos:1986,Torre:2005}. Most methods focus on phase space formulations in the paraxial regime. In those studies, the main focus is on the linear part of the propagation as light passes through an optical device or a medium. The corresponding evolution equation is the paraxial wave equation which takes the same form as the Schr{\"o}dinger equation \cite{Fock:1965,Kogelnik:1965,Kogelnik:1966,Arnaud:1969,Arnaud:1970,Bacry:1981,Wolf:1993sq,Torre:2005,deGosson:2017}. The time independent version is known as the paraxial Helmholtz equation. Those equations represent only classical effects and they do not refer to any quantum behavior.

Here, we present a similar work for light propagation on a generic curved spacetime. The aim is to recover some of the wave-like properties of a light beam by using some phase space quantization techniques.
In our approach, we adhere to the previous assumptions: (i) we focus on only thin light bundles, (ii) we consider sources of monochromatic light.
%%%%%%%%%
\subsection{The operator formalism}\label{eq:The operator formalism}
The operator formalism of quantum mechanics has been adapted by the Newtonian optics community and it has been used in many applications \cite{Nazarathy:1982,Mondragon:1986,Torre:2005}.
We follow a similar procedure here.

In Section \ref{sec:Summary of [30]: Reduced Hamiltonian optics}, we presented a summary of \cite{Uzun:2018} in which 2-dimensional screen components, $\boldsymbol{\xi}$, of the deviation vector, and their derivatives, $\boldsymbol{\dot{\xi}}$, play the role of canonical coordinates and momenta, respectively. This choice follows from the fact that the physical observables such as the size and the shape of a bundle, the wavefront curvature, power contained in the bundle or the distance estimations reside solely on the observational screen-projections of the deviation vector and its derivatives. The dyad basis of the observational screen is known as the Sachs basis, $\vec{s}_a$, with $\{a,b\}=\{1,2\}$. This basis is parallel propagated along the propagation direction, $\vec{k}$. As we chose to construct a phase space via the dyad basis components, the components $\boldsymbol{\xi}$ and $\boldsymbol{\dot{\xi}}$ are raised and lowered by a 2-dimensional Kronecker delta. This means that the Lagrangian sub-spaces of the phase phase is flat and the mathematical problem is analogous to the one of a 2-dimensional oscillator in classical mechanics.

Let us now define the phase space operators of our system. We start with the ones associated with the canonical coordinates, $\boldsymbol{\xi}$, and the canonical momenta, $\boldsymbol{\dot{\xi}}$. We define them through the Schr{\"o}dinger operators acting on $\mathcal{L^2}(\mathbb{R}^2)$ of square-integrable functions. To be more specific, they act on an arbitrary function $f$, as
\begin{align}\label{eq:operator_xi_dot}
\hat{\boldsymbol{\xi}}\left[{f}\right]:= {f} \,\cdot \, \boldsymbol{\xi},\qquad{\rm{and}}\qquad
\hat{\boldsymbol{\dot{\xi}}}\left[{f}\right]:=-i\boldsymbol{\nabla}f,\qquad {\rm{with}}\qquad \boldsymbol{\nabla}:=\frac{\partial\, }{\partial \boldsymbol{\xi}}.
\end{align}
Here,  $\boldsymbol{\nabla}$ represents the gradient operator with respect to the screen components of the deviation vector. Those operators satisfy the commutation rule $\left[\hat{\dot{\xi}}_a,\, \hat{{\xi}}_b\right]=-i\delta _{ab}$.
In order to obtain the Hamiltonian operator, we replace the canonical coordinates and momenta in the Hamiltonian function, (\ref{eq:Hamiltonian_matrix}), with their corresponding operators given in (\ref{eq:operator_xi_dot}), i.e.,
\begin{align}\label{eq:H_operator}
\hat{H}\left(\mathbf{z},t\right)= \frac{1}{2}\hat{\mathbf{z}}^{\dagger}\bm{\Omega}^\intercal{\mathbf{H}}\hat{\mathbf{z}}, \qquad {\rm{with}}\qquad
\hat{\mathbf{z}}=
\left[\begin{array}{c }
 \hat{\boldsymbol{\xi}} \\
 \hat{\boldsymbol{\dot{\xi}}} 
\end{array}\right]
=\left[\begin{array}{c }
 \hat{\xi} ^{{1}} \\
 \hat{\xi} ^{{2}} \\
 \hat{\dot{\xi}} _{{1}} \\
 \hat{\dot{\xi}} _{{2}}
\end{array}\right],
\end{align}
where ``$\dag$'' denotes the conjugate transpose operator.
Note that as we are considering only quadratic Hamiltonians, there is no ordering problem here. The Hamiltonian operator in (\ref{eq:H_operator}) can be written more explicitly as
\begin{equation}\label{eq:H_op_explicit}
\hat{H}=\frac{1}{2}\hat{\boldsymbol{\dot{\xi}}}^\dag\hat{\boldsymbol{\dot{\xi}}}-\frac{1}{2}\hat{\boldsymbol{\xi}}^\dag\mathbf{\mathcal{R}}\hat{\boldsymbol{\xi}},
\end{equation}
in which $\mathbf{\mathcal{R}}$ is the symmetric optical tidal matrix as in (\ref{eq:optical tidal matrix}) that incorporates the information about the spacetime curvature along the propagation direction. 
%%%%%%%%%%%%%
\subsection{Propagator of the wave function associated with the light beam}\label{sec:Propagator of the wave function associated with the light beam}
In the null bundle approach, the information about the trajectory of each geodesic in the bundle is coarse-grained. Here, we assign a wave function to this coarse-grained system of particles. Accordingly, the wave function can be understood as an effective wave function associated with a collection of non-interacting particles, i.e., the null dust. 
In the previous subsection, we presented the Schr{\"o}dinger operators of the phase space. Those will be used in the evolution equation of the light beam wave function in the current section. 

Recall from Section \ref{sec:Summary of [30]: Reduced Hamiltonian optics} that there exists a $4\times 4$ symplectic ray bundle transfer matrix, $\mathbf{T}$, that takes an initial phase space vector, $\mathbf{z}_0$, to a final one, $\mathbf{z}$. Those matrices form a symplectic group which we denote as $\rm{Sp}(4,\mathbb{R})$.
We now introduce the metaplectic group, $\rm{Mp}(4,\mathbb{R})$, which is the unitary representation of the double cover of $\rm{Sp}(4,\mathbb{R})$ \cite{Weil:1963}. This means that for every ray bundle transformation matrix, $\mathbf{T}$, there exist two unitary operators belonging to $\rm{Mp}(4,\mathbb{R})$ which differ by a sign \cite{Moshinsky:1971, Littlejohn:1985}. Metaplectic operators, $\hat{U}$, act on the space of square integrable functions and they are viewed as generalized quadratic Fourier transformers. Accordingly, they act as propagators for our square integrable wave functions.

Following \cite{Wolf:1979}, let us consider the integral transform 
\begin{equation}\label{eq:int_trans}
\Uppsi \left(\boldsymbol{\xi},v\right)=\hat{U}(\mathbf{T})[\Uppsi_0]:=\int_{-\infty}^{\infty} \Uppsi_0\left(\boldsymbol{\xi'},v'\right) K\left(\boldsymbol{\xi},v;\boldsymbol{\xi'},v'\right)d^2\boldsymbol{\xi}'.
\end{equation}
Here, the metaplectic operator, $\hat{U}(\mathbf{T})$, associated with the ray bundle transfer matrix, $\mathbf{T}$, acts as a propagator of the wave function along $\vec{k}$. 
Namely, it is represented by a kernel, $K\left(\boldsymbol{\xi},v;\boldsymbol{\xi '},v'\right)$, which propagates an initial wave function, $\Uppsi _0\left(\boldsymbol{\xi}',v'\right)$, to a final one, $\Uppsi\left(\boldsymbol{\xi},v\right)$. Let us recall one more time that the evolution parameter in question is the affine parameter, $v$, that parametrizes the integral curves of $\vec{k}$ in this work. The explicit form of the kernel can be obtained through the semi-classical method of Van Vleck \cite{VanVleck:1928}, Morette \cite{Morette:1951} and Van Hove \cite{VanHove:1951dra} (See also \cite{Choquard:1996}.). For a system defined on a 4-dimensional phase space, it takes the following form
\begin{equation}\label{eq:VanVleck_Kernel_orig}
K\left(\boldsymbol{\xi},v;\boldsymbol{\xi'},v'\right)=\left(\frac{1}{2\pi i}\right)|\Delta|^{1/2}\exp{\left(iS\right)},\qquad{\rm{where}}\qquad \Delta={\rm{det}}\left[-\frac{\partial ^2 S}{\partial \boldsymbol{\xi} \partial \boldsymbol{\xi}' }\right]
\end{equation}
is known as the \textit{Van Vleck determinant}. Note that in our case, $S$ corresponds to the action functional of the geodesic deviation vector which can be locally expressed through $\boldsymbol{\xi}$ and $\boldsymbol{\xi'}$. 
Then, substitution of 
\begin{equation}\label{eq:gen_fun_second time}
S(\boldsymbol{\xi},\boldsymbol{\xi'};v)=\frac{1}{2}\boldsymbol{\xi}^\intercal \mathbf{D}\mathbf{B}^{-1}\boldsymbol{\xi}-\boldsymbol{\xi'}^\intercal\mathbf{B}^{-1}\boldsymbol{\xi}
+\frac{1}{2}\boldsymbol{\xi'}^\intercal\mathbf{B}^{-1}\mathbf{A}\boldsymbol{\xi'},\nonumber 
\end{equation}
given in (\ref{eq:gen_fun}) into the kernel in (\ref{eq:VanVleck_Kernel_orig}) allows us to write it explicitly through the submatrices of the ray bundle transformation matrix, $\mathbf{T}$. It follows as 
\begin{equation}\label{eq:Kernel_wavization}
K\left(\boldsymbol{\xi},v;\boldsymbol{\xi }',v'\right)=\frac{1}{2\pi i}\left|{\rm{det}\left(\mathbf{B}^{-1}\right)}\right|^{1/2}\exp{\left[iS\left(\boldsymbol{\xi},\boldsymbol{\xi }',v\right)\right]}.
\end{equation}
In the Newtonian optics, the kernel analogous to the one in (\ref{eq:Kernel_wavization}) is known after Collins \cite{Collins:1970}. A similar result can be also found in \cite{Simon:2000b}. 

The wave function in (\ref{eq:int_trans}) that is attributed to a classical light beam here, is known to satisfy a Schr{\"{o}}dinger-like equation \cite{Fock:1965,Kogelnik:1966,Arnaud:1969,Arnaud:1970,Bacry:1981,Wolf:1993sq,Torre:2005,deGosson:2017}
\begin{equation}\label{eq:Sch_classic}
i\frac{\partial \Uppsi}{\partial v}=\hat{H}\Uppsi,
\end{equation}
which is sometimes referred to as the \textit{paraxial wave equation} in Newtonian optics. This follows from the fact that in the paraxial regime, second derivative of the wave function, with respect to the axis parameter, is assumed to be small and the ordinary wave equation reduces to (\ref{eq:Sch_classic})\cite{Torre:2005}. In our case, the problem is not much different, as we consider only thin light bundles which are studied via the linear geodesic deviation equation. The geodesic deviation variables of the bundle are transformed by a matrix transformation, $\mathbf{T}$, whose analogue in the Newtonian paraxial optics is known as the \textit{ABCD matrix}. Nevertheless, (\ref{eq:Sch_classic}) does not capture genuine wave behavior in our case, as the assumptions of the geometric optics regime still apply.

In the next subsection, we identify what the paraxial wave equation (\ref{eq:Sch_classic}) implies for null bundles. This analysis will be important for the following sections once we relate the wavization procedure to physical observables of a light beam.
%%%%%%%%%%%%%%%%%%%%
\subsection{Hydrodynamic analogy and the origin of the wave-like behavior}\label{sec:Hydrodynamic analogy and the origin of the wave-like behaviour}
In order to demonstrate what kind of physical information (\ref{eq:Sch_classic}) incorporates we use an analogy with the hydrodynamic evolution equations. A similar analogy has been considered since the early times of quantum mechanics. This formalism is known due to Madelung's work \cite{Madelung:1927} and it is often referred to as the \textit{hydrodynamic interpretation}. In the quantum case, the hydrodynamic interpretation is mathematically equivalent to the well-known de Broglie-Bohm theory \cite{DeBroglie:1927,Bohm:1952_I,Bohm:1952_II}. Madelung's approach and de Broglie-Bohm theory differ ontologically in the sense that the former is a genuine statistical interpretation while the latter is often interpreted as a deterministic theory. For our classical wavization problem, there is no necessity for an ``interpretation''. We will see in the following sections that those hydrodynamic evolution equations are closely related to the physically measurable quantities of a light beam without referring to any probabilistic framework.

Let us start investigating the paraxial wave equation by reconsidering the wave function that represents the null bundle. First, we write it in its \textit{polar form} as
\begin{equation}\label{eq:cl_wave_polar}
\Uppsi \left(\boldsymbol{\xi },v\right)=\mathscr{R}\exp{\left(i\mathscr{S}\right)},
\end{equation}
where $\mathscr{R}$ and $\mathscr{S}$ are \textit{real} functions. Once $\Uppsi$ is substituted in (\ref{eq:Sch_classic}), one can decompose the paraxial wave equation into its pure imaginary and real parts. We present them respectively as
\begin{align}
\frac{\partial \mathscr{R}}{\partial v}=-\frac{1}{2}\left[\mathscr{R}\boldsymbol{\nabla}^\intercal \boldsymbol{\nabla} \mathscr{S} +\left(\boldsymbol{\nabla} \mathscr{S}\right)^\intercal\left(\boldsymbol{\nabla} \mathscr{R}\right)+\left(\boldsymbol{\nabla} \mathscr{R}\right)^\intercal\left(\boldsymbol{\nabla} \mathscr{S}\right)\right], \label{eq:Sch_imaginary}
\end{align}
and
\begin{align}
\frac{\partial \mathscr{S}}{\partial v}=-\left[\frac{1 }{2}\left(\boldsymbol{\nabla} \mathscr{S}\right)^\intercal\left(\boldsymbol{\nabla} \mathscr{S}\right)-\frac{1}{2}\mathbf{\boldsymbol{\xi}}^\intercal\boldsymbol{\mathcal{R}}\mathbf{\boldsymbol{\xi}}-\frac{1}{2\mathscr{R}}\boldsymbol{\nabla}^\intercal\boldsymbol{\nabla} \mathscr{R}\right],
\label{eq:Sch_real}
\end{align}
where $\boldsymbol{\nabla}=\partial/\partial\boldsymbol{\xi}$ as in (\ref{eq:operator_xi_dot}). 

Let us now define 
\begin{equation}\label{eq:Intensity_and_eff_momenta_defn}
\mathscr{R}^2:=I=\left(\Uppsi^\dagger\Uppsi\right)\qquad {\rm{and}}\qquad \boldsymbol{\nabla}\mathscr{S}:=\mathbf{p}=p^1\vec{s_1}+p^2\vec{s_2},
\end{equation}
and substitute $\mathscr{R}$ and $\boldsymbol{\nabla}\mathscr{S}$ in (\ref{eq:Sch_imaginary}) and (\ref{eq:Sch_real}). Then, we can rewrite those equations respectively as
\begin{equation}\label{eq:Hydro_cont_generic}
\frac{\partial I}{\partial v}+\boldsymbol{\nabla}^\intercal \left(I \,\mathbf{p}\right)=0,
\end{equation}
and
\begin{equation}\label{eq:Hydro_Ham_Jacobi_generic}
\frac{\partial \mathscr{S}}{\partial v}=-\left(H\left(\boldsymbol{\xi},\mathbf{p}\right)+\mathcal{W}\right), \qquad 
\end{equation}
where 
\begin{equation}\label{eq:Wave_potential}
\mathcal{W}:=-\frac{1}{2\mathscr{R}}\boldsymbol{\nabla}^\intercal\boldsymbol{\nabla} \mathscr{R}.
\end{equation}
%Here, the function $H(\boldsymbol{\xi},\boldsymbol{\dot{\xi}})$ refers to the reduced Hamiltonian function of the ray bundle given in (\ref{eq:Red_Hamiltonian}) and we refer to $\mathcal{W}$ as the \textit{wave potential}. 

We realize that the pure imaginary part, (\ref{eq:Hydro_cont_generic}), of the paraxial wave equation corresponds to a continuity equation for  $I$ which represents the intensity of the beam. The real part, (\ref{eq:Hydro_Ham_Jacobi_generic}), corresponds to an equation similar to the Hamilton-Jacobi equation, (\ref{eq:HamiltonJacobi_cl}), of the underlying linear geodesic deviation of the null bundle with an extra term, $\mathcal{W}$. Also , note that in (\ref{eq:Hydro_Ham_Jacobi_generic}), the Hamiltonian function, $H$, inputs $\mathbf{p}$ instead of $\boldsymbol{\dot{\xi}}$. Following the hydrodynamic analogy, the $\mathbf{p}=\boldsymbol{\nabla}\mathscr{S}$ term can be interpreted as the \textit{representative} or \textit{effective momenta} of superposed bundles in our work similar to streamline momenta. 
With this, the associated equations of motion can be written as
\begin{equation}\label{eq:Semiclass_eq_motion}
\frac{d\boldsymbol{p}}{dv}=-\left(\boldsymbol{\nabla}\mathcal{V}+\boldsymbol{\nabla}\mathcal{W}\right),
\end{equation}
due to (\ref{eq:Sch_real}) where $\mathcal{V}=-\mathbf{\boldsymbol{\xi}}^\intercal\boldsymbol{\mathcal{R}}\mathbf{\boldsymbol{\xi}}/2$ is the ray bundle potential which was previously introduced in (\ref{eq:raybundle_potential}). Now, it is easy to see that if $\mathcal{W}\rightarrow0$, then
\begin{align}
\boldsymbol{p}\rightarrow \boldsymbol{\dot{\xi}},\qquad 
H\left(\boldsymbol{\xi},\mathbf{p}\right)\rightarrow H(\boldsymbol{\xi},\boldsymbol{\dot{\xi}}),\qquad 
\frac{d\boldsymbol{p}}{dv}=-\left( \boldsymbol{\nabla}\mathcal{V}+\boldsymbol{\nabla}\mathcal{W} \right)\rightarrow \ddot{\boldsymbol{\xi}}=\mathcal{R}\boldsymbol{\xi}.
\end{align}
This means that the additional potential term, $\mathcal{W}$, is the object that is responsible for the wave-like behavior of the superposed bundles. This is analogous to the case in the Madelung-de Broglie-Bohm formalism where the quantum potential is solely responsible for the quantum behavior of a given system. 

At this point, we should also mention that the hydrodynamic analogy being a useful tool in extracting the wave-like properties of classical and quantum systems was also discussed in \cite{Orefice:2009}. We believe this formalism is relevant in our problem as well since light bundles are represented by an effective stress-energy tensor of a null dust in the geometric optics limit. Most importantly, we will see in Sections \ref{sec:Evolution equations and the power conservation_G.B.} and \ref{sec:Comparison with point sources} that the hydrodynamic analogy helps us in distinguishing light beams which can avoid caustics form the ones which can not, in Section \ref{sec:Gaussian beams}.

In the next section, we study light bundles initiated from a point source. We investigate the evolution of the light bundle wave function and compare its associated observables with the ones of the standard null bundle approach.
%%%%%%%%%%%%%%%%%%%%%%%%%%%%%%%%%%%%%%%%%%%%%%%%%%%%%%%%%%%%%%%%%%%%
\section{Point sources}\label{sec:Point sources}
\subsection{Wave function and its physical relevance}\label{sec:Wave function and its physical relevance}
We now follow the standard assumption of optics in cosmology. That is, we consider a point source which emits spherical waves and only a small section of the wavefront is accessible to an observer who is located at a far distance. One can also reverse the situation and assume that an observer shoots a light beam on the past null cone. In either of the cases, the congruence of null geodesics originates from a vertex point and the initial deviation vector components are zero, i.e., $\boldsymbol{\xi }_0=\mathbf{0}$. 
Accordingly, the initial wave function we assign to a bundle can be written as a delta distribution, i.e.,
\begin{equation}\label{eq:init_wave_point_source}
    \Uppsi _0\left(\boldsymbol{\xi }',v'\right)=C\delta \left(\boldsymbol{\xi }'-\boldsymbol{\xi}_0,v'\right)=C\delta \left(\boldsymbol{\xi }'-\mathbf{0},v'\right),
\end{equation}
where $C$ is some constant. 
Once we substitute this initial wave function into the integral transform (\ref{eq:int_trans}),
we obtain the final wave function of the point source ($P.S.$) as
\begin{equation}\label{eq:wavefunc_point_source}
\Uppsi_{P.S.}\left(\boldsymbol{\xi },v\right)=\frac{C}{2\pi\left|{\rm{det}}\mathbf{B}\right|^{1/2}}\exp\left({\frac{i}{2}\left[\boldsymbol{\xi}^\intercal\mathbf{D}\mathbf{B}^{-1}\boldsymbol{\xi}-\pi\right]}\right),
\end{equation}
in its polar form $\Uppsi_{P.S.} \left(\boldsymbol{\xi },v\right)=\mathscr{R}_{P.S.}\exp{\left(i\mathscr{S}_{P.S.}\right)}$. Meaning, both the amplitude,
\begin{equation}\label{eq:R_P.S.}
\mathscr{R}_{P.S.}=\frac{C}{2\pi\left|{\rm{det}}\mathbf{B}\right|^{1/2}},  
\end{equation}
and the phase,
\begin{equation}\label{eq:S_P.S.}
\mathscr{S}_{P.S.}={\frac{1}{2}\left[\boldsymbol{\xi}^\intercal\mathbf{D}\mathbf{B}^{-1}\boldsymbol{\xi}-\pi\right]},  
\end{equation}
are real in this form. In order to understand what the phase function $\mathscr{S}_{P.S.}$ represents, we consider the generic bundle transformation. Let us recall that the geometry of a null bundle can be studied via two linearly independent solutions, $\vec{\xi}_\perp$ and  $\vec{\tilde{\xi}}_\perp$, of the null geodesic deviation equation projected on a local 2-dimensional screen. Namely, for
\begin{equation}
\mathbf{Q}=
\left[
\begin{array}{c c}
\xi{^1} & \, \, \tilde{\xi}{^1} \\
\xi{^2} & \, \, \tilde{\xi}{^2}
\end{array}
\right] \,\, {\rm{and}} \,\,
\mathbf{P}=
\left[
\begin{array}{c c}
\dot{\xi}{_1} & \, \, \dot{\tilde{\xi}}{_1} \\
\dot{\xi}{_2} & \, \, \dot{\tilde{\xi}}{_2}
\end{array}
\right],
\end{equation}
the wavefront curvature of the bundle can be obtained through $\mathbf{\Gamma}=\mathbf{P}\mathbf{Q}^{-1}$ as outlined in Section \ref{sec:Wavefront curvature matrix and its Riccati evolution equation}. It is easy to check that $\mathbf{Q}$ and $\mathbf{P}$ obey the same phase space transformation rules, (\ref{eq:TransferGr}), as the individual solutions $\boldsymbol{{\xi}}$ and $\boldsymbol{\tilde{\xi}}$. That is,
\begin{align}\label{eq:Q_and_P_generic_sol}
\mathbf{Q}=\mathbf{A}\mathbf{Q'}+\mathbf{B}\mathbf{P'}
\qquad{\rm{and}}\qquad
\mathbf{P}=\mathbf{C}\mathbf{Q'}+\mathbf{D}\mathbf{P'},
\end{align}
where the $2\times 2$ matrices $\mathbf{A},\mathbf{B},\mathbf{C}$ and $\mathbf{D}$ are the sub-blocks of the ray bundle transformation matrix, $\mathbf{T}$, as before. The primed objects indicate the initial values.
For the case of point sources we have $\mathbf{Q'}=\mathbf{0}$, as the initial deviation vectors are zero at a vertex point. Then, 
\begin{equation}\label{eq:Gamma_P.S.}   \mathbf{\Gamma}_{P.S.}=\mathbf{P}\mathbf{Q}^{-1}=\mathbf{D}\mathbf{B}^{-1}
\end{equation}
gives the wavefront curvature of a bundle initiated from a point source. Note that this term is exactly equal to the phase function $\mathscr{S}_{P.S.}$ in (\ref{eq:S_P.S.}) in the so-called  wavized case\footnote{Indeed, there is also a $\pi/2$ phase shift which will discuss in a future work.}. We also  recall that the deformation matrix, $\Gamma_{ab}$, relates the projected solutions of the geodesic deviation equation to their derivatives, i.e., $\dot{\xi}_a=\inner{\vec{s}_a,\xi ^b \mathcal{D} _{\vec{s}_b}\vec{k}}=\Gamma_{ab}\xi ^b$. Then, for null rays initiated from a point source, $\mathbf{\Gamma}_{P.S.}$ in (\ref{eq:Gamma_P.S.}) is exactly equal to the deformation matrix given in (\ref{eq:deform_matrix_defn}). Its components give the expansion and shear scalars of the bundle through (\ref{eq:deform_matrix_components}).

We now investigate what the amplitude, $\mathscr{R}_{P.S.}$, represents. From our previous discussions, we know that the determinant of $\mathbf{Q}$ gives the cross-sectional area of a ray bundle. Then, for a bundle initiated from a point source, i.e., $\mathbf{Q'}=\mathbf{0}$, we have
\begin{equation}\label{eq:Cross_sec_area_P.S.}
\delta \mathcal{X}_{P.S.}={\rm{det}}\mathbf{Q}={\rm{det}}\left(\mathbf{B}\mathbf{P'}\right).
\end{equation}
We remind that $\mathbf{P'}$ is the matrix through which one calculates the solid angles (See Section \ref{sec:Power conservation and distances}.). Then, we have
\begin{equation}
\mathscr{R}_{P.S.}\sim {\left|{\rm{det}}\mathbf{B}\right|^{-1/2}} \sim  \delta \mathcal{X}_{P.S.}^{-1/2}\sim D_{L}^{-1},
\end{equation}
where $D_{L}$ is the luminosity distance as in (\ref{eq:D_L_B}). This means that the amplitude of the wave function is inversely proportional to the luminosity distance as expected.

In summary, for a wavized beam initiated from a point source, the phase function, $\mathscr{S}_{P.S.}$, gives the wavefront curvature matrix and it takes the same form as the one for a standard null bundle. The amplitude, $\mathscr{R}_{P.S.}$, of the wave function is inversely proportional to the square root of the cross-sectional area of the beam. In other words, it is inversely proportional to the luminosity distance.
%%%%%%%
\subsection{The evolution equations and the power conservation}\label{sec:The evolution equations and the power conservation_P.S.}
We now analyze the pure imaginary and the real parts of the paraxial wave equation for a beam initiated at a point source.
We start with the pure imaginary part which takes the form of a continuity equation, (\ref{eq:Hydro_cont_generic}), for the intensity. For this, let us first write the intensity function for the point source case as,
\begin{equation}\label{eq:Intensity_P.S.}
I _{P.S.}=\Uppsi_{P.S.}^\dagger\Uppsi_{P.S.}=\left(\frac{C}{2\pi}\right)^2{\left|{\rm{det}}\mathbf{B}\right|^{-1}}.   
\end{equation}
The term that is interpreted as the effective momenta in the continuity equation follows as
\begin{equation}\label{eq:Phase_Eff_mom_P.S.}
\mathbf{p}_{P.S.}:=\boldsymbol{\nabla}\mathscr{S}_{P.S.}=\mathbf{\Gamma}_{P.S.}\boldsymbol{\xi},
\end{equation}
where $\mathscr{S}_{P.S.}$ is given in (\ref{eq:S_P.S.}). The second equality in the above follows from the fact that $\mathbf{\Gamma}_{P.S.}=\mathbf{D}\mathbf{B}^{-1}$ is symmetric\footnote{This is due to the symplectic symmetry $\mathbf{B}^\intercal\mathbf{D}=\mathbf{D}^\intercal\mathbf{B}$ of the ray bundle transformation matrix, $\mathbf{T}$, presented in (\ref{eq:symp_conds}).}. Then it is easy to show that the continuity equation takes the form,
\begin{equation}\label{eq:Cont_P.S._expansion}
\frac{d I_{P.S.}}{dv}=-{\rm{Tr}}\left(\mathbf{\Gamma}_{P.S.}\right)I_{P.S.}=-2\theta I_{P.S.},
\end{equation}
where $\theta$ is the expansion scalar\footnote{As 
$\mathbf{\Gamma}_{P.S.}$ is equal to the deformation rate matrix of the corresponding ray bundle, the trace of $\mathbf{\Gamma}_{P.S.}$ gives twice the expansion scalar through (\ref{eq:deform_matrix_components}).} of the null bundle. %as in Eq.~(\ref{eq:Gamma_decomp}). 
We also realize that this continuity equation is equivalent to the evolution equation of the squared amplitude of the electromagnetic wave in the leading order approximation of the JWKB-like method. That is, (\ref{eq:Cont_P.S._expansion}) is equivalent to (\ref{eq:evol_ampl_sq}) of the geometric optics limit.  

It is known that in quantum mechanics, the Schr{\"{o}}dinger equation preserves the norm of a wave function which is normalized to one. In our case, it is the paraxial wave equation, (\ref{eq:Sch_classic}), that preserves the non-unit norm. Namely, 
\begin{equation}
\mathcal{P}=\int_{0}^{\boldsymbol{\xi}^*}\left(\Uppsi^\dagger\Uppsi\right)d^2\boldsymbol{\xi},
\end{equation}
is invariant throughout the evolution. In our classical picture, $\mathcal{P}$ corresponds to the integrated intensity over the cross-sectional area. It represents the power contained within a light beam. 

 In order to show that $\mathcal{P}$ is conserved, we first consider the evolution equation for the cross-sectional area, $\delta \mathcal{X}_{P.S.}={\rm{det}}\mathbf{Q}={\rm{det}}\left(\mathbf{B}\mathbf{P'}\right)$, of the beam. Once we take its derivative along  the propagation direction, $\vec{k}$, we obtain
\begin{equation}\label{eq:evol_detQ_P.S.}
\frac{d\left({\rm{det}}\mathbf{Q}\right)}{dv}=\rm{Tr}\left(\mathbf{\Gamma}_{P.S.}\right)\left({\rm{det}}\mathbf{Q}\right),
\end{equation}
by making use of Hamilton's equations, (\ref{eq:four_sets}), in the matrix form.
Then, we observe that the power is indeed conserved, i.e.,
\begin{equation}\label{eq:Power_cons_P.S.}
\frac{d\mathcal{P}_{P.S.}}{dv}
=\frac{dI_{P.S.}}{dv}\delta \mathcal{X}_{P.S.}+I _{P.S.}\frac{d\delta \mathcal{X}_{P.S.}}{dv}=0,
\end{equation}
due to the evolution equations of intensity and the cross-sectional area, presented in (\ref{eq:Cont_P.S._expansion}) and (\ref{eq:evol_detQ_P.S.}) respectively. In other words, the imaginary part of the paraxial wave equation guarantees that the power is constant throughout the propagation. This means that the evolution equation of $I_{P.S.}$ is a genuine continuity equation, rather than being a mere analogy. We should note that this result is also equivalent to the photon number conservation of the bundle which is sometimes referred to as the \textit{area law} in the cosmological context as in (\ref{eq:cons_photon_flux}).

Let us now look into the real part of the paraxial wave equation. First thing to check is the value of $\mathcal{W}$ which is the object that  differentiates the evolution equations of superposed light bundles from the Hamilton-Jacobi equation of a standard null bundle. By substituting the amplitude function $\mathscr{R}_{P.S.}=C/\left(2\pi \left|{\rm{det}}\mathbf{B}\right|\right)^{1/2}$ in the definition of $\mathcal{W}$ (\ref{eq:Wave_potential}), we determine that  $\mathcal{W}_{P.S.}=0$. This follows from the fact that the intensity of the wave is homogeneous on the observational screen and its value is same as the one of the electromagnetic field corresponding to the central null geodesic in the geometric optics limit. 

Moreover, it is easy to show that the effective momenta, $\mathbf{p}_{P.S.}$, of the wavized beam given in (\ref{eq:Phase_Eff_mom_P.S.}) are indeed equal to $\boldsymbol{\dot{\xi}}=\mathbf{D}\mathbf{B}^{-1}\boldsymbol{{\xi}}$ which are the canonical momenta of the standard null bundle. Then, the real part of the paraxial wave equation is equal to the Hamilton-Jacobi equation, (\ref{eq:HamiltonJacobi_cl}), of the ordinary thin null bundle formulation. Accordingly, the equations of motion of the wavized beam are same as the screen projected geodesic deviation equations, $\ddot{\boldsymbol{\xi}}=\mathcal{R}\boldsymbol{\xi}$. 

We should note that this result is directly related to the evolution equation of the wavefront curvature matrix. As $\mathbf{\Gamma}_{P.S.}$ is equal to the deformation matrix, we expect it to satisfy the real, non-linear Riccati equation,
\begin{equation}\label{eq:Riccati_P.S._curvature}
{\mathbf{\dot{\Gamma}}_{P.S.}}+\mathbf{\Gamma}_{P.S.}\mathbf{\Gamma}_{P.S.}-\mathcal{R}=0.
\end{equation}
We observe that the equation above is indeed satisfied once we make use of the matrix Hamiltonian (\ref{eq:four_sets}) that govern the evolution of the ray bundle. Eventually, solving this evolution equation for $\mathbf{\Gamma}_{P.S.}$ is equivalent to solving the Sachs optical equations of a standard bundle as we discussed in Section \ref{sec:Wavefront curvature matrix and its Riccati evolution equation}.

 We complete this section by reaching the following conclusion. In the case of a perfect point source, the evolution equations of a ``wavized'' bundle involve the same information as the evolution equations of a standard ray bundle. No new information can be gained through the so-called wavization procedure. This situation is analogous to the propagation of almost-spherical waves in the paraxial Newtonian optics. In the next section, we explore more interesting solutions of the wavization procedure in which non-trivial results are obtained.
 %%%%%%%%%%%%%%%%%%%%%%%%%%%%%%%%%%%%%%%%%%%%%%%%%%%%%%%%%%%%%%
 \section{Gaussian beams}\label{sec:Gaussian beams}
The connection between the paraxial ray optics and the mathematical formulation of Gaussian beams was established around the 1960s in the Newtonian optics \cite{Kogelnik:1965,Kogelnik:1966,Arnaud:1969,Arnaud:1970}. On the other hand, there is no consensus in the literature on a unique method to obtain a Gaussian beam profile starting from the standard ray theory \cite{Siegman:1986,Cerveny:1990}. The only common ground is the introduction of complex variables. For example, in the early works, it was stated that when the position and the direction of a ray take complex values, the curvature of the wavefront also becomes complex. This results in a transverse profile for the wave amplitude \cite{Arnaud:1970, Deschamps:1971, Arnaud:1985}. In that case, the Riccati equation that govern the evolution of the curvature matrix of the wavefront and which is real in the ray optics scenario becomes complex \cite{Cerveny:1990, Berczynski:2006,Berczynski:2013}. For practical applications, authors independently consider a complex refractive index, a complex wave vector and a complex eikonal in order to study the Gaussian beams \cite{Choudhary:1974,Felsen:1976,Heyman:1983,Nowak:1993}. When the complex geometric optics was developed, it was argued that the light follows trajectories in a genuinely complex space and the initial values of a trajectory take complex values in order to create the wave effects \cite{Keller:1971,Kravtsov:1999,Berczynski:2013}. We should also note that a unified view was established recently through the introduction of a complex Maxwell stress-energy tensor in the Newtonian optics \cite{NietoVesperinas:2022}.

The method we adapt in this work is aligned mostly with the works of Arnaud \cite{Arnaud:1970} and Cerveny \cite{Cerveny:1988,Cerveny:1990} where the authors focus on those rays whose displacement vectors take complex values when measured from an axis. In our case, it is the relative dynamics of the central null geodesic and the outer-most geodesic that shapes the physical observables. Therefore, the location at which the ray pierces a transverse screen is obtained through the geodesic deviation vector.  In this respect, we focus on the complex sum of two real solutions of the geodesic deviation equation. 

In the following sections, we suggest a geometric construction which results in a Gaussian beam profile for a narrow beam initiated from a small, yet a finite sized source. The characteristic parameters of this beam depend on the curvature of the underlying spacetime rather than random processes.
%%%%%%%%%%%%%
\subsection{On the complex wavefront curvature}\label{sec:On the complex wavefront curvature}
 For a generic beam, the amplitude of the intensity of light changes not only in the propagation direction but also in the transverse direction. For a Gaussian beam,  the intensity is maximum at the center and it decays exponentially towards the edges of a $v=\rm{constant}$ surface. In order to have such an exponential decay, one requires a complex curvature matrix, $\mathbf{\Gamma}=\mathbf{\Gamma_R}+i\mathbf{\Gamma_I}$ of the wavefront, where $\mathbf{\Gamma_R}$ and $\mathbf{\Gamma_I}$ are real. To be more specific, the corresponding wave function of the beam should be in the form
\begin{equation}\label{eq:Gaussian_form}
    \Uppsi
    =\frac{\kappa}{\left({\rm{det}}\mathbf{Q}\right)^{1/2}}\exp{\left(\frac{i}{2}\boldsymbol{\xi}^\intercal\mathbf{\Gamma_R}\boldsymbol{\xi}\right)}\exp{\left(\frac{-1}{2}\boldsymbol{\xi}^\intercal\mathbf{\Gamma_I}\boldsymbol{\xi}\right)},
\end{equation}
where $\kappa=\left(I_0\,{\rm{det}}\mathbf{Q'}\right)^{1/2}$ is a constant with $I_0$ being  the initial intensity of the beam and $\mathbf{Q'}$ being the initial value of the complex matrix $\mathbf{Q}$.
Note that without the complex part, $\mathbf{\Gamma_I}$, of the wavefront curvature matrix, the exponential decay of the amplitude is not possible. Here, we should also mention that $\boldsymbol{\xi}$ is not representing the deviation vector of the outer-most geodesic of the bundle any longer. Rather, it is a variable on the 2-dimensional screen space. We also note that ${\rm{det}}\mathbf{Q}$ is independent of $\boldsymbol{\xi}$. Its specification will be clear in a while.

In the previous sections, we only considered those curvature matrices which are real. This was due to choosing real solutions of the geodesic deviation equation to construct $\mathbf{Q}$ and $\mathbf{P}$ matrices. In order to obtain the aforementioned complex curvature matrix, $\mathbf{\Gamma}=\mathbf{P}\mathbf{Q}^{-1}$, one needs to consider complex solutions, or as we do here, complex combinations of real solutions of the geodesic deviation equation. Let us write them explicitly as,
\begin{equation}
\mathbf{Q}=\mathbf{Q_R}+i\mathbf{Q_I}, \qquad{\rm{and}\qquad}\mathbf{P}:=\mathbf{\dot{Q}}=\mathbf{\dot{Q}_R}+i\mathbf{\dot{Q}_I}.
\end{equation}
In general, one can consider any four linearly independent solutions to construct the complex $\mathbf{Q}$ (and $\mathbf{P}$) matrices as long as certain criteria are satisfied. 
To be specific, in order for a Gaussian beam to be well defined throughout the propagation, it has to fulfill the following requirements \cite{Popov:1982}
\begin{itemize}
    \item [i)] The matrix $\mathbf{Q}$ should be regular for all values of the evolution parameter $v$. This means ${\rm{det}}\mathbf{Q}\neq 0\, \forall \,v$,
    \item [ii)] $\mathbf{\Gamma_I}$ should be positive definite,
    \item [iii)] The complex curvature matrix, $\mathbf{\Gamma}$, should be symmetric. In other words, both $\mathbf{\Gamma_R}$ and $\mathbf{\Gamma_I}$ should be symmetric matrices.
 \end{itemize}
It is known that as long as the criteria above are satisfied at the initial point, they are satisfied throughout the evolution \cite{Cerveny:1990}.
%%%%%%%%%%%%%
\subsection{Gaussian beams remain Gaussian}\label{sec:Gaussian beams remain Guassian}
In order for the Gaussian wave function in (\ref{eq:Gaussian_form}) to be a solution of the paraxial wave (\ref{eq:Sch_classic}) for each value of the affine parameter $v$, the solution has to preserve its form throughout the propagation. Then, one requires \cite{Kogelnik:1965,Arnaud:1970}
\begin{equation}\label{eq:evol_comp_wavefront_curv}
\mathbf{\Gamma}=\left(\mathbf{C}+\mathbf{D}\mathbf{\Gamma'}\right)\left(\mathbf{A}+\mathbf{B}\mathbf{\Gamma'}\right)^{-1}
\qquad{\rm{and}}\qquad
\mathbf{Q}= \mathbf{Q'}\left(\mathbf{A}+\mathbf{B}\mathbf{\Gamma'}\right),
\end{equation}
between the initial and final values of $\mathbf{\Gamma}$ and $\mathbf{Q}$.
The relations above are known as the \textit{ABCD law} in the Newtonian optics literature. Even though those results are a well-known, we provide a detailed proof of their derivation in \ref{sec:Evolution of Gamma and Q of a Gaussian beam}, for the completeness of the paper. We also would like to highlight that the relations in (\ref{eq:evol_comp_wavefront_curv}) hold due to the underlying symplectic symmetries of the ray bundle transformation matrix, $\mathbf{T}$.
%%%%%%%%%%
\subsection{The principal curvatures and the widths}\label{sec:The principal curvatures and the widths}
Recall from Section \ref{sec:Wavefront curvature matrix and its Riccati evolution equation} that for light bundles whose deviation vector is in orthogonal correspondence with its propagation vector, the principal curvatures of the wavefront are found through the eigenvalues of a real curvature matrix, $\boldsymbol{\Gamma}$. The wavefront, in that case, is defined by a surface which represents \textit{both} the constant phase \textit{and} the constant amplitude surfaces of a wave field. For Gaussian beams, the situation is different. Namely, the constant amplitude and the constant phase surfaces do not coincide. Accordingly, we need to define both the principal curvatures and the widths in order to characterize the geometry of the beam. This requires the simultaneous diagonalization of the real part, $\mathbf{\Gamma_R}$, and the imaginary part, $\mathbf{\Gamma_I}$, of the complex curvature matrix, $\mathbf{\Gamma}$. Essentially, we are looking for a non-singular, real transformation that takes the Sachs basis into a possibly non-orthonormal basis with respect to which the curvature matrix, $\mathbf{\Gamma}$, is diagonalized. We require this in order for the principal curvatures and the widths to be consistently defined.

Simultaneous diagonalization of two real matrices is not a trivial problem. However, if those two matrices are symmetric and at least one of the matrices is positive definite, then the simultaneous diagonalization is possible \cite{Hildebrand:1952}. In Section \ref{sec:On the complex wavefront curvature}, we mentioned that in order to have a well defined Gaussian beam, both $\mathbf{\Gamma_R}$ and $\mathbf{\Gamma_I}$ have to be symmetric. In addition, $\mathbf{\Gamma_I}$ has to be positive definite. Thus, diagonalization of $\mathbf{\Gamma}$ is possible. 

In \ref{sec:Determination of the principal curvatures and the widths}, we outline the set of transformations one needs to apply on $\mathbf{\Gamma_R}$ and $\mathbf{\Gamma_I}$ in order to diagonalize them simultaneously. Essentially, we transform the Sachs basis, $\vec{s}$, into a new spatial, 2-dimensional, real screen basis, $\Vec{e}$, via $\mathbf{s}=\mathbf{\mathcal{M}}\mathbf{e}$
where $\mathbf{\mathcal{M}}$ is orthonormal. Then, the Sachs basis components, $\boldsymbol{\xi}$, of the screen projected deviation vector are transformed via $\boldsymbol{\xi}=\mathbf{\mathcal{M}}\mathbf{r}$
where $\mathbf{r}$ is the position vector on the 2-dimensional screen in the new basis. The procedure presented in \ref{sec:Determination of the principal curvatures and the widths} allows us to put the wave function of a generic Gaussian beam in (\ref{eq:Gaussian_form}) into the following form,
\begin{equation}\label{eq:Gaussian_form_diag}
    \Uppsi
    =\frac{\kappa}{\left({\rm{det}\mathbf{Q}}\right)^{1/2}}\exp{\left(\frac{i}{2}\boldsymbol{r}^\intercal\mathbf{\Lambda}_{\Gamma_R}\boldsymbol{r}\right)}\exp{\left(\frac{-1}{2}\boldsymbol{r}^\intercal\mathbf{\Lambda}_{\Gamma_I}\boldsymbol{r}\right)}.
\end{equation}
Here, 
\begin{align}\label{eq:Princp_curv_widths_Gamma_diag}
\mathbf{\Lambda}_{\Gamma_R}=\mathbf{\mathcal{M}}^\intercal\Gamma_R\mathbf{\mathcal{M}}=  \left[
\begin{array}{c c}
K_1 & \, \, 0 \\
0 & K_2
\end{array}
\right], \qquad{\rm{and}}\qquad
\mathbf{\Lambda}_{\Gamma_I}=\mathbf{\mathcal{M}}^\intercal\Gamma_I\mathbf{\mathcal{M}}= \left[
\begin{array}{c c}
W_1^{-2} & \, \, 0 \\
0 & W_2^{-2}
\end{array}
\right],
\end{align}
with $\{K_1(v),K_2(v)\}$ being the principal curvatures of the Gaussian beam and $\{W_1(v),W_2(v)\}$ represent its widths given at some value of $v$. 
%%%%%%%%%%%%%%%%%
\subsection{Finite sources}\label{sec:Finite sources}
\subsubsection{The geometric construction:}\label{sec:The geometric construction}
This section presents the core of the paper where we suggest a method to geometrically construct a Gaussian beam. With this construction, the Gaussian beam fulfills the welldefinedness criteria (i)-(iii) of Section \ref{sec:On the complex wavefront curvature}. In addition, it allows one to obtain the physical parameters of the beam explicitly. We should note that the geometric construction we propose here formulates a Gaussian beam which initiates from a finite source. However, it should be noted that this is not a unique way of formulating a Gaussian beam and one might consider other options for different physical scenarios.
\begin{figure}[h]
\centering
\begin{subfigure}{.45\textwidth}
  \centering
  \includegraphics[scale=1.1]{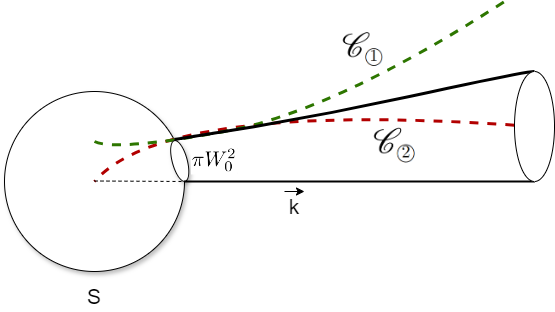}
  \caption{Two null congruences, $\mathscr{C}_{\one}$, and $\mathscr{C}_{\two}$, are initiated from a finite source, $S$, and represented by dashed lines. The Gaussian beam is obtained through their complex sum and it is sketched with solid lines here.}
  \label{fig:Gaussian beam}
\end{subfigure}%
\hspace{1cm}
\begin{subfigure}{.45\textwidth}
  \centering
 \scalebox{0.7}{
\begin{tikzpicture}
    \draw[rotate around={45:(0,0)}, MyRed, thick]  (0,0) ellipse (4cm and 2cm);
    \draw[rotate around={150:(0,0)}, MyGreen, thick]  (0,0) ellipse (4cm and 2cm);
    
    % Sachs basis
    \draw  (0,0) -- (4,0);
    \draw  (0,0) -- (0,4);
    \node[draw=none] at (-0.25,3.8) {$\vec{s}_2$};
    \node[draw=none] at (3.8,-0.25) {$\vec{s}_1$};

    % Semi-major and semi-minor axes of the ellipse
    \draw[thick,dotted]  (0,0) -- (4,4);
    \draw[thick,dotted]  (0,0) -- (-3.5,3.5);
    \node[draw=none] at (4.2,3.8) {$\vec{e}_1$};
    \node[draw=none] at (-3.6,3.2) {$\vec{e}_2$};
    
    % Vectors aligned with the semi-major and semi-minor axes
    \draw[->,thick,black]  (0,0) -- (2.8,2.8);
    \draw[->,thick,black]  (0,0) -- (-1.4,1.4);
    
    % Two arbitrary solutions for point source
    \draw[->,ultra thick,MyRed]  (0,0) -- (3.15,1.5);
    \draw[->,ultra thick,MyRed]  (0,0) -- (-0.9,1.9);
    \node[draw=none,MyRed] at (-1.1,2.2) {$\vec{\boldsymbol{\tilde{\zeta}}}_\perp$};
    \node[draw=none,MyRed] at (3.5,1.7) {$\vec{\boldsymbol{\zeta}}_\perp$};

    % Two arbitrary solutions for extended source
    \draw[->,ultra thick,MyGreen]  (0,0) -- (-3.6,1.2);
    \draw[->,ultra thick,MyGreen]  (0,0) -- (0.9,-2.5);
    \node[draw=none,MyGreen] at (-3.2,1.4) {$\vec{\boldsymbol{\tilde{\eta}}}_\perp$};
    \node[draw=none,MyGreen] at (1.2,-2.3) {$\vec{\boldsymbol{\eta}}_\perp$};
\end{tikzpicture}}
  \caption{Projected solutions of the geodesic deviation equation for congruences $\mathscr{C}_{\one}$ and $\mathscr{C}_{\two}$.}
\label{fig:solutions_two_cong}
\end{subfigure}
\label{fig:Gaussian beams2}
\caption{The sketch of Gaussian beams and the transverse screen.}
\end{figure}

%\begin{figure}[ht]
%\centering
%\includegraphics[scale=1.1]{Gaussian_beam_2_cong.png}
%\caption{Two null congruences, $\mathscr{C}_{\one}$, and $\mathscr{C}_{\two}$, are initiated from a finite source, $S$. Those are represented by dashed lines. The Gaussian beam is obtained through their complex sum and it is sketched with solid lines here.}
%\label{fig:Gaussian beam}
%\end{figure}

Let us begin with considering two null congruences, $\mathscr{C}_{\one}$ and $\mathscr{C}_{\two}$, which share the same central null geodesic with a tangent vector $\Vec{k}$ (See Figure (\ref{fig:Gaussian beam}).). We denote the Sachs basis components of two linearly independent, real solutions of the geodesic deviation equation of $\mathscr{C}_{\one}$ as $\boldsymbol{\eta}$ and $\boldsymbol{\tilde{\eta}}$. Similarly, the Sachs basis components of any two linearly independent, real solutions belonging to $\mathscr{C}_{\two}$ are denoted as $\boldsymbol{\zeta}$ and $\boldsymbol{\tilde{\zeta}}$ (See Figure (\ref{fig:solutions_two_cong}).). We use the deviation vectors of the first congruence, $\mathscr{C}_{\one}$, to construct the matrix $\mathbf{Q_R}$ and the ones of the second congruence, $\mathscr{C}_{\two}$, to construct $\mathbf{Q_I}$, i.e.,
\begin{equation}\label{eq:Q_R_and_Q_I_G.B.}
\mathbf{Q_R}=\left[
\begin{array}{c c}
\eta{^1} & \, \, \tilde{\eta}{^1} \\
\eta{^2} & \, \, \tilde{\eta}{^2}
\end{array}
\right],\qquad
\mathbf{Q_I}=\left[
\begin{array}{c c}
\zeta{^1} & \, \, \tilde{\zeta}{^1} \\
\zeta{^2} & \, \, \tilde{\zeta}{^2}
\end{array}
\right].
\end{equation}
\begin{align}
&\mathbf{Q_R}=\mathbf{A}\mathbf{Q'_R}+\mathbf{B}{\mathbf{P'_R}},\qquad
\mathbf{Q_I}=\mathbf{A}{\mathbf{Q'_I}}+\mathbf{B}\mathbf{P'_I},\nonumber \\
&\mathbf{P_R}=\mathbf{C}\mathbf{Q'_R}+\mathbf{D}{\mathbf{P'_R}},\qquad
\mathbf{P_I}=\mathbf{C}{\mathbf{Q'_I}}+\mathbf{D}\mathbf{P'_I}.\nonumber
\end{align}

Next, we choose different initial conditions for the solutions of the geodesic deviation equation for each congruence. Namely,
\begin{align}\label{eq:initial_conditions}
\{\boldsymbol{\eta '}\neq\mathbf{0}, \boldsymbol{\dot{\eta} '}=\mathbf{0}\} \qquad {\rm{and}}\qquad\{\boldsymbol{\tilde{\eta}'}\neq\mathbf{0},\boldsymbol{\dot{\tilde{\eta}'} }=\mathbf{0}\}\qquad {\rm{for}}\,\,\,  \mathscr{C}_{\one},\nonumber\\
\{\boldsymbol{\zeta '}=\mathbf{0}, \boldsymbol{\dot{\zeta} '}\neq\mathbf{0}\}\qquad {\rm{and}}\qquad \{\boldsymbol{\tilde{\zeta}'}=\mathbf{0},\boldsymbol{\dot{\tilde{\zeta}'}}\neq\mathbf{0}\}\qquad {\rm{for}}\,\,\,  \mathscr{C}_{\two}.
\end{align}
 We will see in a while that having a complex $\mathbf{Q}$, i.e., $\mathbf{Q}=\mathbf{Q_R}+i\mathbf{Q_I}$ as above allows us to extend the formalism from point sources to finite sources. 

 Next, we calculate the complex curvature matrix as 
\begin{align}\label{eq:Comp_curv_ABCD}
&\mathbf{\Gamma}=\mathbf{PQ}^{-1}=\left(\mathbf{P_R}+i\mathbf{P_I}\right)\left(\mathbf{Q_R}+i\mathbf{Q_I}\right)^{-1},\nonumber \\
&\mathbf{\Gamma}=\left(\mathbf{C}{\mathbf{Q'_R}}+i\mathbf{D}\mathbf{P'_I}\right)\left(\mathbf{A}\mathbf{Q'_R}+i\mathbf{B}\mathbf{P'_I}\right)^{-1},
\end{align}
with the initial conditions (\ref{eq:initial_conditions}), i.e., $\{\mathbf{P'_R}=\mathbf{0}, \mathbf{Q'_I}=\mathbf{0}\}$.
In addition, the initial values $\mathbf{Q'_R}$ and $\mathbf{P'_I}$ are chosen in such a way that they satisfy $\mathbf{Q_R'}={W_0}^2\mathbf{I_2}\mathbf{P'_{I}}$, where $W_0$ corresponds to the initial width of the beam that is sometimes referred to as the \textit{waist} or the \textit{spot size}. Such a choice of initial conditions allows one to study a beam which propagates starting from its waist. It incorporates information about the initial area of a finite size from which light rays are initiated.
Substitution of the initial conditions into the complex curvature matrix in (\ref{eq:Comp_curv_ABCD}) gives
\begin{align}\label{eq:complex_curv_finite_G.B.}
\mathbf{\Gamma}_{G.B.}=\left(\mathbf{\tilde{C}}+i{\mathbf{D}}\right)\left(\mathbf{\tilde{A}}+i{\mathbf{B}}\right)^{-1},
\end{align}
where we write $\mathbf{\tilde{A}}={W_0}^2{\mathbf{A}}$ and $\mathbf{\tilde{C}}={W_0}^2{\mathbf{C}}$ for simpler notation.
Then, the curvature matrix can be written in the form $\mathbf{\Gamma}_{G.B.}=\mathbf{\Gamma_R}+i\mathbf{\Gamma_I}$ with\footnote{In order to find the real and the pure imaginary parts of $\mathbf{\Gamma}_{G.B.}$, we first replace $\left(\mathbf{\tilde{A}}+i{\mathbf{B}}\right)^{-1}$ in (\ref{eq:complex_curv_finite_G.B.}) with $\left(\mathbf{\tilde{A}}-i{\mathbf{B}}\right)^{\intercal}\left[\left(\mathbf{\tilde{A}}+i{\mathbf{B}}\right)\left(\mathbf{\tilde{A}}-i{\mathbf{B}}\right)^{\intercal}\right]^{-1}.$
Then, we use the symplectic properties $\mathbf{A}\mathbf{B}^\intercal=\mathbf{B}\mathbf{A}^\intercal$ and $\mathbf{D}\mathbf{A}^\intercal-\mathbf{C}\mathbf{B}^\intercal=\mathbf{I_2}$ of the ray bundle transformation matrix given in (\ref{eq:symp_conds}). 
}
\begin{align}
&\mathbf{\Gamma_R}=\left[\mathbf{\tilde{C}}\mathbf{\tilde{A}}^\intercal+\mathbf{D}\mathbf{B}^\intercal\right]\left[\mathbf{\tilde{A}}\mathbf{\tilde{A}}^\intercal+\mathbf{B}\mathbf{B}^\intercal\right]^{-1},\label{eq:Gamma_R_ext_source} \\
&\mathbf{\Gamma_I}={W_0}^2\left[\mathbf{\tilde{A}}\mathbf{\tilde{A}}^\intercal+\mathbf{B}\mathbf{B}^\intercal\right]^{-1}.\label{eq:Gamma_I_ext_source}
\end{align}
Note that all of the criteria required for a well-defined Gaussian beam propagation are fulfilled in this case. Meaning, 
\begin{itemize}
    \item  [i)] $\mathbf{Q}=\mathbf{A}\mathbf{Q'}+i\mathbf{B}\mathbf{P'}$ is regular everywhere,\label{eq:Q_G.B._ext_source}
    \item [ii)] $\mathbf{\Gamma_I}$ is positive definite, 
    \item [iii)] $\mathbf{\Gamma_R}$ and $\mathbf{\Gamma_I}$ which are respectively given in (\ref{eq:Gamma_R_ext_source}) and (\ref{eq:Gamma_I_ext_source}) are symmetric matrices.
\end{itemize}
The proof of those results are presented in \ref{sec:Welldefinedness of the Gaussian beam for finite sources}. 
%%%%%%%%%%%%%%%%%%
\subsubsection{Evolution equations and the power conservation:}\label{sec:Evolution equations and the power conservation_G.B.}
We now study the evolution equations of the Gaussian beam presented in the previous subsection. 
Let us recall that in the case of a beam initiated from a point source, the wavization procedure does not bring any new information to the standard null bundle evolution equations. Here, we identify certain wave-like properties of a Gaussian beam while keeping track of the differences between beams of point sources and finite sources.

In order to investigate the Gaussian beam evolution within the hydrodynamic analogy, we have to put the Gaussian wave function in (\ref{eq:Gaussian_form}) into its polar form. However, it is not immediately obvious whether this is possible for a generic Gaussian wave function, as the determinant of $\mathbf{Q}$ that appears inside the square root is complex. On the other hand, the geometric construction we presented in Section \ref{sec:The geometric construction} allows us to perform the following manipulations.

We start with decomposing ${\rm{det}}\mathbf{Q}^{1/2}$, that appears on the denominator of the Gaussian wave function (\ref{eq:Gaussian_form}), into its real and pure imaginary parts. For this, we first substitute the initial conditions mentioned earlier into $\mathbf{Q}=\mathbf{A}\mathbf{Q'}+i\mathbf{B}\mathbf{P'}$. Then, its determinant follows as ${\rm{det}}\mathbf{Q}={\rm{det}}\left(\mathbf{\tilde{A}+i\mathbf{B}}\right){\rm{det}}\mathbf{P'}$.
In that case, the Gaussian wave function can be rewritten as
\begin{equation}\label{eq:Gaussian_wavefnc_E.S.}
    \Uppsi
    =\frac{\beta}{\left({\rm{det}}\left[\mathbf{\tilde{A}+i\mathbf{B}}\right]\right)^{1/2}}\exp{\left(\frac{i}{2}\boldsymbol{\xi}^\intercal\mathbf{\Gamma_R}\boldsymbol{\xi}\right)}\exp{\left(\frac{-1}{2}\boldsymbol{\xi}^\intercal\mathbf{\Gamma_I}\boldsymbol{\xi}\right)}.
\end{equation}
Here, the $\left({\rm{det}}\mathbf{P'}\right)^{-1/2}$ is absorbed in the constant $\kappa$ of (\ref{eq:Gaussian_form}), i.e., $\beta=\kappa\left({\rm{det}}\mathbf{P'}\right)^{-1/2}=I_0^{1/2}W_0^2$.  Now, the problem reduces to finding the real and the pure imaginary parts of $({\rm{det}}[\mathbf{\tilde{A}+i\mathbf{B}}])^{1/2}$.
For this, we use the fact that there exists a unitary matrix\footnote{In \ref{sec:Welldefinedness of the Gaussian beam for finite sources}, we reminded the Iwasawa pre-decomposition via which the symplectic matrix, $\mathbf{T}$, can be factorized into three portions: the shearing matrix, the pure magnifier and the fractional Fourier transformer. We highlight that the unitary matrix in (\ref{eq:unitary_Gaussian_beam}) is a slight modification of the unitary matrix, (\ref{eq:u_frac}), that is associated with the fractional Fourier transformer part of $\mathbf{T}$. They are equal when the initial width is set to 1.},
\begin{equation}\label{eq:unitary_Gaussian_beam}
\mathbf{u}= \left[\mathbf{\tilde{A}}\mathbf{\tilde{A}}^\intercal+\mathbf{B}\mathbf{B}^\intercal\right]^{-1/2}\left(\mathbf{\tilde{A}}+i\mathbf{B}\right),   
\end{equation}
associated with $\left(\mathbf{\tilde{A}+i\mathbf{B}}\right)$. It follows from the definition of the unitary matrices that, $\mathbf{u}$ can be uniquely factorized as a product of two unitary matrices, i.e., $\mathbf{u}=\mathbf{{u}_1}\mathbf{{u}_2}$, where $\mathbf{u_1}=\exp{\left(i\alpha/2\right)}\mathbf{I_2}$ and $\mathbf{u_2}\in$ SU(2). Then, taking the determinant of both sides of (\ref{eq:unitary_Gaussian_beam}) gives us\footnote{Here we use (\ref{eq:Gamma_I_ext_source}), and the symplectic condition $\mathbf{A}\mathbf{B}^\intercal=\left(\mathbf{A}\mathbf{B}^\intercal\right)^\intercal$ given in (\ref{eq:symp_conds}).}
\begin{equation}\label{eq:det_tildeA+iB}
{\rm{det}}\left(\mathbf{\tilde{A}+i\mathbf{B}}\right)= W_0^2\exp{\left(i\alpha\right)} \left({\rm{det}}\mathbf{\Gamma_I}\right)^{-1/2}.
\end{equation}
Finally, we substitute equation above into (\ref{eq:Gaussian_wavefnc_E.S.}) and we obtain the Gaussian wave function in its polar form as,
\begin{equation}
\Uppsi_{G.B.} =\mathscr{R}_{G.B.}\exp{\left(i\mathscr{S}_{G.B.}\right)},
\end{equation}
where
\begin{equation}\label{eq:Amplitude_G.B.}
\mathscr{R}_{G.B.}=\frac{I_0^{1/2}W_0}{{\rm{det}}\mathbf{\Gamma_I}^{-1/4}} \exp{\left(-\frac{1}{2}\boldsymbol{\xi}^\intercal\mathbf{\Gamma_I}\boldsymbol{\xi}\right)},  
\end{equation}
and
\begin{equation}\label{eq:Phase_G.B.}
\mathscr{S}_{G.B.}= \frac{1}{2}\boldsymbol{\xi}^\intercal\mathbf{\Gamma_R}\boldsymbol{\xi}-\frac{\alpha}{2}.
\end{equation}
Let us now investigate the physical observables associated with the Gaussian beam. We start with the intensity profile,
\begin{equation}\label{eq:Intensity_G.B.}
    I_{G.B.}= \Uppsi_{G.B.}^\dagger \Uppsi_{G.B.}= \frac{I_0 W_0^2\exp{\left(-\boldsymbol{\xi}^\intercal\mathbf{\Gamma_I}\boldsymbol{\xi}\right)}}{{\rm{det}}\mathbf{\Gamma_I}^{-1/2}}.
    \end{equation}
The first thing we realize is that as opposed to the intensity profile, $I_{P.S.}=I_{P.S.}(v)$, of a point source given in (\ref{eq:Intensity_P.S.}), the Gaussian beam intensity, $I_{G.B.}(\boldsymbol{\xi},v)$, changes along the transverse plane. Once we integrate the Gaussian beam intensity in ({\ref{eq:Intensity_G.B.}}) along the transverse plane, we find the power contained in the Gaussian beam, i.e., 
\begin{equation}\label{eq:Power_G.B.}
\mathcal{P}_{G.B.}=\int_{-\infty}^{\infty} I_{G.B} d^2\boldsymbol{\xi}=\mathcal{P}_{0},
\end{equation}
where $\mathcal{P}_{0}=I_0\left(\pi W_0^2\right)$ is a constant that gives the initial power contained within the beam.
Thus, similar to the bundle initiated from a point source, the power is conserved within our Gaussian beam initiated at a finite extend. This is the consequence of defining the beam in the geometric optics limit.

Next, we look into the evolution equations and we start our analysis with the imaginary part of the so-called paraxial wave equation. For this, we first calculate $\mathbf{p}_{G.B.}$ which is interpreted as the representative effective momenta of the beam. Through its definition in (\ref{eq:Intensity_and_eff_momenta_defn}) it follows as
\begin{equation}\label{eq:Eff_mom_G.B.}
\mathbf{p}_{G.B.}=\boldsymbol{\nabla}\mathscr{S}_{G.B.}=\mathbf{\Gamma_{R}}\boldsymbol{\xi}.
\end{equation}
Substitution of $I_{G.B.}$ %in Eq.~(\ref{eq:Intensity_G.B.}) 
and $\mathbf{p}_{G.B.}$ into the imaginary part of the paraxial wave equation, (\ref{eq:Hydro_cont_generic}), gives an evolution equation for $\mathbf{\Gamma_I}$ as,
\begin{equation}\label{eq:Sch_imag_G.B._evol_Gamma_I}
\mathbf{\dot{\Gamma}_I}+\mathbf{\Gamma_I}\mathbf{\Gamma_R}+\mathbf{\Gamma_R}\mathbf{\Gamma_I}=0.
\end{equation}
Note that this equation is equivalent to,
\begin{equation}
\frac{d\left({\rm{det}}\mathbf{\Gamma_I}\right)}{dv}=-2\left({{\rm{Tr}}}\mathbf{\Gamma_R}\right)\left({\rm{det}}\mathbf{\Gamma_I}\right).
\end{equation}
As the imaginary part of the wavefront curvature, $\mathbf{\Gamma_I}$, represents the widths of the beam after a proper diagonalization process, one reaches the conclusion that the imaginary part of the paraxial wave equation determines how the widths of the beam evolve.

We now consider the real part of the paraxial wave equation, that is in the form of a Hamilton-Jacobi equation with an extra potential, $\mathcal{W}$. Accordingly, we first obtain the extra potential, by substituting $\mathscr{R}_{G.B.}$ into its definition given in (\ref{eq:Wave_potential}). Then, we get
\begin{equation}\label{eq:wave_potential_G.B.}
    \mathcal{W}_{G.B.}=\frac{1}{2}\left({{\rm{Tr}}}\mathbf{\Gamma_I}+\boldsymbol{\xi}^\intercal \mathbf{\Gamma_I}\mathbf{\Gamma_I}   \boldsymbol{\xi}\right).
\end{equation}
Finally, substitution of this result, $\mathscr{S}_{G.B.}$ and $\mathbf{p}_{G.B.}$
into the real part of the paraxial wave equation, (\ref{eq:Hydro_Ham_Jacobi_generic}), gives us
\begin{equation}\label{eq:Sch_real_G.B._evol_Gamma_R}
\mathbf{\dot{\Gamma}_R}+\mathbf{\Gamma_R}\mathbf{\Gamma_R}+\mathbf{\Gamma_I}\mathbf{\Gamma_I}-\boldsymbol{\mathcal{R}}=0,
\end{equation}
with $\dot{\alpha}={{\rm{Tr}}}\mathbf{\Gamma_I}$.
Here, $\alpha$ is the argument of the determinant of the unitary matrix defined in (\ref{eq:unitary_Gaussian_beam}).
As the principal curvatures of the wavefront are obtained via the diagonalization of the real part of the curvature matrix, $\mathbf{\Gamma_R}$, one concludes that the real part of the paraxial wave equation provides an evolution equation for the principal curvatures of the beam.
%%%%%%%%%%%%%
\subsubsection{Comparison with point sources:}\label{sec:Comparison with point sources}
The Gaussian beam of a finite source demonstrates some wave-like properties as opposed to the wave function of a beam initiated at a point source. We list those properties as the following:
\begin{itemize}
    \item [(i)] One of the fundamental trait of wave-like phenomena in nature is that the amplitude and the phase functions of a wave are coupled. When we look into the evolution equation (\ref{eq:Sch_imag_G.B._evol_Gamma_I}) and (\ref{eq:Sch_real_G.B._evol_Gamma_R}) we realize that $\mathbf{\Gamma_I}$ and $\mathbf{\Gamma_R}$, which are essentially responsible for the intensity profile and the phase of the wave function respectively, are indeed coupled. The evolution equations of $\mathbf{\Gamma}_{R}$ and $\mathbf{\Gamma}_{I}$ can be studied under a single \textit{complex} Riccati equation 
\begin{equation}\label{eq:Complex_Riccati_G.B.}
{\mathbf{\dot{\Gamma}}_{G.B.}}+\mathbf{\Gamma}_{G.B.}\mathbf{\Gamma}_{G.B.}-\mathcal{R}=0,
\end{equation}
analogous to the Newtonian optics \cite{Berczynski:2006}. To be more specific, once we substitute $\mathbf{\Gamma}_{G.B.}=\mathbf{\Gamma}_{R}+i\mathbf{\Gamma}_{I}$ in (\ref{eq:Complex_Riccati_G.B.}), its imaginary part gives (\ref{eq:Sch_imag_G.B._evol_Gamma_I}) and its real part gives (\ref{eq:Sch_real_G.B._evol_Gamma_R}). At this point, we recall from Section \ref{sec:Point sources} that the wavefront curvature matrix, $\mathbf{\Gamma}_{P.S.}$, of the point source wave function also satisfies a non-linear Riccati equation, ${\mathbf{\dot{\Gamma}}_{P.S.}}+\mathbf{\Gamma}_{P.S.}.\mathbf{\Gamma}_{P.S.}-\mathcal{R}=0$. Only that this equation is real. 
Such an argument is also in line with the fact that constant-phase and constant-amplitude surfaces differ in Gaussian beams. Whereas, for a light bundle initiated from a point source, those two surfaces coincide.

\item [(ii)] In Section \ref{sec:Hydrodynamic analogy and the origin of the wave-like behaviour}, we outlined that once the wavefunction of the beam is written in its polar form, the evolution equation of the phase function, (\ref{eq:Hydro_Ham_Jacobi_generic}), takes the form of a Hamilton-Jacobi equation with an extra potential-like term, $\mathcal{W}$. We observe that for a point source wave function, $\mathcal{W}_{P.S.}=0$ holds. On the other hand, for a Gaussian wave function, $\mathcal{W}_{G.B.}\neq 0$ and it is given via (\ref{eq:wave_potential_G.B.}). Recall that this extra potential is solely responsible for the wave-like behavior of a bundle via the modification it brings to the equations of motion. In this respect, the equations of motion of a point source bundle are same as the projected geodesic deviation equation of standard bundles. Whereas, the equations of motion of a Gaussian beam differs from the projected geodesic deviation equation. This brings us to the next argument.

\item[(iii)] In this work, we denoted the momentum vector associated with the central geodesic of a bundle as, $\vec{k}$. On the other hand, null congruences are composed of many geodesics. The position of the outermost geodesic is represented by the canonical coordinate, $\boldsymbol{\xi}$, and the associated canonical momenta are represented by $\boldsymbol{\dot{\xi}}$ for a standard null congruence. In addition, the wavization procedure introduces a new object, $\mathbf{p}$, that acts as an effective momenta of the beam. For point sources, we observe that  $\mathbf{p}_{P.S.}=\boldsymbol{\dot{\xi}}$ holds. However, for the case of a Gaussian beam, $\mathbf{p}_{G.B.}$ is not equal to $\boldsymbol{\dot{\xi}}$ and it is obtained through (\ref{eq:Eff_mom_G.B.}). Mathematically speaking, this is due to the non-vanishing $\mathcal{W}_{G.B.}$ term that acts like an extra potential in our harmonic oscillator type problem. However, one should be careful in this analysis. Namely, the Gaussian beam is composed of two bundles, $\{\mathscr{C}_{\one}$, $\mathscr{C}_{\two}\}$, and $\mathbf{p}_{G.B.}$ keeps track of neither of the physical bundle trajectories. Rather, it is an object of statistical nature which emerges due to the superposition of two bundles. It acts like a streamline momenta in the hydrodynamic analogy. Accordingly, we prefer to use the term \textit{wave-like} effects to signal the underlying coarse-graining rather than claiming that those are associated with genuine wave effects.
\end{itemize}
There are also some similarities between the point source bundles and the Gaussian beams. For instance, the evolution of the term ${\rm{det}\mathbf{Q}}$ along the beam follows the same relation as in the point source case. That is,
\begin{equation}\label{eq:complex_evol_detQ_G.B.}
\frac{d\left({\rm{det}}\mathbf{Q}_{G.B.}\right)}{dv}=\rm{Tr}\left(\mathbf{\Gamma}_{G.B.}\right)\left({\rm{det}}\mathbf{Q}_{G.B.}\right),
\end{equation}
is similar to the evolution equation,~(\ref{eq:evol_detQ_P.S.}), of ${\rm{det}}\mathbf{Q}_{P.S.}$. The difference is that the equation for the Gaussian beam is complex and the one for the point source is real.
Nevertheless, the cross-sectional area of a beam can be obtained by the same object in either of the cases. That is  
\begin{equation}\label{eq:X_area_general_detQ}
 \delta \mathcal{X}:= {\rm{det}}\left(\mathbf{Q}^\dag \mathbf{Q} \right)^{1/2}.
\end{equation}

This brings us to the discussion of the point source limit of a Gaussian beam. This limit is obtained by taking the initial width of the beam to zero. Namely, when $W_0\rightarrow0$, we have
\begin{align}
    \mathbf{\Gamma}_I &\rightarrow 0,\nonumber \\
    \mathcal{W}_{G.B.} &\rightarrow 0,\nonumber\\
    \mathbf{p}_{G.B.} &\rightarrow \boldsymbol{\dot{\xi}},\nonumber\\
    \delta \mathcal{X}_{G.B.} &\rightarrow  \delta \mathcal{X}_{P.S.},\nonumber\\
    \psi_{G.B.} &\rightarrow  \psi_{P.S.},\nonumber\\
    {\rm{Complex}}\,\mathbf{\Gamma}_{G.B.}  &\rightarrow {\rm{Real}}\, \mathbf{\Gamma}_{P.S.},\nonumber\\
    {\rm{Complex\, Riccati\,Equation\,}} (\ref{eq:Complex_Riccati_G.B.}) &\rightarrow {\rm{Real\, Riccati\,Equation\,}}(\ref{eq:Riccati_P.S._curvature}).\nonumber
\end{align}

We then conclude with the following arguments. When a small yet a finite size source emits light, the corresponding wavefront is not expected to be perfectly spherical. Accordingly, a distant observer, who has access to a very small portion of this wavefront, is not anticipated to estimate a homogeneous intensity profile on his/her observational screen. Rather, the local wavefront can be modeled via the superposition of at least two congruences initiated from different points of the source. In a hypothetical scenario, the associated intensity then takes a Gaussian profile  which depends on the curvature of the underlying spacetime. Once the initial width of the beam is taken to be zero, the standard null bundle geometry is retrieved and the intensity profile is homogeneous on the observational screen.
%%%%%%%%%%%%%
\subsubsection{Comparison with Sbierski's work:}\label{sec:Comparison with Sbierski's work}
In the introductory Sections~\ref{sec:Work in the literature} and \ref{sec:Current work} we tried to clarify the differences of our work and the work in the literature. Among those, Sbierski's work \cite{Sbierski:2013mva} is the most relevant one for comparison of our results. Let us give a more detailed discussion here and list the differences between the two methods.

\begin{itemize}
\item [(i)] In Sbierski's work, one is after the approximate solutions of the Maxwell equation around a bicharacteristic whose projection to the spacetime is a null geodesic. One uses a \textit{Gaussian ansatz} for the wavefunction defined on the points of this curve. In our case, the Gaussian beams are constructed by a collection of rays. The wave function on each of the curve that makes up the bundle is defined through the \textit{locally plane wave ansatz} as for the standard null bundles in gravity. 
\item [(ii)] The Hamiltonian function and the choice of the canonical coordinates presented in the extended version of Sbierski's work \cite{Sbierski:2013arxiv}, are different in nature when compared to the ones introduced here. Namely, in \cite{Sbierski:2013arxiv}, for a null curve $\gamma(s)$, one defines the components of the tangent vector $d\phi:=\dot{\gamma}$ through $\partial _\mu \phi$. The canonical coordinates are chosen through the \textit{spacetime coordinates}, $x^\mu$ and the 4-\textit{momenta of the photon}, $p_\mu=\partial _\mu \phi$. The Hamiltonian in question is then $H(x^\mu,p_\nu)=g^{\mu \nu}p_\mu p_\nu/2$. Note that $d\phi.d\phi$ vanishes at the higher orders within the Gaussian ansatz. However, in the geometric optics limit, this function vanishes at the lowest order. What is more important is that the Hamiltonian system we solve in our work is the one of the null geodesic \textit{bundle}. Essentially, we are taking the local limit of an extended object which should in principal be treated bilocally. Our Hamiltonian involve the projections of the Riemann tensor along the line of sight.  Our phase space is 4-dimensional, rather than 8, where the phase space vector is defined through $\mathbf{z}=[\xi^a , \dot{\xi}_b]^\intercal$. Our Hamiltonian, $H(\boldsymbol{\xi}, \boldsymbol{\dot{\xi}};v)$, is constructed through the dyad components of the screen-projected deviation vector and their derivatives along the propagation. This makes it covariant under transformations that keep the propagation vector unchanged. This also allows for the evolution equations of the Gaussian beam to be directly related to the observer's local measurements, such as cross-sectional areas, distances, image distortions, etc.

\item [(iii)] In Sbierski's method, $H(x^\mu,p_\nu)$ is expected to satisfy $\partial _\alpha \partial _\beta \Big|_\gamma H=0$ at the second order. The corresponding equation is equivalent to a \textit{coordinate dependent} Riccati equation (Equation 2.27 of \cite{Sbierski:2013arxiv}). On the other hand, the evolution equations of our Gaussian beam are equivalent to a complex Riccati equation (\ref{eq:Complex_Riccati_G.B.}) which is covariant. As the Hamiltonians considered in the two cases are inherently different, there seems to be no direct relation between those two sets of equations. 

\item [(iv)] In Sbierski's method, the local conservation of the photon
flux density is satisfied approximately. In our case, the local flux density is conserved exactly. Accordingly, the power contained in the beam is conserved both for the standard point source bundles and our Gaussian beams.
\end{itemize}
In this respect, Siberski's Gaussian wave packets capture genuine wave effects and our Gaussian beams represent a collective behavior of superposed bundles.
%%%%%%%%%%%%%%%%%%%%%%%%%%%%%%%%%%%%%%%%%%%%%%%%%%%%%%%%%%
\section{Cosmological distances and caustic avoidance}\label{sec:Cosmological distances and caustic avoidance}
\subsection{Qualitative analysis}\label{sec:A qualitative analysis}
In cosmology, the cross-sectional areas, through which the distances are estimated, are obtained through thin null bundles initiated at a point source. However, as we discussed earlier, point sources are hypothetical objects and their intensities are by definition divergent at the source point. 

Previously, we showed that the cross-sectional area, $\delta \mathcal{X}$, of a beam is obtained through $\delta \mathcal{X}:= {\rm{det}}\left(\mathbf{Q}^\dag \mathbf{Q} \right)^{1/2}$ both for point source bundles and for the Gaussian beams. We remind that $\mathbf{Q}$ is a $2\times 2$ matrix which involves the screen-projections of the null geodesic deviation equation. Once the initial conditions in Section \ref{sec:Point sources} and in Section \ref{sec:Gaussian beams} are implemented, the cross-sectional areas of point source beams, $\delta \mathcal{X}_{P.S.}$, and the finite source Gaussian beams, $\delta \mathcal{X}_{G.B.}$, are respectively found as
\begin{align}
\delta \mathcal{X}_{P.S.}&={\rm{det}}\mathbf{P_i}|{\rm{det}}\mathbf{B}|,\label{eq:Compare_X-sec-area_P.S.}\\
\delta \mathcal{X}_{G.B.}&={\rm{det}}\mathbf{P'}{\rm{det}}\left({W_0}^4\mathbf{{A}}\mathbf{{A}}^\intercal+\mathbf{B}\mathbf{B}^\intercal\right)^{1/2},\label{eq:Compare_X-sec-area_G.B.}
\end{align}
where $W_0$ is the initial width of the beam.
In order to compare the two situations, we consider a case where the solid angle calculated at the vertex of a standard null bundle is equal to the divergence angle of the Gaussian beam of a finite source, i.e., ${\rm{det}}\mathbf{P_i}={\rm{det}}\mathbf{P'}$.
In that case, the Gaussian beam extends to a larger cross sectional area at a given $v$ value when compared to a bundle initiated from a vertex. This follows from ${W_0}^4\mathbf{{A}}\mathbf{{A}}^\intercal$ being positive definite.  Accordingly, by following the definition of the luminosity distance in (\ref{eq:distances_X-sec_angle}), we expect that $D_{L}^{G.B.}>D_{L}^{P.S.}$ holds, i.e., the luminosity distance estimated through a Gaussian beam coming from a finite source is larger than the one obtained through the standard point source bundle approach. Another way of putting this result is the following. If the power contained in a bundle initiated from a vertex is assumed to be equal to the power contained in a Gaussian beam, then the average intensity of a Gaussian beam at the observer's screen is smaller than the homogeneous intensity of a vertex bundle. The magnitude of this effect, of course depends on the spacetime curvature through matrix $\mathbf{{A}}$ and the initial value of the beam width, $W_0$. Given the fact that the typical sizes of coherent sources are very small compared to the distance traveled by light, the magnitude of this effect is expected to be tiny. 

This brings us to the discussion of caustics which are studied in two main categories regarding their global and local properties in general relativity: (i) null cone caustics, (ii) null bundle caustics. In the current work, we focus on the latter case. By null bundle caustics we refer to the caustics of an instantaneous wavefront which is defined through the intersection of the actual wavefront with a spacelike surface associated with the observer \cite{Hasse:1996,Perlick:2010}. In our case, the instantaneous wavefront is given via the locally defined, transverse, observational screen. Let us recall from Section \ref{sec:Hamilton-Jacobi equations of the ray bundle} that in the standard thin null bundle framework, caustics are those catastrophes where the cross-sectional area of the bundle collapses to a point or a line. This means that caustics are defined via the singular points of the so-called Jacobi matrix. This matrix corresponds to the upper right block of our ray bundle transformation matrix, $\mathbf{T}$. Then, ${\rm{det}}\mathbf{B}=0$ is the required condition in order to define a caustic in the standard literature. In that case, one can not estimate cosmological distances to caustics points as they are defined through ${\rm{det}}\mathbf{B}$ as in (\ref{eq:D_A_B}) and (\ref{eq:D_L_B}).

Coming back to the current work, we now compare the intensity of a bundle initiated from a vertex (point source) and the one of the Gaussian beam (finite source). We write them respectively as
\begin{align}\label{eq:Compare_wavefncs}
I_{P.S.}\sim |{\rm{det}}\mathbf{B}|^{-1},\qquad{\rm{and}}\qquad
I_{G.B.}\sim {\rm{det}}\left({W_0}^4\mathbf{{A}}\mathbf{{A}}^\intercal+\mathbf{B}\mathbf{B}^\intercal\right)^{-1/2},
\end{align}
through (\ref{eq:Intensity_P.S.}) and (\ref{eq:Intensity_G.B.}). We realize that the intensity of the point source beam, i.e, $I_{P.S.} \rightarrow \infty$ when ${\rm{det}}\mathbf{B}=0$. Accordingly, the wave function, $\psi_{P.S.}$, in (\ref{eq:wavefunc_point_source}) is also singular. On the other hand, the wave function and the intensity profile of a Gaussian beam naturally avoid caustics. Meaning, we can study the propagation of light beams through the caustic points of standard bundles and calculate their finite intensity profiles. Another way of seeing this result is that the cross-sectional area of a Gaussian beam, (\ref{eq:Compare_X-sec-area_G.B.}), does not have singularities as opposed to the cross-sectional area of a point source bundle, (\ref{eq:Compare_X-sec-area_P.S.}). This follows from the facts that: (i) The determinant of symplectic matrices are unity. Therefore, we can never have a symplectic ray bundle transformation matrix, $\mathbf{T}$, with both $\mathbf{A}$ and $\mathbf{B}$ sub-blocks being zero; (ii) Symplectic property of $\mathbf{T}$ also dictates that  
$\mathbf{A}\mathbf{\mathbf{D}}^{\intercal}-\mathbf{B}\mathbf{\mathbf{C}}^{\intercal}=\mathbf{I_2}$ should hold. Therefore, the case with $\mathbf{A}=\mathbf{0_2}$ and ${\rm{det}}\mathbf{B}=0$ is not allowed as long as $\mathbf{T}$ remains symplectic;
(iii) Since matrices ${W_0}^4\mathbf{{A}}\mathbf{{A}}^\intercal$ and $\mathbf{B}\mathbf{B}^\intercal$ are both positive definite, one can never have ${W_0}^4\mathbf{{A}}\mathbf{{A}}^\intercal=-\mathbf{B}\mathbf{B}^\intercal$ either. Then, $\delta \mathcal{X}_{G.B.}\neq 0$ should hold for each value of an affine parameter $v$.

We also would like to emphasize a profound property of metaplectic operators, which is at the core of the current work. For this, let us now consider even a stricter condition on the submatrix $\mathbf{B}$ than ${\rm{det}}\mathbf{B}=0$. We consider the case where $\mathbf{B}=\mathbf{0_2}$. It is known that even in such a case, the metaplectic operator used in this work acts on a wave function continuously. Namely, the kernel in (\ref{eq:Kernel_wavization}), takes the form \cite{Littlejohn:1985,Lopez:2019}
\begin{equation}\label{eq:Kernel_B=0}
K\left(\boldsymbol{\xi},v;\boldsymbol{\xi }',v'\right)=\frac{1}{\sqrt{{\rm{det}}\mathbf{A}^{-1}}}\delta \left(\boldsymbol{\xi}-\mathbf{A}\boldsymbol{\xi'}\right)\exp{\left(\frac{i}{2}\boldsymbol{\xi'}^\intercal\mathbf{A}^\intercal\mathbf{C}\boldsymbol{\xi'}\right)}.
\end{equation}
Thus, if we consider some arbitrary wave function, $\tilde{\Uppsi}\left(\boldsymbol{\xi'}\right)$, that successfully represents a light beam, its integral transform via the kernel above is given by \cite{Wolf:1979,Simon:2000b}
\begin{equation}\label{eq:Psi_trans_B=0}
\Uppsi\left(\boldsymbol{\xi};v\right)=\frac{1}{\sqrt{{\rm{det}}\mathbf{A}}}  \tilde{\Uppsi}\left(\mathbf{A}^{-1}\boldsymbol{\xi}\right)\exp\left(\frac{i}{2}\boldsymbol{\xi}^\intercal\mathbf{C}\mathbf{A}^{-1}\boldsymbol{\xi} \right).  
\end{equation}
If we also have $\mathbf{C}=\mathbf{0_2}$, then the phase factor in (\ref{eq:Psi_trans_B=0}) vanishes. If we further have $\mathbf{A}=\mathbf{I_2}$, then the kernel becomes $\delta \left(\boldsymbol{\xi}-\boldsymbol{\xi'}\right)$ and the integral transform is an identity transform.
%\footnote{
%We remind that this situation corresponds to the initial conditions of the bundle transfer matrix which is $\mathbf{T}\left(v_0,v_0\right)=\mathbf{I_4}$, as discussed in Section \ref{sec:Symplectic evolution}.}
%\footnote{We remind that this situation corresponds to the initial conditions of the null bundle transfer matrix being an identity.}.
Meaning, the integral transform (\ref{eq:int_trans}) never ceases to propagate the wave function. This property was first introduced and discussed in Littlejohn's seminal work \cite{Littlejohn:1985}. For a rigorous proof of the caustics avoidance of more general wave functions, we urge the reader to analyze Section 5 of Littlejohn's paper.

We finish this discussion by adding that the treatment of caustics via generic phase space rotations has been revived lately by Lopez \textit{et al.} \cite{Lopez:2019,Lopez:2020,Lopez:2021,Lopez:2022} in a different context. We believe that the application of those caustic avoidance methods to cosmological scenarios might shed new light on distance estimations in realistic inhomogeneous universe models \cite{Ellis:1998ha,Ellis:1998qga}. 
%%%%%%%%%%%%%%%%%%%%%%%%%%%%
\subsection{An example: Barriola-Vilenkin monopole}\label{sec:An example: The Barriola-Vilenkin monopole}
We consider light bundle propagation in Barriola and Vilenkin's global monopole spacetime \cite{Barriola:1989hx} as an application for caustic avoidance of Gaussian beams. In the literature, the global monopole in question is considered to be sourced by a global symmetry breaking. Such topological defects are important for studying phase transitions and the structure formation in the early universe. We should also mention that the lensing of global monopoles were previously studied as a stand alone in \cite{Barriola:1989hx, Durrer:1993tti, Perlick:2003vg, Jusufi:2017, Ono:2019} and within the context of black hole and white hole spacetimes in \cite{Cheng:2010nd, Man:2015, Ahmed:2023dvc}.

Our reason for analyzing light propagation in Barriola-Vilenkin spacetime in this paper is two-fold: (i) In this spacetime,  the solutions of the null geodesics and their geodesic deviation vectors can be explicitly written with respect to the affine parameter of the bundle. This makes our analysis analytically tractable. (ii) We would like to compare the geometric construction we propose here to a previous work of Perlick \cite{Perlick:2003vg} so that the caustic avoidance of the Gaussian beams can be explicitly shown.

The metric of Barriola-Vilenkin monopole in spherical coordinates is written as
\begin{equation}
    ds^2 = -dt^2 + dr^2 + \kappa^2\,r^2\left(d\theta ^2 + \sin^2\theta d\phi^2\right).
\end{equation}
This is one of the simplest non-flat solutions of the Einstein equations with the only independent component of the Riemann tensor being given by $\tensor{R}{^\theta _\phi _\theta _\phi} = \left(1-\kappa^2\right)\sin^2\theta$ where $1-\kappa^2$ is a measure of deficit solid angle of space when compared to the surface area of a 2-sphere \cite{Barriola:1989hx}.
As we would like to compare our result with the one in \cite{Perlick:2003vg}, we follow the same choices made by Perlick. Namely, we consider cases where $\kappa > 0$ and we constrain the null geodesics to follow $\theta=\pi/2$ equatorial paths. In addition, we consider the parametrization of the observer's sky by an angle $\Theta$. One can think of this parameter as the observer's viewing angle with respect to the equatorial plane. It corresponds to a direction of a source on the sky. Accordingly, it allows one to parameterize the initial conditions of a set of null geodesics $\vec{k} = \dot{t} \partial _t + \dot{r} \partial _r + \dot{\phi} \partial _\phi $ which are chosen as
\begin{align}\label{eq:Monopole_geod_ICs}
    t(0)&=0,\qquad\dot{t}(0)=-1,\nonumber\\
    r(0)&=r_0,\qquad \dot{r}(0)=\cos(\Theta), \nonumber \\
    \phi(0)&=0,\qquad \dot{\phi}(0)=\frac{\sin(\Theta)}{\kappa r_0}.
\end{align}
Here, the overdot represents the derivative operator with respect to the affine parameter, $v$. Then, by considering the null geodesic Lagrangian $\mathcal{L}=\left(-\dot{t}^2+\dot{r}^2+\kappa^2r^2\dot{\phi}^2\right)/2$ and the initial conditions (\ref{eq:Monopole_geod_ICs}), the Euler-Lagrange (geodesic) equations can be solved explicitly for the affine parameter, $v$. In that case, the affinely parametrized tangent of the null geodesic is found as
\begin{equation}
    \vec{k} = - \partial _t + \frac{\left(r^2-r_0^2\sin^2\Theta\right)^{1/2}}{r} \partial _r + \frac{r_0\sin\Theta}{\kappa r^2} \partial _\phi,
\end{equation}
with
\begin{equation}\label{eq:Monopole_sol_geod_r}
r(v) = \left(v^2 + 2r_0\cos{\Theta v} + r_0^2\right)^{1/2}
\end{equation}
being the solution of the null geodesic equation.
This is a past directed, $\inner{\,\vec{k}, \vec{u}}\,>0$, and outgoing, $\inner{\,\vec{k}, \vec{d}\,}>0$, null vector for an observer with normalized 4-velocity $\vec{u}=\partial _t$ and spatial observation direction vector $\vec{d}={\left(r^2-r_0^2\sin^2\Theta\right)^{1/2}}/(r)\partial _r  + {r_0\sin\Theta}/\left(\kappa r^2\right)\partial _\phi $.  The null tangent can be written as $\vec{k}=-\omega\left(\vec{u}-\vec{d}\right)$ with the value of the frequency, $\omega$, being unity. In this configuration, the value of $\omega$ is preserved throughout the propagation, i.e., observers do not measure any redshift.

In this example, we consider a slightly different method than the one used in Perlick's work \cite{Perlick:2003vg}. Nevertheless, we will observe that the results match for the bundles initiated from a vertex. Firstly, 
in order to obtain the screen-projected geodesic deviation vectors, we must obtain the optical tidal matrix components, $\tensor{{\mathcal{R}}}{_{\,}_{a}_{b}}:=\tensor{R}{_{\alpha}_{\mu}_{\nu}_{\beta}}s^\alpha_a k^\mu k^\nu s^\beta_b$. For this, we choose the following as the orthonormal basis of the observational screen
\begin{equation}
\vec{s}_1 = \frac{1}{\kappa r} \partial _\theta, \qquad \vec{s}_2 = \frac{r_0 \sin{\Theta}}{r}\partial _r -\frac{\left(r^2-r_0^2\sin{\Theta}\right)^{1/2}}{\kappa r^2}\partial _\phi.
\end{equation}
It is easy to show that $\{\vec{s}_1, \vec{s}_2\}$ satisfy the conditions of a Sachs basis, i.e., 
$\inner{\vec{s}_a, \vec{s}_b}=\delta _{a b}$, $\inner{\,\vec{u},\vec{s}_{a}\,}=0$, $\inner{\,\vec{d}, \vec{s}_{a}\,}=0$ and $\mathcal{D} _{\vec{k}}{\vec{s}}_{a}=0$ hold.

Now, we project the Riemann tensor on the screen and we make use of the solution of the geodesic $r(v)$ in (\ref{eq:Monopole_sol_geod_r}). Then, we find the optical tidal matrix components written explicitly as a function of the evolution parameter, $v$, as 
\begin{align}
\mathcal{R}_{11}&=\frac{\left(\kappa ^2 -1\right)r_0^2\sin^2\Theta}{\kappa^2\left(v^2 + 2r_0\cos{\Theta v} + r_0^2\right)^{2}},\\
\mathcal{R}_{12}&=\mathcal{R}_{21}=0, \qquad \mathcal{R}_{22} =0.
\end{align}
With this, the matrix screen-projected deviation equation, $\ddot{\boldsymbol{\xi}}=\boldsymbol{\mathcal{R}}\boldsymbol{\xi}$, reduces to two uncoupled  equations for $\xi^1$ and $\xi^2$ as $\ddot{\xi}^1=\mathcal{R}_{11}\xi^1$ and $\ddot{\xi}^2=0$. Both of those equations have exact analytical solutions. In order to solve for ${\xi}^1(v)$, we apply the following transformation:
\begin{equation}\label{eq:Monopole_geod_dev_tr}
\zeta := \frac{\xi^1}{\left(v^2 + 2r_0\cos{\Theta v} + r_0^2\right)^{1/2}}\qquad {\rm{and}} \qquad z:= \int \frac{dv}{v^2 + 2r_0\cos{\Theta v} + r_0^2}.
\end{equation}
Then, the deviation equation for $\xi ^1$ transforms into a constant coefficient linear equation for $\zeta$ as
\begin{equation}
\frac{d^2\zeta (z)}{dz^2}+\left(\frac{r_0\sin{\Theta}}{\kappa}\right)^2 \zeta (z) =0.
\end{equation}
Solving the above equation for $\zeta(z)$ and making use of the transformation (\ref{eq:Monopole_geod_dev_tr}) finally gives us
\begin{equation}\label{eq:Monopole_xi_1}
\xi ^1 = g(v)\left(C_1\sin{\left[{f(v)}\right]} + C_2\cos{\left[{f(v)}\right]}\right),
\end{equation}
with
\begin{equation}
g(v):=\left(v^2 + 2r_0\cos{\Theta v} + r_0^2\right)^{1/2}, \qquad
f(v):=\frac{\sin{\Theta}}{|\sin{\Theta}|}\frac{1}{\kappa}\arctan \left(\frac{v+r_0\cos{\Theta}}{r_0|\sin{\Theta}|}\right),
\end{equation}
and $\{C_1,C_2\}$ are constants. 

The solution for the second deviation vector component, $\xi^2$, is simply 
\begin{equation}\label{eq:Monopole_xi_2}
\xi^2(v)=C_3v+C_4,
\end{equation}
where $\{C_3,C_4\}$ are also constants which will be determined through the initial conditions one imposes on the light beam.

Now, we are going to analyze the magnitude of the intensity as a function of the viewing angle, $\Theta$, for two cases: (i) vertex beams initiated from point sources. This is the case explored in detail by Perlick \cite{Perlick:2003vg} as well to study the caustics of spherically symmetric spacetimes  in general, (ii) the Gaussian beams initiated from finite sized sources to show how the caustics of vertex beams are avoided. 

Let us first recall that the null bundle transformation matrix, $\mathbf{T}$, with sub matrices $\{\mathbf{A}, \mathbf{B}, \mathbf{C}, \mathbf{D}\}$ transforms the initial screen-projected deviation vector and its derivative along $\vec{k}$ to their final values.
A vertex beam with the initial conditions $\{\boldsymbol{\xi}_o = \boldsymbol{0}, \boldsymbol{\dot{\xi}}_o \neq \boldsymbol{0}\}$ allows us to find the components of $\mathbf{B}$ and $\mathbf{D}$ matrices. Whereas, the components of $\mathbf{A}$ and $\mathbf{C}$ matrices are obtained through a beam initiated at a finite extend with initial conditions $\{\boldsymbol{\xi}_o \neq \boldsymbol{0}, \boldsymbol{\dot{\xi}}_o = \boldsymbol{0}\}$. Once we impose those conditions on the newly found solutions $\xi^1$, $\xi^2$ and their derivatives, we find the components of the submatrices as
\begin{align}
A_{11} &= \frac{g(v)}{r_0}\left(\cos{[f(0)]}\cos{[f(v)]}+\sin{[f(0)]}\sin{[f(v)]}\right)-\frac{\cos{\Theta}}{r_0}B_{11},\label{eq:Monopole_A_11}\\
B_{11} &= \frac{\kappa g(v)}{\sin{\Theta}}\left(\cos{[f(0)]}\sin{[f(v)]}-\sin{[f(0)]}\cos{[f(v)]}\right)\label{eq:Monopole_B_11},
\end{align}
\begin{align}
A_{12}&=0=A_{21}, \qquad A_{22}= 1,\qquad C_{11} = \dot{A}_{11},\qquad C_{12}=0=C_{21}, \qquad C_{22}= 0. \label{eq:Monopole_A_12_21_22}\\
B_{12}&=0=B_{21}, \qquad B_{22}= v, \qquad D_{11} = \dot{B}_{11},\qquad D_{12}=0=D_{21}, \qquad D_{22}= 1.\label{eq:Monopole_B_12_21_22}
\end{align}
Now, we are going to demonstrate that the caustics of vertex bundles can be avoided with the Gaussian beams. Let us start with recalling that the caustics we are referring to are those points where the cross-sectional area, $\delta \mathcal{X}$, of infinitesimal light bundles becomes zero. Accordingly the apparent magnitude, $m = 2.5 \log _{10}D_U^2+M$ diverges to infinity. Here, distance $D_U$ is the uncorrected luminosity distance measured by astronomers 
%\footnote{In Section \ref{sec:Hamilton-Jacobi equations of the ray bundle}, we denoted the relativistic luminosity distance by $D_L$ which involves a factor of redshift correction as compared to $D_U$. Those distances are estimated via the cross-sectional area through $D_U=\left(1+z\right)D_L=\left(1+z\right)\left({\delta \mathcal{X}_o}/{\delta \Omega_s}\right)^{1/2}$. Note that, for the zero-redshift observers, those two distance measures become equal.}
and $M$ is chosen in such a way that $m$ is zero at $\Theta = 0$. We remind that, in this work, the relativistic luminosity distance is denoted by $D_L$ which involves a factor of redshift correction as compared to $D_U$. Namely, $D_U=\left(1+z\right)D_L=\left(1+z\right)\left({\delta \mathcal{X}_o}/{\delta \Omega_s}\right)^{1/2}$. Note that, for the zero-redshift observers, those two distance measures become equal.  
As the cross-sectional areas of vertex bundles and Gaussian beams are given through (\ref{eq:Compare_X-sec-area_P.S.}) and (\ref{eq:Compare_X-sec-area_G.B.}) respectively, we can compute the value of $D_U$ both for vertex bundles and the Gaussian beams. Accordingly, the respective apparent magnitudes follow as
\begin{align}
m_{\rm{vertex}} &= 2.5 \log _{10}|{\rm{det}}\mathbf{B}|+M,\\
m_{\rm{Gaussian}} &= 2.5 \log _{10}\left[{\rm{det}}\left({W_0}^4\mathbf{{A}}\mathbf{{A}}^\intercal+\mathbf{B}\mathbf{B}\right)^{1/2}\right]+M,
\end{align}
for a fixed solid angle, $\Omega_s$, at the source point. Then, by inserting the components of $\mathbf{{A}}$ and $\mathbf{{B}}$ matrices from the equation set (\ref{eq:Monopole_A_11})-(\ref{eq:Monopole_B_12_21_22}) we find the absolute magnitudes that would be measured by different types of beams in a global monopole spacetime. 
\begin{figure}[ht]
\centering
\includegraphics[width=0.6\textwidth]{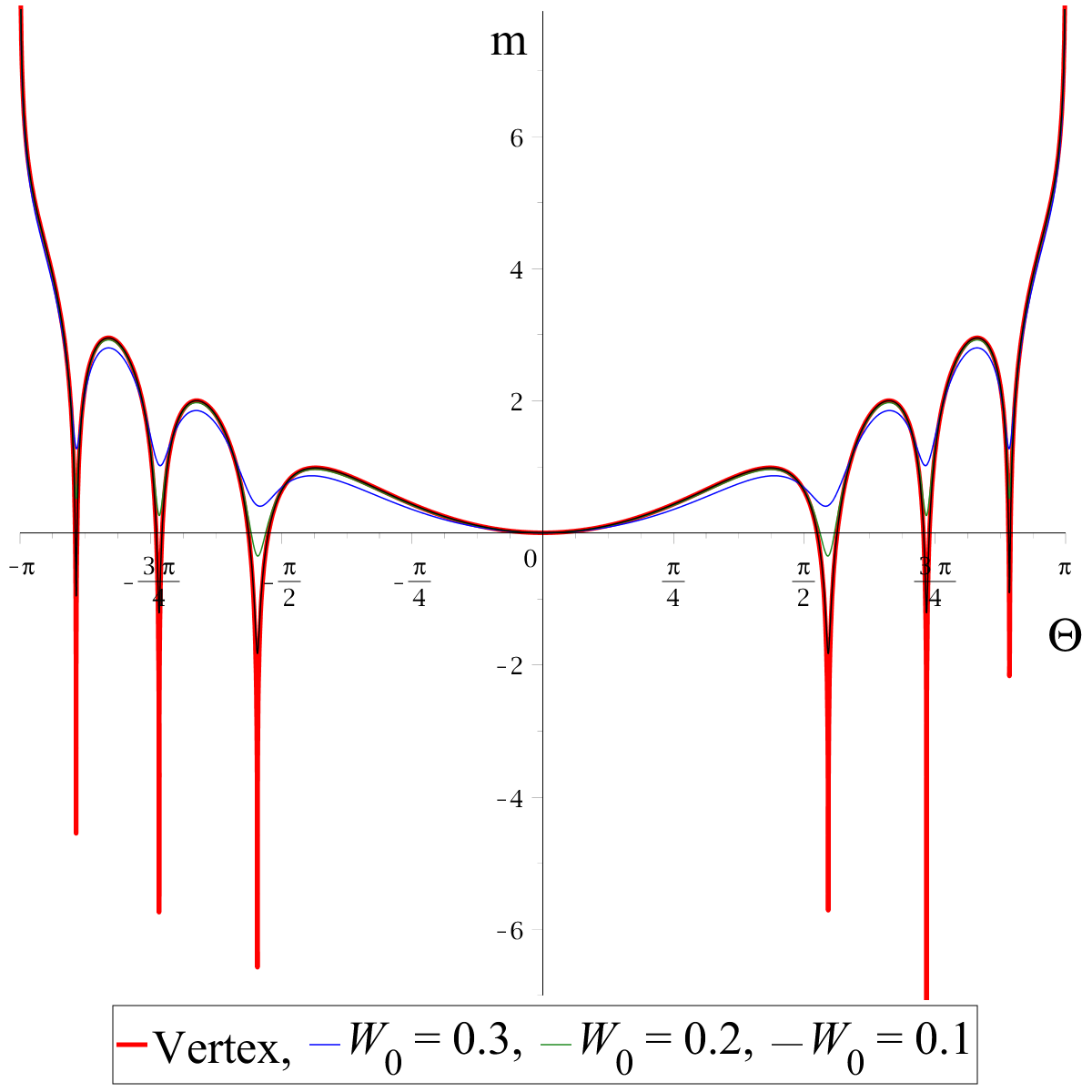}
\caption{Apparent magnitude, m, versus observation direction, $\Theta$. For an observer located at $r_0=0.77$, we consider sources at $r=1$ distributed on the observer's sky for each  $\Theta$ value. Affine distances to those fixed sources are obtained through the solution of the geodesic given in (\ref{eq:Monopole_sol_geod_r}). The strength of the monopole deficit solid angle is chosen as $\kappa=1/3.7$. The red curve represents the magnitude calculated for a vertex beam. The blue, green and black curves are those which correspond to Gaussian beams with initial widths $0.3, 0,2$ and $0.1$ respectively. The smaller the initial width is, the closer the Gaussian beam magnitude is to the one of a vertex bundle.} 
\label{fig:Monopole_mag_vs_angle_all}
\end{figure}
\begin{figure}[h]
\centering
\includegraphics[width=0.6\textwidth]{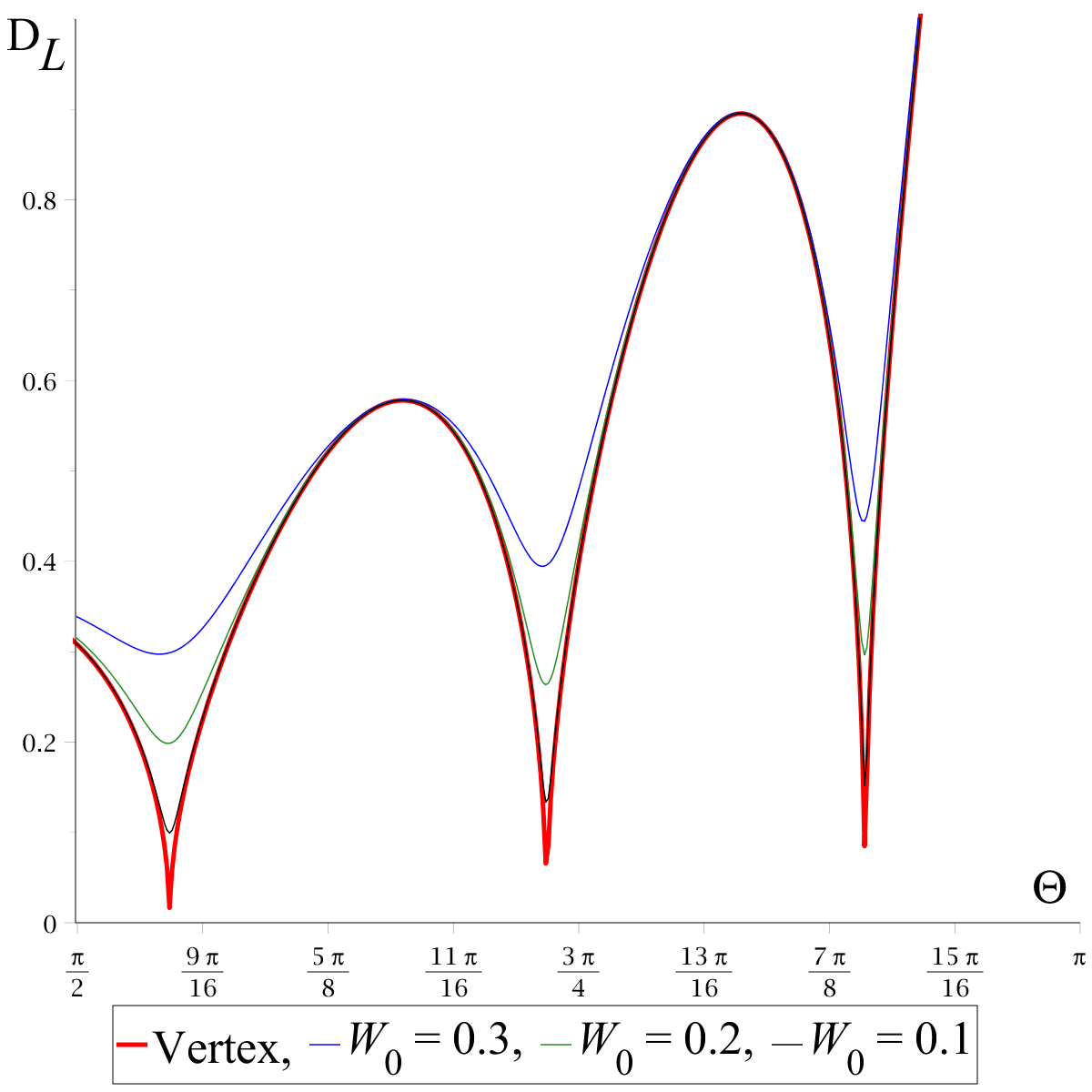}
\caption{Luminosity distance, $D_L$, versus observation direction, $\Theta$. The locations of the observer and the source in addition to the value of $\kappa$ are same as in the caption of Figure \ref{fig:Monopole_mag_vs_angle_all}. The red curve represents the estimated $D_L$ for a standard vertex bundle. Blue, green and black curves are those which correspond to the $D_L$ of Gaussian beams with initial widths $0.3, 0.2$ and $0.1$ respectively. The smaller the initial width is, the smaller the estimated luminosity distance becomes.}. 
\label{fig:D_lum_vs_angle_zoom}
\end{figure}
In Figure {\ref{fig:Monopole_mag_vs_angle_all}} we plot $m_{\rm{vertex}}$ and $m_{\rm{Gaussian}}$ as a function of the observation angle $\Theta$ for a source located at a fixed affine distance. It is observed that for standard point source light bundles, sources aligned with certain directions in the observer's sky are infinitely magnified. Our apparent magnitude plot of the vertex bundle  matches the one presented in Perlick's work \cite{Perlick:2003vg} which was obtained via a slightly different method. On the other hand, Gaussian beams avoid those singularities, allowing the distance calculations possible in every direction (See the caption for more details.). 
In Figure \ref{fig:D_lum_vs_angle_zoom}, we focus on the patch of the observer's sky in range $\pi /2<\Theta<\pi$. Here, the aim is to estimate the luminosity distance, $D_L$, by using a Gaussian beam initiated from a source located at $r=1$ while staying in the same physical set up. We observe that the larger the initial width of a Gaussian beam is, the larger the estimated $D_L$ becomes for a fixed direction on the sky. This result is inline with the discussion in Section \ref{sec:A qualitative analysis}. We should also note that the initial widths chosen in this example are numerically exaggerated in order to demonstrate the effect on our plots. For realistic scenarios, one would expect much smaller initial width to source distance ratios. Therefore, the effect of a single caustic on the luminosity distance estimated by using a Gaussian beam is expected to be small. In order to have a reasonable evaluation, one should consider more realistic spacetimes with inhomogeneous and anisotropic matter distributions where the effect of multiple caustics on the line of sight can be studied. At this point, we leave such an investigation for a future work.
%%%%%%%%%%%%%%%%%%%%%%%%%%%%%%%%%%%%%%%
\section{Some key insights}\label{sec:Key Insights and Limitations}
\subsection{Coherence vs. incoherence: Point, finite and extended sources}\label{sec:Coherence vs. incoherence: Point, finite and extended sources}
In astrophysical and cosmological applications, one usually considers two types of sources in general: (i) point sources which emit coherent light, such as massive stars, pulsars, supernova, etc. (ii) extended sources which are incoherent light sources, such as galaxies, clusters, or interstellar clouds. Observed images are put under one of those classes depending on the angular resolution of a given telescope.

The standard thin null bundle formalism of general relativity is relevant for point sources. In the current work, we suggested a framework to extend it to \textit{finite sources}. The Gaussian beam introduced in Section \ref{sec:Gaussian beams} initiates at a small yet finite extend at which the intensity is non-divergent. The idea was to model the local wavefronts of monochromatic, coherent sources while tackling the unrealistic mathematical singularities of hypothetical point sources. 

The mathematical construction we provide here is directly relevant when one studies collimated high energy sources such as pulsars. The presented construction of the intensity profiles represents the \textit{fundamental Gaussian mode} of emission from such sources. We note that this mode might not be observed directly. On the other hand, we will see in the next subsection that this method is (indirectly) relevant when modeling generic wavefronts or approximating certain wave effects in generic spacetimes. 
%%%%%%%%%%%%
\subsection{Fourier transform vs. Gaussian beam decomposition}\label{sec:Fourier transform vs. Gaussian beam decomposition}
In gravitational lensing studies, the effect of finite and extended sources in addition to the applications of wave optics phenomena have been inspiring interest for the last few decades \cite{Peters:1974, Herlt:1978, Schneider:1992, Nakamura:1999, Suyama:2005mx, Matsunaga:2006uc, Yoshida:2006, Lee:2009, Nambu:2013, Yoo:2013cia, Nambu:2016, Takahashi:2016jom, Turyshev:2017,  Fleury:2017we, Turyshev:2018, Tsukamoto:2018, Fleury:2018cro, Fleury:2019, Jow:2020rcy}. In the standard gravitational lensing theory, one determines the properties of intervening lenses (or matter distributions) via the distortions they create on the cross-sectional area of an image  with respect to a pre-defined background. Calculations of the magnification matrix and the flux of images are well studied for point sources. The underlying framework can be traced back to the standard thin bundle approximation. 

On the other hand, enlarging the theory of gravitational lensing for extended sources is not an easy task. For this, one usually assumes a \textit{weak field} regime such that the angular position of the unlensed source can be linked to the one of the image by adding the contributions associated with each point source that makes up the extended source in question \cite{Schneider:1992}. Similarly, the magnification of an extended source can be obtained via the integration of point source magnifications on the source plane weighted by some surface brightness profile of liking. Those methods were criticized on the basis that the thin bundle approximation is an over-simplified assumption to mimic the light propagation in the real universe \cite{Fleury:2017we,Fleury:2018cro,Fleury:2019}. The authors claim that one needs to consider the strong field regime of gravity and study thick bundles in order to model the extended source effect. We should note that since an extended source is usually considered as a collection of point sources, most formalisms break down when the caustics are introduced into the picture. Then, one has to beyond the ray picture and introduce the wave effects into the problem.

The proper inclusion of wave effects to standard lensing studies is possible due to the assumption that the background is (conformally) flat. When combined with the conformal invariance property of the Maxwell field, this allows one to write the physical electromagnetic field as a Fourier transform of another field which is the solution of the \textit{flat-space} Helmholtz equation \cite{Schneider:1992}. Then, the field at the observer's plane is obtained via the Kirchhoff's integral of the flat-space solution \cite{Born:1970} between the observer and the lens plane \cite{Schneider:1992}. Note that this method would not work for generic backgrounds.

In the Fourier transform framework, an input field is decomposed into a set of plane waves that move in different directions. Accordingly, the input waveform can be decomposed into the ones of the point sources. Those are mathematically modeled by delta distributions which act as the elementary waveforms of the Fourier method. However, this framework becomes problematic when one wants to adapt it to generic curved spacetimes. On the other hand, the construction suggested in the current work might find its use in the aforementioned areas. This follows from a technique which is sometimes referred to as the \textit{Gaussian beam decomposition} method in the Newtonian optics. 

Gaussian beam decomposition can be traced back to the work of Gabor \cite{Gabor:1946}. Superposition methods that involve Gaussian beams were mainly developed in \cite{Bastiaans:1980,  Popov:1982,  Cerveny:1983, Nowack:1984, Greynolds:1986a, Greynolds:1986b, White:1987, Cerveny:1988, Greynolds:2014} and recent improvements were given by \cite{Harvey:2015,Worku:2018, Worku:2020}. Loosely speaking, in this method, one (i) decomposes the input field into a set of (possibly overlapping) Gaussian beams which act as elementary waveforms, (ii) propagates each Gaussian beam through the optical geometry, (iii) superposes the set of beams at the output plane to determine the propagated field in question. Note that this method allows for \textit{any} wave field to be decomposed into Gaussian beams. Even though those superposition techniques are essentially approximation schemes, they are highly accurate in modeling generic wavefonts in the curved geometries of non-gravitational optics \cite{Cerveny:1990,Worku:2020}. In the current work, we established the Gaussian beams of thin null bundles on generic curved spacetimes. Even though those beams do not capture genuine wave effects, and thus should not be referred to as elementary wavelets in our case, they can still be used in similar decomposition techniques to reconstruct generic waveforms in an approximate manner. Accordingly, the aforementioned summation methods of the Newtonian theory can in practice be implemented for coarse-grained field estimations of extended sources in gravity. What is more profound is that this method allows for propagating fields through caustics of the standard thin null bundle theory as outlined here. Therefore, caustics can be avoided without solving the Maxwell equations in the wave optics regime. 
\subsection{Real universe, observations and simulations}\label{sec:Real universe, observations and simulations}
According to the standard concordance model of cosmology, the Universe can be modeled by a homogeneous and isotropic background spacetime and small perturbations around it. On the other hand, the real universe is known to be highly inhomogeneous and anisotropic especially at low redshifts. This is reflected in the geometry of light bundles as they are affected by the matter distribution. The effect is two fold: (i) dominated by the Ricci term, $\Phi_{00}$, causing convergence, (ii) dominated by the Weyl term, $\Psi_0$, causing shear. The latter is neglected in homogeneous and isotropic background models.

On small angular scales, where the light beams are expected to pass close to clumps of matter, strong lensing becomes relevant. The distorted light bundles form caustics which have an impact on cross-sectional areas, distances and accordingly the estimated expansion rate of the Universe. In \cite{Holz:1997ic} a flat FLRW background ($\Omega=1,\,\Lambda=0$) with Newtonian perturbations was considered. It was shown with a Monte-Carlo simulation that the percentage
of light beams which pass through caustics ranges from 5$\%$ to $35\%$ in the redshift range $z=0.5-3$. By comparing the cross-sectional areas of various beams in the perturbed model to the ones in the background, the authors conclude that the $50\%$ of all point source intensities would be lowered by a factor of two with respect to the intensities estimated within the background spacetime for $z=3$. This is when both the underdensities and caustics were taken into account in the strong lensing scenario.

In \cite{Ellis:1998ha}, authors argued that many astrophysical sources are relevant for strong lensing. From stars to star forming regions and galaxies various sizes of cusp angles are observed in range $3''-30''$ for real astrophysical objects. It is emphasized that even though cusp angles might be small for each lens, their cumulative effect is significant on cosmological distances. According to the authors, the effects of caustics should be estimated by keeping in mind their hierarchical structure, i.e., a beam might form multiple caustics throughout the propagation. In the same work, around $10^{22}$ caustics were estimated on the observer's past light cone until the surface of last scattering ($z\approx 1200$). It was argued that the all-sky-average of distances does not match the one of the homogenous and isotropic background model.

Here, we should note that the standard strong lensing formalism is formulated on (conformally) flat backgrounds. Accordingly, distances estimated within the standard formalism might not capture the realistic effects of caustics on cosmological distances. On the other hand, the strong lensing formalism have been extended to involve generic backgrounds by making use of the null geodesic deviation equation and the Jacobi map \cite{Frittelli:2000a, Frittelli:2000b}. We believe this formulation is highly underappreciated. Our formalism, combined with the authors', provides a means to estimate the effect of caustics on distances in the most general set up. We plan to focus on this problem in our future work.
%%%%%%%%%%%%%%%%%%%%%%%%%%%%%%%%%%%%%%%%%%%%%%%%%%%%
\section{Conclusions}\label{sec:Conclusions}
In the current work, we extended the thin null bundle formalism, that is associated with point sources, to light beams initiated from a finite extend. The framework we present here follows from our previous work in which thin light bundles were studied on a reduced phase space \cite{Uzun:2018}. In the aforementioned work, canonical coordinates and momenta were chosen as the observational screen projections of the geodesic deviation vector of the null bundle and their derivatives along the propagation direction. The underlying Hamiltonian formalism and the symplectic ray bundle transformation matrix are analogous to the ones in the Newtonian optics in the paraxial regime. This follows from the fact that only the linear part of the deviation is considered in almost all practical considerations, i.e., deviation vectors are represented by the Jacobi fields.

In the current work, we associated a classical wave function with a light beam and introduced the Schr{\"{o}}dinger operators associated with the canonical coordinates and momenta. We then showed that the wave function is propagated by a unitary transformation. This integral transform propagates the wave function along the null propagation direction and it is defined via the metaplectic operator associated with the underlying symplectic transformation matrices of the phase space. We remind that this procedure is already well-know in the Newtonian optics and it is sometimes referred to as \textit{wavization} \cite{Castanos:1986, Torre:2005}. With this method, it can be shown that the classical wave function in question satisfies an equation which is in the form of a Schr{\"{o}}dinger equation \cite{Fock:1965,Kogelnik:1965,Kogelnik:1966,Arnaud:1969,Arnaud:1970,Bacry:1981,Wolf:1993sq,Torre:2005,deGosson:2017}. 

We then used this construction to derive the wave function of light bundles initiated at a point source and at a finite source. We showed that the point source wave function and its evolution equation provide the same set of information as the standard thin bundle approach. Whereas, the information contained in a beam initiated from a finite sized source is richer. Namely, we found the Gaussian beam solutions of the so-called wavization procedure. This was achieved via a complex superposition of two null congruences which obey certain initial conditions. We showed that Gaussian beam solutions satisfy evolution equations similar to the Hamilton-Jacobi equation of the geodesic deviation vector, with an extra term. Due to this extra term, the beam is now represented by a modified momenta analogous to a streamline momenta. Namely, it represents the momenta of the superposed bundles rather than tracing the location of the outer-most geodesic of the bundle. This is one of the fundamental traits of our method which reflects its coarse-grained nature. We should note that the Gaussian intensity profile of our beam is directly related to the underlying spacetime curvature and not to random processes. The welldefinedness of the Gaussiam beam follows from the symplectic symmetries of the phase space. 

Finally, we showed that the Gaussian beams avoid caustics. To be specific, the intensity of a Gaussian beam does not diverge to infinity at the caustics of a standard  null bundle as the cross-sectional area of a Gaussian beam never collapses to a point (or a line). A specific analytically tractable example is demonstrated in a Barriola-Vilenkin monopole spacetime. The results were compared with other works in the literature. 

In summary, we eliminated the mathematical singularities of the thin null bundle formalism as we replaced the hypothetical point sources with more realistic finite sources. In practice, the Gaussian beams presented in this work can be used to study coherent sources. On the other hand, the Gaussian beam decomposition techniques \cite{Bastiaans:1980,  Popov:1982,  Cerveny:1983, Nowack:1984, Greynolds:1986a, Greynolds:1986b, White:1987, Cerveny:1988, Greynolds:2014}, have been improved to a level that any wavefront can now be decomposed into its fundamental Gaussian modes \cite{Harvey:2015,Worku:2018, Worku:2020}. If those methods can be successfully implemented to our framework, the extended source effect can be studied in generic curved spacetimes as well. We plan to investigate the validity and the applicability of those claims elsewhere.   
%%%%%%%%%%%%%%%%%%%%%%%%%%%%%%%%%%%%%%%%%%%%%%%%%%%%%%%%%%%%%%%%%%%%
\section*{Acknowledgments}
Many thanks to Volker Perlick, Pierre Fleury, Abraham Harte, Marius Oancea and Miko{\l{}}aj Korzy{\'n}ski for the useful discussions in the making of this paper. The author also thanks to Nicolas A. Lopez for a technical clarification on the various notations of metaplectic operators.  

This research is part of the project No. 2021/43/P/ST2/01802 co-funded by the
National Science Center and the European Union Framework Program for Research
and Innovation Horizon 2020 under the Marie Skłodowska-Curie grant agreement No.
945339. For the purpose of Open Access, the author has applied a CC-BY public
copyright license to any Author Accepted Manuscript (AAM) version arising from this submission.
%%%%%%%%%%%%%%%%%%%%%%%%%%%%%%%%%%%%%%%%%%%%%%%%%%%%%%%%%%%%%%%%%%%%
%\section*{References}
\bibliography{references} 
%%%%%%%%%%%%%%%%%%%%%%%%%%%%%%%%%%%%%%%%%%%%%%%%%%%%%%%%%%%%%%%%%%%%
\appendix
%%%%%%%%%%%%%%%%%%%%%%%%%%%%%%%%%%%%%%%%%%%%%%%%%%%
\section{Evolution of \texorpdfstring{$\mathbf{\Gamma}$}{Gamma} and \texorpdfstring{$\mathbf{Q}$}{Q} of a Gaussian beam}\label{sec:Evolution of Gamma and Q of a Gaussian beam}
In this section we prove the relationships in (\ref{eq:evol_comp_wavefront_curv}). For this, let us substitute the initial Gaussian wave function, 
\begin{equation}
    \Uppsi_0\left(\boldsymbol{\xi',v'}\right)
    =\frac{\kappa}{\left({\rm{det}}\mathbf{Q'}\right)^{1/2}}\exp{\left(\frac{i}{2}\boldsymbol{\xi'}^\intercal\mathbf{\Gamma'}\boldsymbol{\xi'}\right)}
\end{equation}
in the linear transform (\ref{eq:int_trans}) to study its evolution. Then, we obtain the final wave function as 
\begin{equation}\label{eq:Subs_Gaussian_into_prop}
    \Uppsi= \frac{\kappa}{2\pi i}\frac{\exp{\frac{i}{2}\left(\boldsymbol{\xi}^\intercal \mathbf{D}\mathbf{B}^{-1}\boldsymbol{\xi}\right)}}{\left(\rm{det}[\mathbf{Q'B}]\right)^{1/2}}\int \exp{\frac{i}{2}\left(\boldsymbol{\xi'}^\intercal\mathbf{m}\boldsymbol{\xi'}+2\boldsymbol{\xi'}^\intercal\mathbf{n}\right)}d^2\boldsymbol{\xi'},
\end{equation}
where 
\begin{equation}\label{eq:matrices_m_n}
\mathbf{m}=\left(\mathbf{B}^{-1}\mathbf{A}+\mathbf{\Gamma'}\right)\qquad {\rm{and}}\qquad \mathbf{n}=-\left(\mathbf{B}^{-1}\right)\boldsymbol{\xi}.
\end{equation}
Note that $\mathbf{m}$ is symmetric as both $\mathbf{\Gamma'}$ and $\mathbf{B}^{-1}\mathbf{A}$ are symmetric matrices. The latter follows from the symplectic symmetry $\mathbf{A}\mathbf{B}^\intercal=\left(\mathbf{A}\mathbf{B}^\intercal\right)^\intercal$ in (\ref{eq:symp_conds}). Then, the exponent term in the integrand of (\ref{eq:Subs_Gaussian_into_prop}) can be put into a more convenient form, i.e.,
\begin{equation}
\boldsymbol{\xi'}^\intercal\mathbf{m}\boldsymbol{\xi'}+2\boldsymbol{\xi'}^\intercal\mathbf{n}=\boldsymbol{\xi''}^\intercal \mathbf{m}\boldsymbol{\xi''}-\mathbf{n}^\intercal\mathbf{m}^{-1}\mathbf{n},
\end{equation}
where  $\boldsymbol{\xi''}=\boldsymbol{\xi'}+\mathbf{m}^{-1}\mathbf{n}$. This allows us to rewrite the integral (\ref{eq:Subs_Gaussian_into_prop}) as,
\begin{align}\label{eq:Subs_Gaussian_into_pro2}
    \Uppsi&=\frac{\kappa \exp{\frac{i}{2}\left[\boldsymbol{\xi}^\intercal \left(\mathbf{D}\mathbf{B}^{-1}\right)\boldsymbol{\xi}-\mathbf{n}^\intercal\mathbf{m}^{-1}\mathbf{n}\right]}}{2\pi i\left(\rm{det}[\mathbf{Q'B}]\right)^{1/2}}\int \exp{\frac{i}{2}\left(\boldsymbol{\xi''}^\intercal\mathbf{m}\mathbf{\boldsymbol{\xi''}}\right)}d^2\boldsymbol{\xi''}\nonumber \\
    &=\frac{\kappa\exp{\frac{i}{2}\left[\boldsymbol{\xi}^\intercal \left(\mathbf{D}\mathbf{B}^{-1}\right)\boldsymbol{\xi}-\mathbf{n}^\intercal\mathbf{m}^{-1}\mathbf{n}\right]}}{\left(\rm{det}[\mathbf{Q'B\mathbf{m}}]\right)^{1/2}}.
\end{align}
Substitution of $\mathbf{m}$ and $\mathbf{n}$ given in (\ref{eq:matrices_m_n}) back into (\ref{eq:Subs_Gaussian_into_pro2}) gives,
\begin{equation}\label{eq:Subs_Gaussian_into_pro3}
    \Uppsi=\frac{\kappa\exp{\frac{i}{2}\boldsymbol{\xi}^\intercal \left(\mathbf{D}\mathbf{B}^{-1}-\mathbf{B}^{\intercal-1}\left(\mathbf{A}+\mathbf{B}\mathbf{\Gamma'}\right)^{-1}\right)\boldsymbol{\xi}}}{\left(\rm{det}[\mathbf{Q'\left(\mathbf{A}+\mathbf{B}\mathbf{\Gamma'}\right)}]\right)^{1/2}}.
\end{equation}
Recall that   $\mathbf{B}^{-1}\mathbf{A}=\left(\mathbf{B}^{-1}\mathbf{A}\right)^\intercal$ and the second symplectic relation in (\ref{eq:symp_conds}) gives $\mathbf{D}\mathbf{A}^\intercal-\mathbf{C}\mathbf{B}^\intercal=\mathbf{I}$. Then, it is easy to show that $\mathbf{B}^{\intercal-1}=\mathbf{D}\mathbf{B}^{-1}\mathbf{A}-\mathbf{C}$ holds. We use this relation in order to simplify the term in the exponent of (\ref{eq:Subs_Gaussian_into_pro3}). Namely,
\begin{align}
   &\mathbf{D}\mathbf{B}^{-1}-\mathbf{B}^{\intercal-1}\left(\mathbf{A}+\mathbf{B}\mathbf{\Gamma'}\right)^{-1}\nonumber \\
&\qquad=\mathbf{D}\mathbf{B}^{-1}+\left(\mathbf{C}-\mathbf{D}\mathbf{B}^{-1}\mathbf{A}\right)\left(\mathbf{A}+\mathbf{B}\mathbf{\Gamma'}\right)^{-1}\nonumber \\
&\qquad=\left[\mathbf{D}\mathbf{B}^{-1}\left(\mathbf{A}+\mathbf{B}\mathbf{\Gamma'}\right)+\left(\mathbf{C}-\mathbf{D}\mathbf{B}^{-1}\mathbf{A}\right)\right]\left(\mathbf{A}+\mathbf{B}\mathbf{\Gamma'}\right)^{-1}\nonumber \\
&\qquad=\left(\mathbf{C}+\mathbf{D}\mathbf{\Gamma'}\right)\left(\mathbf{A}+\mathbf{B}\mathbf{\Gamma'}\right)^{-1}.
\end{align}
With this result, we rewrite the wave function in (\ref{eq:Subs_Gaussian_into_pro3}) as,
\begin{equation}\label{eq:Subs_Gaussian_into_pro4}
    \Uppsi=\frac{\kappa\exp{\frac{i}{2}\boldsymbol{\xi}^\intercal \left(\left(\mathbf{C}+\mathbf{D}\mathbf{\Gamma'}\right)\left(\mathbf{A}+\mathbf{B}\mathbf{\Gamma'}\right)^{-1}\right)\boldsymbol{\xi}}}{\left(\rm{det}[\mathbf{Q'\left(\mathbf{A}+\mathbf{B}\mathbf{\Gamma'}\right)}]\right)^{1/2}}.
\end{equation}
Comparing (\ref{eq:Subs_Gaussian_into_pro4}) with the generic form of the Gaussian wave function in (\ref{eq:Gaussian_form}) shows that the relationship between the initial (primed) and the final (unprimed) values of $\mathbf{\Gamma}$ and ${\mathbf{Q}}$ are respectively given by,
\begin{equation}
\mathbf{\Gamma}=\left(\mathbf{C}+\mathbf{D}\mathbf{\Gamma'}\right)\left(\mathbf{A}+\mathbf{B}\mathbf{\Gamma'}\right)^{-1},\qquad{\rm{and}}\qquad
{\mathbf{Q}}=\mathbf{Q'}\left(\mathbf{A}+\mathbf{B}\mathbf{\Gamma'}\right)\nonumber.
\end{equation}
Hence, relations in (\ref{eq:evol_comp_wavefront_curv}) are satisfied.
%%%%%%%%%%%%%%
\section{Determination of the principal curvatures and the widths}\label{sec:Determination of the principal curvatures and the widths}
Here, we present the diagonalization procedure of the complex curvature matrix $\mathbf{\Gamma}=\mathbf{\Gamma_R}+i\mathbf{\Gamma_I}$ which is equivalent to the simultaneous diagonalization of the two quadratic forms $\boldsymbol{\xi}^\intercal\mathbf{\Gamma_R}\boldsymbol{\xi}$ and $\boldsymbol{\xi}^\intercal\mathbf{\Gamma_I}\boldsymbol{\xi}$. Some of the steps we present here can be found in \cite{Hildebrand:1952}. We remind that those quadratic forms define ellipses in 2-dimensions. One can refer to the pictorial representation in \cite{Aravind:1989} for the visualization of the diagonalization procedure. We list the steps as follows:
\begin{itemize}

\item [(i)] Start with the positive definite matrix $\mathbf{\Gamma _I}$. Apply the following transformation on the transverse positions on the screen, $\boldsymbol{\xi}$, and on $\mathbf{\Gamma _I}$,
\begin{equation}\label{eq:transform_1}
\boldsymbol{\xi}=\mathbf{F}\,\boldsymbol{\tilde{\xi}}\qquad{\rm{and}}\qquad
\mathbf{\tilde{\Gamma} _I}=\mathbf{F}^\intercal \mathbf{\Gamma _I}\mathbf{F},
\end{equation}
where $\mathbf{F}$ is the modal matrix of $\mathbf{\Gamma _I}$ satisfying $\mathbf{F}^\intercal=\mathbf{F}^{-1}$. Then, $\mathbf{\tilde{\Gamma} _I}$ is diagonal and $\boldsymbol{\xi}^\intercal\mathbf{\Gamma_I}\boldsymbol{\xi}=\boldsymbol{\tilde{\xi}}^\intercal\mathbf{\tilde{\Gamma}_I}\boldsymbol{\tilde{\xi}}$ holds. In the mean time, the real part of the quadratic form becomes $\boldsymbol{\xi}^\intercal\mathbf{\Gamma_R}\boldsymbol{\xi}=\boldsymbol{\tilde{\xi}}^\intercal\mathbf{\tilde{\Gamma}_R}\boldsymbol{\tilde{\xi}}$, where $\mathbf{\tilde{\Gamma} _R}=\mathbf{F}^\intercal \mathbf{\Gamma _R}\mathbf{F}$ is in general not in the diagonal form.

\item [(ii)] Consider the matrix $\mathbf{E}:={\rm{diag}}\left(\lambda_{1}^{-1/2},\lambda_{2}^{-1/2}\right)$,
where $\lambda_{1}$ and $\lambda_{2}$ are the eigenvalues of $\mathbf{\Gamma_I}$. They are real and positive. This is due to $\mathbf{\Gamma_I}$ being a real and positive definite matrix. Thus, one can always define such a matrix as $\mathbf{E}$.
Now, apply the following scaling transformation,
\begin{equation}\label{eq:transform_2}
\boldsymbol{\tilde{\xi}}=\mathbf{E}\,\boldsymbol{\breve{\xi}},\qquad{\rm{and}}\qquad
\mathbf{\breve{\Gamma} _I}=\mathbf{E}^\intercal \mathbf{\tilde{\Gamma} _I}\mathbf{E}. 
\end{equation}
As $\mathbf{\tilde{\Gamma} _I}$ is already the matrix with eigenvalues of $\mathbf{\Gamma _I}$ in its diagonal entities, transformation (\ref{eq:transform_2}) results in $\mathbf{\breve{\Gamma} _I}=\mathbf{I_2}$. Thus, the imaginary part of the quadratic form becomes $\boldsymbol{\tilde{\xi}}^\intercal\mathbf{\tilde{\Gamma}_I}\boldsymbol{\tilde{\xi}}=\boldsymbol{\breve{\xi}}^\intercal\mathbf{\breve{\Gamma}_I}\boldsymbol{\breve{\xi}}=\boldsymbol{\breve{\xi}}^\intercal \boldsymbol{\breve{\xi}}$. In the mean time, the real part of the quadratic form also scales as $\boldsymbol{\tilde{\xi}}^\intercal\mathbf{\tilde{\Gamma}_R}\boldsymbol{\tilde{\xi}}=\boldsymbol{\breve{\xi}}^\intercal\mathbf{\breve{\Gamma}_R}\boldsymbol{\breve{\xi}}$ where $\mathbf{\breve{\Gamma}_R}=\mathbf{E}^\intercal \mathbf{\tilde{\Gamma}_R}\mathbf{E}$.

\item [(iii)] Next step requires the diagonalization of $\mathbf{\breve{\Gamma} _R}$. For this, apply the following transformation,
\begin{equation}\label{eq:transform_3}
\boldsymbol{\breve{\xi}}=\mathbf{G}\boldsymbol{\acute{\xi}}\qquad{\rm{and}}\qquad
\mathbf{\acute{\Gamma} _R}=\mathbf{G}^\intercal \mathbf{\breve{\Gamma} _R}\mathbf{G},   
\end{equation}
where $\mathbf{G}$ is the orthonormal modal matrix of $\mathbf{\breve{\Gamma} _R}$. Then, $\mathbf{\acute{\Gamma} _R}$ is a diagonal matrix and the real part of the quadratic form satisfies $\boldsymbol{\breve{\xi}}^\intercal\mathbf{\breve{\Gamma}_R}\boldsymbol{\breve{\xi}}=\boldsymbol{\acute{\xi}}^\intercal\mathbf{\acute{\Gamma}_R}\boldsymbol{\acute{\xi}}$. In the mean time, the imaginary part of the quadratic form transforms as $\boldsymbol{\breve{\xi}}^\intercal\mathbf{\breve{\Gamma}_I}\boldsymbol{\breve{\xi}}=\boldsymbol{\acute{\xi}}^\intercal\mathbf{\acute{\Gamma}_I}\boldsymbol{\acute{\xi}}$ where $ \mathbf{\acute{\Gamma} _I}=\mathbf{G}^\intercal \mathbf{\breve{\Gamma} _I}\mathbf{G}$. However, as $\mathbf{\breve{\Gamma} _I}$ is a unit matrix and $\mathbf{G}^\intercal =\mathbf{G}^{-1}$ holds, the imaginary part of the quadratic form is unaffected by this transformation. At this point, both the real and the imaginary part of the quadratic form are diagonalized.

\item [(iv)] Apply an additional re-scaling in order to undo the effect of Step (ii), i.e.,
\begin{equation}\label{eq:transform_4}
\boldsymbol{\acute{\xi}}=\mathbf{E}^{-1}\boldsymbol{r},\qquad
\mathbf{\Lambda}_{\Gamma_I}=\left(\mathbf{E}^{-1}\right)^\intercal \mathbf{\acute{\Gamma} _I}\mathbf{E}^{-1},\qquad
\mathbf{\Lambda}_{\Gamma_R}=\left(\mathbf{E}^{-1}\right)^\intercal \mathbf{\acute{\Gamma} _R}\mathbf{E}^{-1}.
\end{equation}
With this transformation we have 
\begin{equation}
\mathbf{\Lambda}_{\Gamma_I}=\mathbf{\tilde{\Gamma}_I}= 
\left[
\begin{array}{c c}
\lambda_1 & \, \, 0 \\
0 & \lambda_2
\end{array}
\right]
=
\left[
\begin{array}{c c}
W_1^{-2} & \, \, 0 \\
0 & W_2^{-2}
\end{array}
\right],
\end{equation}    
meaning the widths, $\{W_1, W_2\}$, of the Gaussian beam are obtained through the eigenvalues of $\mathbf{\Gamma_I}$. The principal curvatures, $\{K_1, K_2\}$, are obtained through
\begin{equation}
\mathbf{\Lambda}_{\Gamma_R}=
\left[
\begin{array}{c c}
K_1 & \, \, 0 \\
0 & K_2
\end{array}
\right],
\end{equation}   
as in transformation (\ref{eq:transform_4}).
\end{itemize}
Following steps (i)-(iv) we can collect the given transformation in each step to define a  total transformation matrix $\mathbf{\mathcal{M}}$ introduced in Section \ref{sec:The principal curvatures and the widths} as
\begin{equation}
\mathbf{\mathcal{M}}=\mathbf{F}\mathbf{E}\mathbf{G}\mathbf{E}^{-1}.
\end{equation}
%%%%%%%%%%%%%%%
\section{Welldefinedness of the Gaussian beam for finite sources}\label{sec:Welldefinedness of the Gaussian beam for finite sources}
In this section, we show that the criteria required for a welldefined Gaussian beam evolution outlined in Section \ref{sec:On the complex wavefront curvature} are fulfilled for the Gaussian beam wave function suggested in (\ref{eq:Gaussian_form}). Those Gaussian beams are represented by a complex curvature matrix, $\mathbf{\Gamma}$, whose real and imaginary parts are respectively given by (\ref{eq:Gamma_R_ext_source}) and (\ref{eq:Gamma_I_ext_source}). 

The first criterion of a welldefined Gaussian wave function is $\,{\rm{\det}\mathbf{Q}}\neq 0\,\,\forall \,\, v$. We satisfy this by having a complex $\mathbf{Q}$ matrix satisfying ${\rm{det}}\mathbf{Q}={\rm{det}}\left(\mathbf{\tilde{A}+i\mathbf{B}}\right){\rm{det}}\mathbf{P'}$ with the initial conditions chosen as in Section \ref{sec:The geometric construction}. The determinant of $\mathbf{Q}$ is never singular as long as the ray bundle transformation matrix $\mathbf{T}$ is symplectic. Namely, one can not have a symplectic transformation, with both $\mathbf{A}$ and $\mathbf{B}$ submatrices being zero. Also, according to the symplectic condition~(\ref{eq:symp_conds}), we have $\mathbf{A}\mathbf{\mathbf{D}}^{\intercal}-\mathbf{B}\mathbf{\mathbf{C}}^{\intercal}=\mathbf{I_2}$. Meaning, the case with $\mathbf{A}=\mathbf{0_2}$ and ${\rm{det}}\mathbf{B}=0$ is not allowed either due to the symplectic symmetries. This fact can be used to argue the welldefinedness from another point of view as well. Recall from (\ref{eq:four_sets}) that the initial values of $\mathbf{A}$ and $\mathbf{B}$ satisfy $\mathbf{A'}=\mathbf{I_2}$ and $\mathbf{B'}=\mathbf{0_2}$. Thus, the initial singularity that exists for the point source intensity, $I' _{P.S.}\propto{\left|{\rm{det}}\mathbf{B'}\right|^{-1}}$, and the wavefront curvature, $\mathbf{\Gamma'}_{P.S.}=\mathbf{D'B'}^{-1}$, does not appear for a Gaussian beam initiated at a finite waist. 

According to the second criterion, the pure imaginary part, $\mathbf{\Gamma_I}={W_0}^2\left[{W_0}^4\mathbf{{A}}\mathbf{{A}}^\intercal+\mathbf{B}\mathbf{B}^\intercal\right]^{-1}$, of the complex curvature matrix should be positive definite. This is required in order to have an exponential decay of intensity in the transverse plane. Note that a matrix multiplied with its transpose is a positive definite matrix as matrices $\mathbf{{A}}\mathbf{{A}}^\intercal$ and $\mathbf{{B}}\mathbf{{B}}^\intercal$ are. Scaling the result with a positive scalar or summing the result with another positive definite matrix does not change this property.  Moreover, the inverse of a positive definite matrix is also positive definite. Then, $\mathbf{\Gamma_I}$ given in (\ref{eq:Gamma_I_ext_source}) is also positive definite.

The third criterion requires both $\mathbf{\Gamma_I}$ and $\mathbf{\Gamma_R}$ to be symmetric matrices. Indeed, it is easy to show that $\mathbf{\Gamma_I}$ is symmetric. Namely, multiplication of a matrix with its transpose results in a symmetric matrix. Moreover, summation of two symmetric matrices is symmetric and inverse of symmetric matrices are also symmetric. Thus, $\mathbf{\Gamma_I}$ is a symmetric matrix. On the other hand, the analysis of the symmetry properties of $\mathbf{\Gamma_R}$ given in (\ref{eq:Gamma_R_ext_source}) is more involved.

Let us first consider the case where the initial width is set to unity, i.e., consider a situation where the waist size is given by $W_0=1$. In that case, the real part of the curvature matrix becomes $\mathbf{\Gamma_R}=\left[\mathbf{C}\mathbf{A}^\intercal+\mathbf{D}\mathbf{B}^\intercal\right]\left[\mathbf{A}\mathbf{A}^\intercal+\mathbf{B}\mathbf{B}^\intercal\right]^{-1}$. We now show that this matrix is symmetric given the ray bundle transformation matrix, $\mathbf{T}$, is symplectic. We use certain results of \cite{deGosson:2006} in our proof. To show this, let us introduce the Iwasawa factorization first.

In \cite{Iwasawa:1949}, Iwasawa showed that any symplectic matrix belonging to $Sp(2, \mathbb{R})$ can be decomposed into three parts belonging to: a nilpotent subgroup, an abelian subgroup and a maximally compact subgroup. The effects of those matrices on the phase space are respectively shearing, magnification and rotation. Accordingly, such transformations are interpreted as lenses, magnifiers and fractional Fourier transformers in optics. 

The Iwasawa decomposition was generalized for higher dimensions in which case it is named as a factorization or a pre/modified-decomposition. For a symplectic matrix $\mathbf{T}\in Sp(4, \mathbb{R})$ like ours, this pre-decomposition is given as \cite{Arvind:1995,Wolf:2004}
\begin{align}\label{eq:Iwasawa}
\mathbf{T}=
\left[
\begin{array}{c|c}
\mathbf{A} & \mathbf{B} \\
\hline
\mathbf{C} & \mathbf{D}
\end{array}
\right]
&=\left[
\begin{array}{c|c}
\mathbf{I_2} & \mathbf{0_2} \\
\hline
\mathbf{-g} & \mathbf{I_2}
\end{array}
\right]
\left[
\begin{array}{c|c}
\mathbf{s} & \mathbf{0_2} \\
\hline
\mathbf{0_2} & \mathbf{s^{-1}}
\end{array}
\right]
\left[
\begin{array}{c|c}
\rm{Re}\,\mathbf{u} & \rm{Im}\,\mathbf{u} \\
\hline
-\rm{Im}\,\mathbf{u} & \rm{Re}\,\mathbf{u}
\end{array}
\right]\nonumber\\
\nonumber\\
&=\qquad\mathbf{L(g)}\qquad \,\,\,\,\,\mathbf{M(s)}\qquad \qquad \,\,\mathbf{F(u)}, 
\end{align}
where $\mathbf{L(g)}$ represents the shearing in phase space, $\mathbf{M(s)}$ represents pure magnifications and $\mathbf{F(u)}$ is a fractional Fourier transformer representing rotation-like effects. Here, the $2\times 2$ matrices that appear in (\ref{eq:Iwasawa}) are given in terms of the sub-blocks of the symplectic matrix, $\mathbf{T}$, as
\begin{align}
%\label{eq:Iwasawa_submatrices}
\mathbf{g}&=-\left(\mathbf{C}\mathbf{A}^{\intercal}+\mathbf{D}\mathbf{B}^{\intercal}\right)\left(\mathbf{A}\mathbf{A}^{\intercal}+\mathbf{B}\mathbf{B}^{\intercal}\right)^{-1},\label{eq:g_lens}\\
\mathbf{s}&=\left(\mathbf{A}\mathbf{A}^{\intercal}+\mathbf{B}\mathbf{B}^{\intercal}\right)^{1/2},\label{eq:s_mag}\\
\mathbf{u}&=\left(\mathbf{A}\mathbf{A}^{\intercal}+\mathbf{B}\mathbf{B}^{\intercal}\right)^{-1/2}\left(\mathbf{A}+i\mathbf{B}\right)\in U(2).\label{eq:u_frac}
\end{align}
Now we realise that the negative of the matrix $\mathbf{g}$ given in (\ref{eq:g_lens}) is equal to the real part, $\mathbf{\Gamma_R}$, of our complex wavefront curvature matrix when the initial width of the beam is set to unity, i.e., $W_0=1$. This matrix is responsible for the lensing effect in the Iwasawa factorization through $\mathbf{L(g)}$ which also obeys the symplectic symmetry conditions given in (\ref{eq:symp_conds}). When those symplectic symmetry conditions are imposed, we see that four of those conditions are trivial. The remaining two conditions impose  $\mathbf{g}=\mathbf{g}^\intercal$. This means that $\mathbf{\Gamma_R}$ is symmetric when $W_0=1$. 

A similar connection with the Iwasawa factorization can be formed with the imaginary part of the complex curvature matrix when the initial width is set to one. Namely, $\mathbf{\Gamma_I}={W_0}^2\left[{W_0}^4\mathbf{A}\mathbf{A}^\intercal+\mathbf{B}\mathbf{B}^\intercal\right]^{-1}$ becomes equal to $\mathbf{s}^{-2}$ when $W_0=1$, where $\mathbf{s}$ is given by (\ref{eq:s_mag}). Note that $\mathbf{s}$ is also symmetric. This can be proven either though a similar argument we presented for $\mathbf{g}$, or more trivially, it being composed of addition of two symmetric matrices. In summary, the real part of the curvature matrix is responsible for the shearing effect in the phase space and the imaginary part is responsible for the phase space magnifications through the matrix $\mathbf{M}(s)$.

Now, let us return to the generic case when the initial widths are not necessarily set to unity. In that case, $\mathbf{\Gamma_I}$ is still a symmetric matrix due to the aforementioned arguments. However, in order to show that $\mathbf{\Gamma_R}$ is symmetric even when $W_0\neq1$, we need to take into account the following properties:
\begin{itemize}
\item [(i)] $\mathbf{\Gamma_R}=\left[{W_0}^4\mathbf{C}\mathbf{A}^\intercal+\mathbf{D}\mathbf{B}^\intercal\right]\left[{W_0}^4\mathbf{A}\mathbf{A}^\intercal+\mathbf{B}\mathbf{B}^\intercal\right]^{-1}$ is a $2\times 2$ matrix.
\item [(ii)] $\left(\mathbf{C}\mathbf{A}^{\intercal}+\mathbf{D}\mathbf{B}^{\intercal}\right)\left(\mathbf{A}\mathbf{A}^{\intercal}+\mathbf{B}\mathbf{B}^{\intercal}\right)^{-1}$ is symmetric.
\item [(iii)] Even though $\mathbf{C}\mathbf{A}^\intercal$ and $\mathbf{D}\mathbf{B}^\intercal$ are not necessarily symmetric, they can be written as multiplication of symmetric matrices, i.e.,  
\begin{equation}
\mathbf{C}\mathbf{A}^\intercal=\left( \mathbf{C}\mathbf{A}^{-1}\right)\left(\mathbf{A}\mathbf{A}^\intercal\right),\qquad{\rm{and}}\qquad
\mathbf{D}\mathbf{B}^\intercal=\left( \mathbf{D}\mathbf{B}^{-1}\right)\left(\mathbf{B}\mathbf{B}^\intercal\right).
\end{equation}
Here, $\mathbf{C}\mathbf{A}^{-1}$ is symmetric due to the symplectic condition $\mathbf{A}^\intercal\mathbf{C}=\left(\mathbf{A}^\intercal\mathbf{C}\right)^\intercal$ and $\mathbf{D}\mathbf{B}^{-1}$ is symmetric due to the symplectic condition $\mathbf{B}^\intercal\mathbf{D}=\left(\mathbf{B}^\intercal\mathbf{D}\right)^\intercal$ in (\ref{eq:symp_conds}). Matrices $\mathbf{A}\mathbf{A}^\intercal$ and $\mathbf{B}\mathbf{B}^\intercal$ are obviously symmetric due to them being composed of multiplication of matrices with their transposes.
\end{itemize}
Finally, by taking (i), (ii) and (iii) into account, it can be shown with simple algebra that $\mathbf{\Gamma_R}$ is symmetric for arbitrary values of the initial width.
%%%%%%%%%%%%%%%%%%%%%%%%%%%%%%%%%%%%%%%%%%%%%%%%%%%
\end{document}